\documentclass[12pt,twoside,english]{extarticle}
\usepackage[T1]{fontenc}
\usepackage[latin9]{inputenc}
\usepackage{geometry}
\usepackage{amsmath}
\usepackage{amstext}
\usepackage{amssymb}
\usepackage{graphicx}
\usepackage{esint}
\usepackage{url}
\usepackage{epstopdf}
\usepackage{dcolumn}
\newcolumntype{d}{D{.}{.}{-1} } 

\usepackage{xspace}

\makeatletter

\providecommand{\tabularnewline}{\\}

\newcommand{\ninej}[9]
{
  \left\{ \begin{array}{ccc} #1 & #2 & #3 \\
    #4 & #5 & #6 \\ #7 & #8 & #9 \end{array} \right\}
}
\newcommand{\sixj}[6]
{
  \left\{ \begin{array}{ccc} #1 & #2 & #3 \\
    #4 & #5 & #6 \end{array} \right\}
}

\makeatother

\usepackage{babel}

\begin{document}

\title{Long-range interactions \\ between ultracold atoms and molecules}

\author{Maxence Lepers and Olivier Dulieu \\ Laboratoire Aim\'e Cotton, CNRS, Universit\'e Paris-Sud, ENS Paris-Saclay, \\ Universit\'e Paris-Saclay, 91405 Orsay, France}

\maketitle

\begin{abstract}
The term \textit{long-range interactions} refers to electrostatic and magnetostatic potential energies between atoms and molecules with mutual distances ranging from a few tens to a few hundreds Bohr radii. The involved energies are much smaller than the usual chemical bond energies. However, they are comparable with the typical kinetic energies of particles in an ultracold gas ($T\ll 1K$), so that the long-range interactions play a central role in its dynamics. The progress of research devoted to ultracold gases shed a new light on the well-established topic of long-range interactions, because: (i) the interacting atoms and molecules can be prepared in a well-defined quantum (electronic, vibrational, rotational, fine or hyper-fine), ground or excited level; and (ii) long-range interactions can be tailored at will using external electromagnetic fields.

In this chapter, we present the essential concepts and mathematical relations to calculate long-range potential energies. We start with deriving the multipolar expansion of the electrostatic interaction energy between classical charge distributions, both in Cartesian coordinates for pedagogical purpose, and in spherical coordinates for practical use. Then we combine multipolar expansion and quantum perturbation theory, to obtain the general first- and second-order energy corrections, including the well-known van der Waals energy. We consider two central examples in the current context of ultracold gases: (i) a pair of alkali-metal atoms and (ii) a pair of alkali-metal heteronuclear diatomic molecules submitted to an electric field. We highlight the key role of the total angular momenta of the interacting particles and of the complex, irrespective of their electronic or nuclear, orbital or spin nature.
\end{abstract}

\section{Introduction}
\label{sec:introduction}

For historical reasons\footnote{It is worthwhile to quote here the excellent historical survey of intermolecular forces proposed by Ilya G. Kaplan \cite{kaplan2006}.}, intermolecular interactions are often referred to as "van der Waals forces", a well-known name for students familiar with classical thermodynamics. Indeed, Johannes Diderick van der Waals, from the University of Leiden, was the first scientist to propose (in 1873) a model \cite{vanderwaals1873} describing the modification to the ideal gas law -- \textit{i.e.}~a gas of non-interacting neutral particles -- taking into account the interactions between particles in an effective way. He was awarded the 1910 Nobel Prize in Physics for this work. For one mole of a gas of volume $V$, temperature $T$, and pressure $P$, the well-known law $PV = \cal{R} T$ has to be modified into $(P+a/V^2)(V-b)=\cal{R}T$ to reproduce the experimental observations (where $\cal{R}$ is the ideal gas constant). The effective volume available to the particles in a container is reduced due to their own volume, parametrized with $b$: in other words the particles feel, at very short distances, a repulsive force mimicking hard spheres.  The $a$ parameter holds for the --assumed-- isotropic attractive interaction between the particles.

\begin{figure}
\begin{centering}
\includegraphics[scale=0.6]{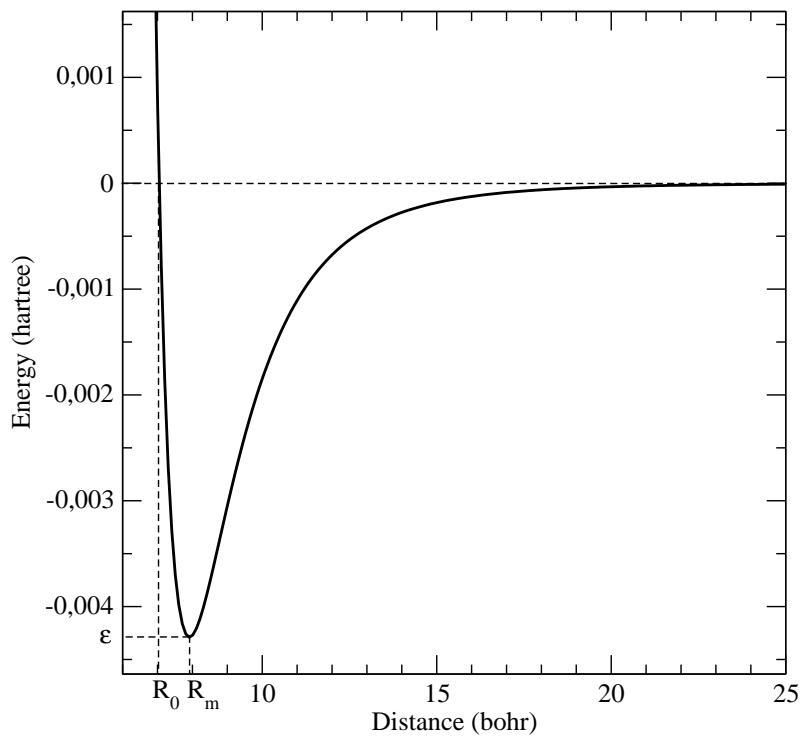}
\par\end{centering}
\caption{Example of a Lennard-Jones potential energy curve of equation (\ref{eq:LJ}) for a realistic neutral diatomic molecule. Values of parameters are $\epsilon=-0.00429$~hartrees, and $R_0=7.05$~bohrs. The minimum of the potential well is found at $R_m=7.88$~bohr.}
\label{fig:LJ}
\end{figure}

Later, as novel experimental methods became available (such as spectroscopy, scattering, cristallography,...) a large activity was devoted to modeling the intermolecular potential energy through effective expressions depending on the intermolecular distance $R$. The pioneering work of John Edward Lennard-Jones \cite{lennard-jones1924} resulted in the well known expression:
\begin{equation}
  V^{\textrm{LJ}}(R) = 4\epsilon
  \left[ \left( \frac{R_0}{R}\right)^{12}-\left( \frac{R_0}{R}\right)^{6}\right]
  \label{eq:LJ}
\end{equation}
where $\epsilon$ is the minimal potential energy in absolute value, and $R_0$ the distance where the two terms of equation \eqref{eq:LJ} compensate each other (see Fig.~\ref{fig:LJ}).

The expression of equation (\ref{eq:LJ}) is enlightening in several respects:
\begin{itemize}
\item it reveals that --in most situations--, the intermolecular potential energy displays a well;
\item the position $R_m$ of this minimum roughly defines the limit between the short-range ($R<R_m$) repulsive and the long-range ($R>R_m$) attractive domains of distances;
\item the long-range part of the interaction varies as $R^{-6}$, so that such an interaction will be indeed qualified as "van der Waals" interaction, as it will be explained in the course of this chapter.
\end{itemize}

Obviously, the intermolecular potential energy vanishes for infinite distances. But as usual in physics, the distance beyond which it can be neglected is defined by comparison with some characteristic quantity of the considered problem or system. Let us consider a pair of molecules approaching each other from infinity. If their relative kinetic energy has the same magnitude as the well depth ($\epsilon$ in Fig.\ref{fig:LJ}) then their collision will not be sensitive to the long-range tail of the potential energy. In contrast, if they have a negligible relative kinetic energy compared to $\epsilon$, their approach will be mainly influenced by this long-range tail: this is the purpose of studying ultracold gases and ultracold collisions!

\begin{figure}
\begin{centering}
\includegraphics[scale=1.2]{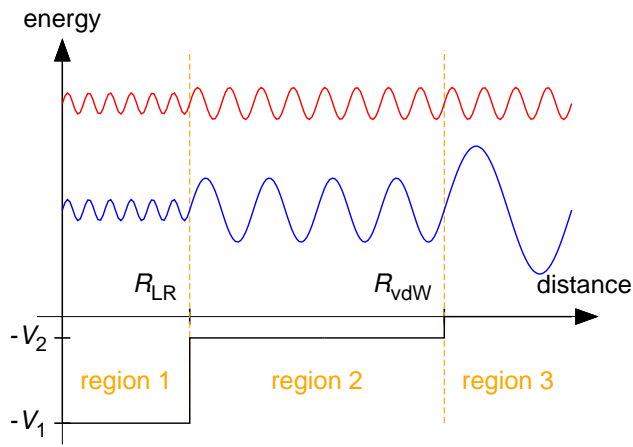}
\par\end{centering}
\caption{Schematic view of the potential energy of a pair of colliding particles with the radial wavefunctions of their relative motion. The predominance of long-range interactions in ultracold collisions occurs in the region $R_{\textrm{LR}}<R<R_{\textrm{vdW}}$. }
\label{fig:squarewell}
\end{figure}

The scheme reported in Figure \ref{fig:squarewell} depicts a schematic one-dimensional quantum representation of the colliding particles involving relevant characteristic distances, where $R$ is the distance between the center-of-mass of each particle. The interaction potential energy is schematized as a series of squared steps. No interaction is present beyond a van der Waals distance $R_{vdW}$ (region 3), while the strongest interaction takes place below a given distance $R_{LR}$ (region 1). In between, the long-range tail of the interaction dominates over region 2 ($R_{\textrm{LR}}<R<R_{\textrm{vdW}}$). Two examples of the radial wave functions associated to the relative motion of the particles are displayed. For high colliding energy (red curve) the local ($R$-dependent) de Broglie wavelength of the wave function is almost insensitive to the nature of the long-range interaction, and is fully determined by the short-range part. In contrast, at low collision energy (blue curve), the wave function is sensitive to any variation of the mutual interaction. The approach of the particles is fully controlled by the long-range interactions: at large distances the interaction is weak, and the particles spend a long time in region 2, resulting in large probability density.

In real quantum systems at ultracold energies, there are generally many more degrees of freedom besides $R$ (vibrations and rotations of the molecules, relative rotation), and the interaction energy is of course continuous with respect to all relevant coordinates. Nevertheless, it is possible to define the characteristic distances $R_{\textrm{LR}}$ and $R_{vdW}$. It is easy to realize that long-range interactions are well defined when the interacting particles conserve their identity, \textit{i.e.} when their electronic cloud do not overlap. The renowned Canadian molecular spectroscopist Robert J.~LeRoy proposed the definition $R_{\textrm{LR}}=2 \left[\langle r_A^2 \rangle^{1/2}+ \langle r_B^2 \rangle^{1/2} \right]$, where $r_A$ and $r_B$ are the mean radii of the particles A and B \cite{leroy1974}.

In the well-known case of an attractive van der Waals interaction varying as $C_6/R^6$ (see subsection \ref{ssec:pert-2nd} for details), where $C_6<0$ is referred to as the van der Waals coefficient, one can define the associated energy scale \cite{jones2006} 
\begin{equation}
E_{vdW}= \frac{1}{2} \left(\frac{\hbar^2}{2\mu R_{vdW}^2}\right)
\label{eq:EvdW}
\end{equation}
where $\mu$ is the reduced mass of the interacting particles. The length scale is defined from dimensional analysis by \cite{jones2006}
\begin{equation}
R_{vdW}= \frac{1}{2} \left(\frac{2\mu |C_6|}{\hbar^2}\right)^{1/4}
\label{eq:RvdW}
\end{equation}
Note that this characteristic distance, relevant in the context of ultracold collisions, should not be mixed up with the well-known van der Waals radius (see for instance \cite{bondi1964}) entering in the interpretation of the van der Waals equation of state for gases. Furthermore, a definition involving slightly different numerical coefficients has been proposed in Ref.~\cite{gribakin1993}. This distance characterizes the range of the van der Waals interaction. As an illustration, we reproduce in Table \ref{tab:vdW} the values of $E_{vdW}$ and $R_{vdW}$ displayed by Chin \textit{et al.} \cite{chin2010} for several atomic species used in experiments at ultracold temperatures. It is striking to see that $E_{vdW}$ is larger than the typical kinetic energies reached with ultracold gases. When $R \gg R_{vdW}$, the interaction between the particles is vanishingly small, so that the characteristic length is the de Broglie wavelength $2\pi/k$ determined by the initial (weak) relative kinetic energy $E=(\hbar k)^2/(2\mu)$. When $R<R_{vdW}$, the de Broglie wavelength is well defined by the local momentum $\hbar k(R)=\sqrt{2\mu(E-C_6/R^6)}$. As it will be invoked in other chapters, the energy $E_{vdW}$ determines the kinetic energy threshold below which the connection between the long-range and the short-range parts of the relative wave function of the colliding particles motion around $R_{vdW}$ cannot be described semi-classically, and is rather governed by purely matter-wave concepts like quantum reflection.

\begin{table}
\begin{center}
\begin{tabular}{cccccc}
atom      &mass      &$C_6$   &$R_{vdW}$&$E_{vdW}/k_B$ \\ 
          &(a.m.u)   &(a.u.)  &(a.u.)   &mK             \\ \hline \hline
$^{6}$Li  &6.0151223 &-1393.39&31.26    &29.47\\
$^{23}$Na &22.9897680&-1156   &44.93    &3.732\\
$^{40}$K  &39.9639987&-3897   &64.90    &1.029\\
$^{40}$Ca &39.962591 &-2221   &56.39    &1.363\\
$^{87}$Rb &86.909187 &-4698   &82.58    &0.2922\\
$^{88}$Sr &87.905616 &-3170   &75.06    &0.3497\\
$^{133}$Cs&132.905429&-6860   &101.0    &0.1279\\ \hline \hline
\end{tabular}
\end{center}
\caption{Masses $m$, $C_6$ coefficients, van der Waals lengths $R_{vdW}$ and energies $E_{vdW}$ for several atomic species reported in Ref.~\cite{chin2010}, and references therein. The quantities are expressed in units commonly used in atomic and molecular physics. For distances, 1~a.u. is one Bohr radius $a_0$, and one Hartree $E_h$ for energies. Consequently, 1 a.u. for $C_6$ is equivalent to  1$E_h a_0^6$. 1 atomic-mass unit (a.m.u) equals $1/12$ the mass of a $^{12}$C atom.}
\label{tab:vdW}
\end{table}

Up to now we have not yet invoked the origin of van der Waals interaction. This generic name actually refers to several phenomena --all associated to the name of an outstanding scientist--, induced by different kinds of interactions between electric charge distributions, but all leading to the same $1/R^6$ variation with their mutual distance $R$:
\begin{itemize}
\item the London interaction, or \textit{dispersion} interaction \cite{london1930,london1937} relies on the instantaneous charge distribution of a neutral composite particle which is not uniform, as electrons are permanently moving around the nuclei, thus leading to instantaneous dipoles. When approaching each other, two such distributions constantly adjust their instantaneous dipoles by inducing a dipole on each other, resulting into an attractive force. This is the only van der Waals interaction that can occur between neutral atoms, which cannot have permanent electric dipoles.
\item the Debye interaction, or \textit{induction} interaction \cite{debye1920,debye1921} concerns a molecule \textit{with a permanent dipole moment} in free space inducing a dipole moment on another \textit{apolar} particle. In this respect it describes an effect similar to the London interaction, but it is \textit{induced} by a permanent dipole moment instead of an instantaneous one. 
\item the Keesom interaction \cite{keesom1921a,keesom1921b} occurs between two molecules that possess a permanent dipole moment. Just like for two magnets, the interaction between two polar molecules is in principle \textit{anisotropic}, that is to say attractive or repulsive according to the respective orientation of the two molecules. However at room temperatures, and especially in liquids, molecules rotate due to thermal agitation, and all their relative orientations are equally probable. The resulting potential energy is attractive and scales with temperature as $T^{-1}$. In the ultracold regime, where $T\to 0$, the latter picture is obviously not valid any more: firstly because thermal agitation is almost reduced to zero, and secondly because the rotational motion of the molecules is \textit{quantized}.
\end{itemize}

The description above already suggests that long-range interactions between composite particles depend on the --permanent or instantaneous-- inhomogeneities of their charge distributions. When $T \rightarrow 0$ other interactions than the van der Waals ones can occur, varying as $C_n/R^n$ with $n \ne 6$. This will be the central objective of the present chapter, aiming at systematically deriving the analytical expressions of these interactions. Indeed, as suggested above, the intermolecular interactions have a small influence in gases at room temperature, as the related kinetic energy of the particles largely exceeds the potential energy of their long-range interactions. With the achievement of ultracold gases in the laboratory, their dynamics is dominated by the long-range interactions. In many applications to particular systems, the main challenge will be the precise evaluation of the intensity of such interactions, related to the $C_n$ coefficient above. Pioneering works in this matter includes Refs. \cite{fontana1961a, fontana1961b, fontana1962, buckingham1965, meath1966, chang1967, hirschfelder1967, gray1968, meath1968, tang1969, gray1976a, gray1976b}. Two recent books  \cite{kaplan2006, stone1996} and a review \cite{tao2013} are also devoted to this broad area of intermolecular forces.

The basic principle of the formalism developed in this chapter relies on the \textit{perturbative} treatment of the interaction between two distant electric charge distributions A and B -- or two ensembles of electric charges occupying two bounded regions of space --, in free space or in the presence of an external electric field. They are placed at a fixed distance $R$ between their respective center-of-mass, \textit{i.e.}~at a distance much larger than their individual extension. When they do not interact, their energy spectrum is described by the Hamiltonian operators $\hat{\mathrm{H}}_A$ and $\hat{\mathrm{H}}_B$, with eigenvalues and eigenvectors $\{E_q^{(0)}(A),\, |\Psi_q^{(0)}(A)\rangle\}$ and $\{E_q^{(0)}(B),\, |\Psi_q^{(0)}(B)\rangle\}$ respectively, where $q$ is an integer labeling the energy level, and the superscript $(0)$ standing for unperturbed ($0^{\textrm{th}}$-order) levels. The potential energy of interaction between A and B is associated with the operator $\hat{\textrm{V}}$, whose matrix elements are {}``small" in the sense of perturbation theory, \textit{i.e.}~$\langle\hat{\textrm{V}}\rangle \equiv \langle \Psi_q^{(0)} |\hat{\mathrm{V}}| \Psi_q^{(0)} \rangle\ll |E_{q+1}^{(0)}(A)-E_q^{(0)}(A)|,\, |E_{q+1}^{(0)}(B)-E_q^{(0)}(B)|$. We use $\hat{\textrm{V}}$ as a generic notation to recall the essential results of perturbation theory; in the next sections, we will give explicit forms of $\hat{\textrm{V}}$.

The unperturbed ($0^{\textrm{th}}$-order) states of the A$+$B complex are the eigenstates of the Hamiltonian $\hat{\mathrm{H}}_0 = \hat{\mathrm{H}}_A + \hat{\mathrm{H}}_B$. Its eigenvalues are the sum of individual eigenvalues $E_q^{(0)} = E_q^{(0)}(A) + E_q^{(0)}(B)$, and its eigenvectors are the outer products of individual eigenvectors |$\Psi_q^{(0)}\rangle = |\Psi_q^{(0)}(A)\rangle \otimes |\Psi_q^{(0)}(B)\rangle$. In this chapter, we will calculate the $1^{\textrm{st}}$-order $E_q^{(1)}$ and $2^{\textrm{nd}}$-order $E_q^{(2)}$ energy corrections generated by $\hat{\mathrm{V}}$.

We recall that for a non-degenerate level $q$, the $1^{\textrm{st}}$-order correction to its energy is the diagonal matrix element of the perturbation operator,
\begin{equation}
  E_q^{(1)} = \langle \Psi_q^{(0)} 
    |\hat{\mathrm{V}}| \Psi_q^{(0)} \rangle ,
\end{equation}
while the $2^{\textrm{nd}}$-order energy correction implies the same matrix element but for all other levels $r \ne q$
\begin{equation}
  E_q^{(2)} = -\sum_{r\ne q}
    \frac{\left|\langle\Psi_r^{(0)}|
      \hat{\mathrm{V}} |\Psi_q^{(0)} \rangle \right|^2}
      {E_r^{(0)}-E_q^{(0)}}
   = -\sum_{r\ne q} \frac{
    \langle\Psi_q^{(0)}|\hat{\mathrm{V}}|\Psi_r^{(0)}\rangle
    \langle\Psi_r^{(0)}|\hat{\mathrm{V}}|\Psi_q^{(0)}\rangle}
    {E_r^{(0)}-E_q^{(0)}} ,
  \label{eq:Eq2}
\end{equation}
where we assume that the matrix elements $\langle\Psi_r^{(0)}|\hat{\mathrm{V}}|\Psi_q^{(0)}\rangle$ are real numbers.

If the level $q$ is $n$-fold degenerate, we introduce the projection operator $\hat{\mathrm{P}}_q$ onto the subspace of degeneracy associated with $q$, built from the degenerate unperturbed eigenvectors $|\Psi_{q,i}^{(0)}\rangle$, $1\le i\le n$
\begin{equation}
  \hat{\mathrm{P}}_q = \sum_{i=1}^n
    |\Psi_{q,i}^{(0)}\rangle \langle\Psi_{q,i}^{(0)}| .
\end{equation}
The first-order energy corrections $E_{q,i}^{(1)}$ are the eigenvalues of the operator $\hat{\mathrm{P}}_q \hat{\mathrm{V}} \hat{\mathrm{P}}_q$, which is the restriction of the perturbation operator to the subspace of degeneracy associated with $q$. As for the second-order energies, they are the eigenvalues of the effective operator \cite{landau1967}
\begin{equation}
  \hat{\mathrm{W}} = -\sum_{r\ne q} \frac{
    \hat{\mathrm{P}}_q \hat{\mathrm{V}}
    |\Psi_{r}^{(0)}\rangle \langle\Psi_r^{(0)}|
    \hat{\mathrm{V}} \hat{\mathrm{P}}_q}
    {E_r^{(0)}-E_q^{(0)}} \,.
  \label{eq:W_AB}
\end{equation}
The operator $\hat{\mathrm{W}}$ is effective in the sense that it depends on the unperturbed energy spectrum.

The treatment above is general as no indication has been given yet on the way to define the $|\Psi_{q,i}^{(0)}(A)\rangle$ and $|\Psi_{q,i}^{(0)}(B)\rangle$ eigenvectors. If the charge distributions A and B are (possibly Rydberg \cite{gallagher1994}) atoms or rotating molecules, we can assume, without loss of generality, that their quantum levels are at least determined by their total angular momenta $\mathbf{J}_A$ and $\mathbf{J}_B$. We will see throughout this chapter that, whatever their orbital or spin, electronic or nuclear characters, those angular momenta have a crucial influence on the very nature of the long-range interactions. In other words, the mean values of $\mathbf{J}_A$ and $\mathbf{J}_B$ determine which interaction vanishes ($C_n=0$), attracts ($C_n<0$) or repels ($C_n>0$) the two atoms or molecules. The scope of this chapter is the detailed study of the influence of these angular momenta, while the discussion of the precise calculation of $C_n$ coefficients will be only briefly addressed.

The rest of the chapter is organized as follows. In section \ref{sec:multi-exp} the expressions of the $\hat{\mathrm{V}}$ perturbation operators are obtained through a multipolar expansion expressed in cartesian coordinates (Section \ref{ssec:mult-cartes}), useful to understand the physical origin of the various terms, and in spherical coordinates (Section \ref{ssec:mult-spher}), providing an expression more adapted to quantum systems with well-defined angular momenta. These expressions are inserted in the above perturbative expansion in Section \ref{sec:pert-calc}. The matrix elements of $\hat{\mathrm{V}}$ are calculated first in free space (Section \ref{ssec:pert-qlm}), and then in the presence of an external static electric field (Section \ref{ssec:efield}) to evaluate the relevant Stark shift of the levels. The $1^{\textrm{st}}$-order and the $2^{\textrm{nd}}$-order corrections to the unperturbed level energies are evaluated in Sections \ref{ssec:pert-1st} and \ref{ssec:pert-2nd}. We illustrate these developments by applying them to an exemplary case of the research on ultracold atomic gases: the long-range interaction between two alkali-metal atoms (Section \ref{ssec:pert-alk-ato}) which possess a simple structure with a single valence electron and thus a spin 1/2. At the publication date of the present book, diatomic molecules composed of two different alkali-metal atoms are at the heart of the investigations on ultracold molecules: they are readily created by associating ultracold atoms using advanced laser and magnetic field technologies \cite{carr2009, dulieu2009}. They exhibit a permanent electric dipole moment in their own frame, and they are closed-shell molecules (zero spin). Therefore they are appealing for a complete calculation of their long-range interaction in the context of this tutorial (Section \ref{sec:hetero}). Two complementary points of view are addressed: the one of the body-fixed frame (Section \ref{ssec:mol-bf}), and the one of the laboratory-fixed frame (Section \ref{ssec:mol-sf}) in which the mutual rotation of the two molecules can be easily implemented. Several extensions of the present formalism are discussed in the concluding section (Section \ref{sec:conclusion}) in terms of both the methodology as well as the treatment of more complex systems (open-shell molecules, long-range interaction with a surface,...).

\section{Multipolar expansion of the potential energy between two distant charge distributions}
\label{sec:multi-exp}

In this section we consider two distant charge distributions A and B, which interact through Coulombic forces that give rise to an electrostatic potential energy. Assuming that the charges move along closed orbits inside each distribution, creating so-called stationary currents, one can also define a magnetostatic potential energy. In atomic and molecular systems, the latter roughly scales as $\alpha^{2}\approx5\times10^{-5}$ times the electrostatic energy ($\alpha$ being the fine-structure constant), so that it will be most often neglected except in some particular situations which will be mentioned at the end of this section. From now on, the interaction potential energy $V_{AB}$ relies on classical point-like charges. In the next sections, using the equivalence principle of quantum mechanics, we will transform $V_{AB}$ into an operator $\hat{\mathrm{V}}_{AB}$, reflecting the perturbation exerted on each other by the two interacting atoms or molecules.

\begin{figure}
  \begin{centering}
  \includegraphics[scale=0.6]{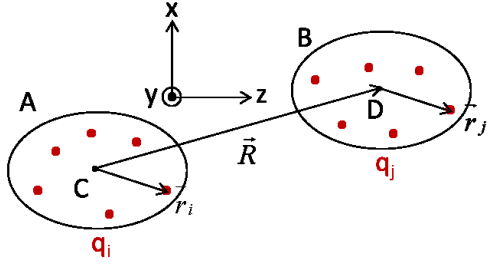}
  \par\end{centering}
  \caption{\label{fig:distr}Scheme of the charge distributions A and B, and of the coordinate system $xyz$.}
\end{figure}

Figure \ref{fig:distr} depicts the studied system. The positions $\mathbf{r}_{i}$ ($\mathbf{r}_{j}$) of the charges $q_{i}$ ($q_{j}$) in A (B) are given with respect to the center of mass C (D) of the distribution A (B). The vector joining C and D is denoted $\mathbf{R}$. For convenience, we chose C to be the origin of our $xyz$ coordinate frame, and we assume C to be fixed in the laboratory. Note that this choice is made for convenience, the interaction energy does not depend on it.

The total electrostatic potential energy $V_{\mathrm{tot}}$ of the system A$+$B, hereafter denoted as the \textit{complex}, is the sum of potential energies between all pairs of charges
\begin{equation}
V_{\mathrm{tot}} = \frac{1}{4\pi\epsilon_{0}} \left( 
  \sum_{\substack{i,i'\in A \\ i\ne i'}}
    \frac{q_{i}q_{i'}}{\left|\mathbf{r}_{i'}-\mathbf{r}_{i}\right|}
  + \sum_{\substack{j,j'\in B \\ j\ne j'}}
    \frac{q_{j}q_{j'}}{\left|\mathbf{r}_{j'}-\mathbf{r}_{j}\right|}
  + \sum_{\substack{i\in A \\ j\in B}}
  \frac{q_{i}q_{j}}
    {\left|\mathbf{R}+\mathbf{r}_{j}-\mathbf{r}_{i}\right|}\right),
  \label{eq:mult-vtot}
\end{equation}
where $\epsilon_{0}$ is the vacuum permittivity. The first two terms correspond to the electrostatic potential energy inside each charge distribution. The third term is the interaction potential energy between the two charge distributions. which will be called \textit{interaction energy} and written $V_{AB}(\mathbf{R})$, since it explicitly depends on $\mathbf{R}$. The aim of this section is to rewrite $V_{AB}(\mathbf{R})$, considering that the two charge distributions are so far away from each other that they do not overlap, \textit{i.e.}
\begin{equation}
  \left|\mathbf{R}\right|\gg\left|\mathbf{r}_{i}\right|,\left|\mathbf{r}_{j}\right|\,\forall i\in A,j\in B.\label{eq:mult-cond}
\end{equation}
This will enable us to expand $V_{AB}(\mathbf{R})$ as a series of terms proportional to powers of $\left|\mathbf{R}\right|^{-1}$, and depending on global properties of A and B: their \textit{multipole moments}.

In Section \ref{ssec:mult-cartes}, we derive its expression in Cartesian coordinates; in Section \ref{ssec:mult-spher} we use spherical coordinates in order to derive the most suitable expressions for practical calculations with atoms and molecules.

\subsection{The interaction energy in Cartesian coordinates}
\label{ssec:mult-cartes}

The calculations in Cartesian coordinates are performed in three steps, following the lines of reference \cite{cohen-tannoudji1973}, Chaps.~X and XI:
\begin{itemize}
  \item taking only the distribution A, and calculating the electrostatic potential that A exerts at a distant point;
  \item taking only distribution B, and calculating its interaction energy with an external weakly-varying electrostatic potential;
  \item considering that B is subject to the electrostatic potential created by A.
\end{itemize}

\subsubsection{Electrostatic potential created by the distribution A at a distant point}

We consider a point M and we call $\mathbf{R}$ the vector joining C and M. The electrostatic potential $\Phi_{A}(\mathbf{R})$ created by A at M reads
\begin{equation}
  \Phi_{A}(\mathbf{R}) = \frac{1}{4\pi\epsilon_{0}}
    \sum_{i\in A}\frac{q_{i}}{\left|\mathbf{R}-\mathbf{r}_{i}\right|}\,,
  \label{eq:mult-phiA}
\end{equation}
where we assume that $\Phi_{A}(\mathbf{R})=0$ when $\left|\mathbf{R}\right|\to+\infty$. The point M is located so far from the distribution A, namely $\left|\mathbf{R}\right|\gg\left|\mathbf{r}_{i}\right|$, $\forall i\in A$, that we can express $\Phi_{A}(\mathbf{R})$ with a Taylor expansion. To that end we write
\begin{equation}
  \left|\mathbf{R}-\mathbf{r}_{i}\right|
    = \sqrt{\left(\mathbf{R}-\mathbf{r}_{i}\right)^{2}}
    = R\sqrt{1-2\frac{\mathbf{u}\cdot\mathbf{r}_{i}}{R}
      +\frac{r_{i}^{2}}{R^{2}}},
  \label{eq:mult-R-ri}
\end{equation}

where we set $\left|\mathbf{r}_{i}\right|=r_{i}$ and $\mathbf{R}\equiv R\mathbf{u}$, $\mathbf{u}$ being a unit vector. We also recall that
\begin{equation}
  \frac{1}{\sqrt{1-x}}=1+\frac{x}{2}+\frac{3x^{\text{2}}}{8}+\mathcal{O}(x^{3})\,.
  \label{eq:mult-R-ri-Tayl}
\end{equation}

In order to apply equation (\ref{eq:mult-R-ri-Tayl}) properly, we note the following point. If we consider that the ratio $r_{i}/R$ is on the order of a small quantity $\varepsilon$, it means that in equation (\ref{eq:mult-R-ri}), $\mathbf{u}\cdot\mathbf{r}_{i}/R$ is on the order of $\varepsilon$ and $r_{i}^{2}/R^{2}$ on the order of $\varepsilon^{2}$. Therefore to calculate equation (\ref{eq:mult-R-ri}) up to $O(\varepsilon^{2})$, we apply equation (\ref{eq:mult-R-ri-Tayl}) with $x=2\frac{\mathbf{u}\cdot\mathbf{r}_{i}}{R}+\frac{r_{i}^{2}}{R^{2}}$, but drop the terms scaling as $\varepsilon^{3}$ and $\varepsilon^{4}$. This gives
\begin{equation}
\frac{1}{\left|\mathbf{R}-\mathbf{r}_{i}\right|}=\frac{1}{R}+\frac{\mathbf{u}\cdot\mathbf{r}_{i}}{R^{2}}+\frac{3\left(\mathbf{u}\cdot\mathbf{r}_{i}\right)^{2}-r_{i}^{2}}{2R^{3}}+\mathcal{O}\left(\frac{r_{i}^{3}}{R^{4}}\right).\label{eq:mult-R-ri-2}
\end{equation}
Finally, plugging equation (\ref{eq:mult-R-ri-2}) into equation (\ref{eq:mult-phiA}) and distributing the sum on charges $q_{i}$ in each term, we obtain
\begin{equation}
\Phi_{A}(\mathbf{R})=\frac{1}{4\pi\epsilon_{0}}\left(\frac{\sum_{i}q_{i}}{R}+\frac{\sum_{i}\mathbf{u}\cdot\left(q_{i}\mathbf{r}_{i}\right)}{R^{2}}+\frac{\sum_{i}q_{i}\left[3\left(\mathbf{u}\cdot\mathbf{r}_{i}\right)^{2}-r_{i}^{2}\right]}{2R^{3}}\right)+\mathcal{O}\left(R^{-4}\right).\label{eq:mult-phiA-2}
\end{equation}

In equation (\ref{eq:mult-phiA-2}) the electrostatic potential $\Phi_{A}(\mathbf{R})$ appears as a sum of terms proportional to inverse powers of $R$. Each term depends on a property of the whole charge distribution A: for the term scaling as $R^{-1}$, on the total charge
\begin{equation}
  q(A) = \sum_{i\in A}q_{i}\,;
\end{equation}
for the term scaling as $R^{-2}$, on the dipole moment
\begin{equation}
  \mathbf{d}(A)=\sum_{i\in A}q_{i}\mathbf{r}_{i}\,,
  \label{eq:mult-dip}
\end{equation}
and more precisely its component $d_{u}(A)=\mathbf{u\cdot}\mathbf{d}(A)$ along the $\mathbf{u}$ direction. Treating the $R^{-4}$ term is slightly more involved: to that end, we first expand the square of the scalar product in equation (\ref{eq:mult-phiA-2}) as
\begin{equation}
  \left(\mathbf{u}\cdot\mathbf{r}_{i}\right)^{2}
    = \sum_{\alpha,\beta=x,y,z}u_{\alpha}u_{\beta}r_{i\alpha}r_{i\beta}\,.
\end{equation}
Then to obtain a similar form for the term $r_{i}^{2}$ we also introduce $u_{\alpha}$ and $u_{\beta}$ using the property of unit vectors $\sum_{\alpha}u_{\alpha}^{2}=u_{x}^{2}+u_{y}^{2}+u_{z}^{2}=1$, 
\begin{equation}
  r_{i}^{2} = r_{i}^{2}\sum_{\alpha=x,y,z}u_{\alpha}^{2}
    = r_{i}^{2}\sum_{\alpha,\beta=x,y,z} 
      u_{\alpha}u_{\beta}\delta_{\alpha\beta}\,,
\end{equation}
where $\delta_{\alpha\beta}$ is the Kronecker $\delta$ which is equal to 1 for $\alpha=\beta$ and 0 otherwise. This allows writing the third term of equation (\ref{eq:mult-phiA-2}) as
\begin{equation}
  \frac{\sum_{i}q_{i}\left[3\left(\mathbf{u}\cdot\mathbf{r}_{i}\right)^{2}
         - r_{i}^{2}\right]}{2R^{3}}
  = \sum_{\alpha\beta}\left[u_{\alpha}u_{\beta} \sum_{i\in A}
    \frac{q_{i}\left(3r_{i\alpha}r_{i\beta}
         -r_{i}^{2}\delta_{\alpha\beta}\right)}{2R^{3}}\right]
  = \frac{\sum_{\alpha\beta}u_{\alpha}u_{\beta}Q_{\alpha\beta}(A)}{R^{3}}
  \label{eq:mult-quadrup}
\end{equation}
where $Q_{\alpha\beta}(A)$ is the element in the $\alpha$ and $\beta$ directions of the two-dimensional tensor representing the quadrupole moment $\mathbf{Q}(A)$. It follows from equation (\ref{eq:mult-quadrup}) that $\mathbf{Q}(A)$ is a symmetric tensor, i.e.~$Q_{\alpha\beta}(A)=Q_{\beta\alpha}(A)$, with a vanishing trace, \textit{i.e.}~$\sum_{\alpha}Q_{\alpha\alpha}(A)=0$. Finally equation (\ref{eq:mult-phiA-2}) can be written in a condensed form
\begin{equation}
  \Phi_{A}(\mathbf{R})=\frac{1}{4\pi\epsilon_{0}}\left(\frac{q(A)}{R}+\frac{\mathbf{u}\cdot\mathbf{d}(A)}{R^{2}}+\frac{\mathbf{u}\cdot\left(\mathbf{Q}(A)\cdot\mathbf{u}\right)}{R^{3}}\right)+\mathcal{O}\left(R^{-4}\right).
  \label{eq:mult-phiA-3}
\end{equation}
Note that equation (\ref{eq:mult-phiA-3}) can be expanded to higher orders in $R^{-1}$, which imply higher-rank multipole moments, such as the octupole moment.

\begin{figure}
\begin{centering}
\includegraphics[scale=0.8]{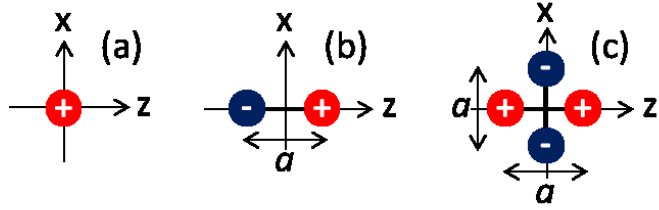}
\par\end{centering}
\caption{Schematic representation of the lowest order multipole moments appearing in equation \ref{eq:mult-phiA-3}: (a) a charge $+q$, (b) a dipole, and (c) a quadrupole. In these examples, each red (blue) particle carries a $+q$ ($-q$) charge. \label{fig:mult-schema}}
\end{figure}

A simple representation of the total charge, dipole and quadrupole moments is displayed in figure \ref{fig:mult-schema}. In equation (\ref{eq:mult-phiA-3}) the first term is equal to the electrostatic potential created by a point charge $q(A)$ located in C: this is why the total charge is sometimes called ``electric monopole''. The dipole moment is non-zero if the barycenter of the positive charges differs from the barycenter of the negative charges. The dipole moment is a vector pointing from the negative charges towards the positive charges. In figure \ref{fig:mult-schema}, the dipole is represented as two charges of opposite sign, $+q$ and $-q$, located along the $z$ axis, and separated by the distance $a$. The components of the corresponding dipole moment are then $d_{x}(A)=d_{y}(A)=0$ and $d_{z}(A)=qa$. The simplest way to represent a quadrupole is to draw four charges of opposite signs on the corner of a square (figure \ref{fig:mult-schema}(c)). In this case, the total charge and dipole moment are zero, and the only non-vanishing components of the quadrupole moment are $Q_{zz}(A) = -Q_{xx}(A) = 3qa^{2}/4$.

It is important to stress that, if the total charge is non-zero, the dipole and quadrupole moments, see equations (\ref{eq:mult-dip}) and (\ref{eq:mult-quadrup}), \textit{depend on the origin of the coordinate system}. A consequence is, for example, that the dipole moment of a charged diatomic heteronuclear molecule increases linearly with the interatomic distance, but with a slope depending on the origin of the coordinate system.
To illustrate that, let us assume that the multipole moments of A are calculated with respect to the point C' which, in the frame $Cxyz$, is given by the vector $\mathbf{r}_{C'}$. For instance the dipole moment of A is then $\mathbf{d}'(A)=\sum_{i}q_{i}\mathbf{r}'_{i}=\sum_{i}q_{i}(\mathbf{r}_{i}-\mathbf{r}_{C'})$. However the electrostatic potential $\Phi_{A}(\mathbf{R})$ created by the whole distribution A \textit{is invariant with respect to the choice of origin}. If we call $\mathbf{R}'$ the vector joining C' and M, the distance between the charge $q_{i}$ and the point M is now expressed as $\left|\mathbf{R}'-\mathbf{r}'_{i}\right|=\left|\mathbf{R}-\mathbf{r}_{C'}-\left(\mathbf{r}{}_{i}-\mathbf{r}_{C'}\right)\right| \equiv \left|\mathbf{R}-\mathbf{r}{}_{i}\right|$, which leads to the same electrostatic potential as in equation (\ref{eq:mult-phiA}). By making a similar Taylor expansion as equation (\ref{eq:mult-R-ri-2}) around C' (with $\left|r_{C'}\right|\ll R$), replacing $\mathbf{R}$, $\mathbf{d}(A)$ and $\mathbf{Q}(A)$ by their modified values $\mathbf{R}'$, $\mathbf{d}'(A)$ and $\mathbf{Q}'(A)$, we end to the same expression of the electrostatic potential term by term than in equation (\ref{eq:mult-phiA-3}). Note that similarly, for a non-vanishing dipole moment, the quadrupole moment also depends on the origin of the coordinate system.

\subsubsection{Interaction energy between distribution B and an external weakly-varying electrostatic potential}

We now consider that the charge distribution B is submitted to an external electrostatic potential $\Phi(\mathbf{r})$. The interaction potential energy $V_{B}$ is
\begin{equation}
V_{B}=\sum_{j\in B}q_{j}\Phi(\mathbf{r}_{j})\,.\label{eq:mult-vb}
\end{equation}
Assuming that $\Phi(\mathbf{r})$ does not vary significantly over the whole charge distribution B, we can write the Taylor expansion of equation (\ref{eq:mult-vb}) around the center of mass D
\begin{eqnarray}
V_{B} & = & \sum_{j\in B}q_{j}\left\{ \Phi(\mathbf{r}_{D})+\sum_{\alpha=x,y,z}\frac{\partial\Phi(\mathbf{r}_{D})}{\partial r_{\alpha}}\left(r_{\alpha j}-r_{\alpha D}\right)\right.\nonumber \\
 &  & \left.+\frac{1}{2}\sum_{\alpha,\beta=x,y,z}\frac{\partial^{2}\Phi(\mathbf{r}_{D})}{\partial r_{\alpha}\partial r_{\beta}}\left(r_{\alpha j}-r_{\alpha D}\right)\left(r_{\beta j}-r_{\beta D}\right)+...\right\} .\label{eq:mult-vb-2}
\end{eqnarray}
We can distribute the sum over the charges in each term of equation (\ref{eq:mult-vb-2}). The first term gives $q(B)\Phi(\mathbf{r}_{D})$, where $q(B)$ is the total charge of B, and the second one gives
\begin{equation}
  \sum_{\alpha=x,y,z}\frac{\partial\Phi(\mathbf{r}_{D})}{\partial r_{\alpha}}
  \sum_{j\in B}q_{j}\left(r_{\alpha j}-r_{\alpha D}\right)
    = \mathbf{\nabla}\Phi(\mathbf{r}_{D})\cdot\mathbf{d}(B)
    = -\mathbf{E}(\mathbf{r}_{D})\cdot\mathbf{d}(B)\,,
\end{equation}
where $\mathbf{\nabla}\Phi(\mathbf{r}_{D})$ is the gradient of $\Phi(\mathbf{r})$ taken at $\mathbf{r}=\mathbf{r}_{D}$, $\mathbf{E}(\mathbf{r}_{D})=-\mathbf{\nabla}\Phi(\mathbf{r}_{D})$ is the electric field at D, and $\mathbf{d}(B)$ is the dipole moment of B.

For the third term of equation (\ref{eq:mult-vb-2}), we reorder equation (\ref{eq:mult-quadrup}),
\begin{equation}
  \frac{1}{2}\sum_{j\in B}q_{j}\left(r_{\alpha j}-r_{\alpha D}\right)
    \left(r_{\beta j}-r_{\beta D}\right) = \frac{1}{3}
    Q_{\alpha\beta}(B)+\frac{\delta_{\alpha\beta}}{6}
    \sum_{j\in B}q_{j}\left(\mathbf{r}_{j}-\mathbf{r}_{D}\right)^{\text{2}}\,,
\end{equation}
and we obtain
\begin{eqnarray}
 &  & \frac{1}{2}\sum_{\alpha,\beta=x,y,z}\frac{\partial^{2}\Phi(\mathbf{r}_{D})}{\partial r_{\alpha}\partial r_{\beta}}\sum_{j\in B}q_{j}\left(r_{\alpha j}-r_{\alpha D}\right)\left(r_{\beta j}-r_{\beta D}\right)\nonumber \\
 & = & \frac{1}{3}\sum_{\alpha,\beta=x,y,z}\frac{\partial^{2}\Phi(\mathbf{r}_{D})}{\partial r_{\alpha}\partial r_{\beta}}Q_{\alpha\beta}(B)+\frac{1}{6}\sum_{\alpha=x,y,z}\frac{\partial^{2}\Phi(\mathbf{r}_{D})}{\partial r_{\alpha}^{2}}\sum_{j\in B}q_{j}\left(\mathbf{r}_{j}-\mathbf{r}_{D}\right)^{\text{2}}\nonumber \\
 & = & \frac{1}{3}\sum_{\alpha,\beta=x,y,z}\frac{\partial^{2}\Phi(\mathbf{r}_{D})}{\partial r_{\alpha}\partial r_{\beta}}Q_{\alpha\beta}(B)+\frac{1}{6}\mathbf{\nabla}^{2}\Phi(\mathbf{r}_{D})\sum_{j\in B}q_{j}\left(\mathbf{r}_{j}-\mathbf{r}_{D}\right)^{\text{2}}
\label{eq:mult-vb-quadr}
\end{eqnarray}
where $\mathbf{\nabla}^{2}\Phi(\mathbf{r}_{D})$ is the Laplacian of $\Phi(\mathbf{r})$ taken at $\mathbf{r}=\mathbf{r}_{D}$. Since the electrostatic potential $\Phi$ exists independently from the presence of the charge distribution B, it satisfies the Laplace equation $\mathbf{\nabla}^{2}\Phi(\mathbf{r})=0$, so that the last term of equation (\ref{eq:mult-vb-quadr}) vanishes. Finally the interaction energy $V_{B}$ can be written in the compact form
\begin{equation}
V_{B}=q(B)\Phi(\mathbf{r}_{D})-\mathbf{d}(B)\cdot\mathbf{E}(\mathbf{r}_{D})+\frac{1}{3}\mathbf{\nabla}\cdot\left(\mathbf{Q}(B)\cdot\mathbf{\nabla}\Phi(\mathbf{r}_{D})\right)+...
\label{eq:mult-vb-3}
\end{equation}

As in equation (\ref{eq:mult-phiA-3}) the interaction energy $V_{B}$ appears as a sum of terms involving the multipole moments of the charge distribution B. Again the dipole $\mathbf{d}(B)$ and quadrupole moments $\mathbf{Q}(B)$ can depend on the point from which they are calculated for a charged system. However, it can be easily shown that the energy $V_{B}$ does not depend on that origin. This can be easily understood since the choice of this specific point does not change the positions of the charge $q_{j}$ in the external electrostatic potential $\Phi(\mathbf{r})$.

\subsubsection{Interaction energy between the distributions A and B}

\begin{figure}
\begin{centering}
\includegraphics[scale=0.8]{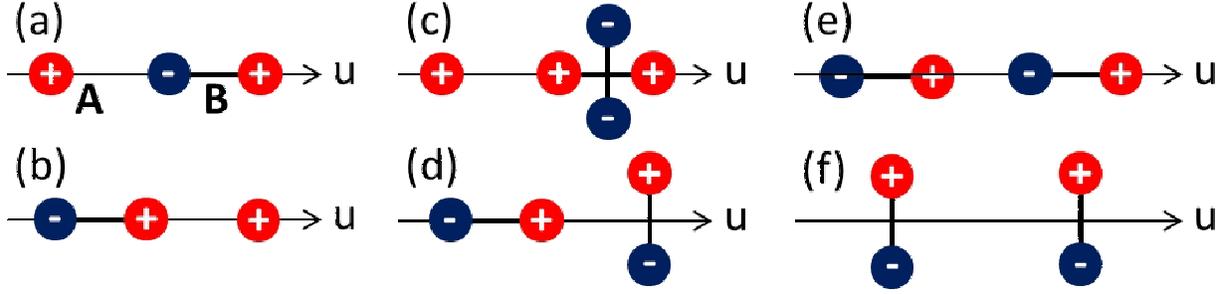}
\par\end{centering}
\caption{Schematic representation of some selected types of interactions between various multipole moments of the charge distributions A and B (on each scheme, A (B) is the left (right) charge distribution). (a) and (b) are charge-dipole terms $\propto R^{-2}$ of opposite signs: the charge $q(A)>0$ is closer to the {}`-' of B in case (a), which induces an attraction and a negative sign in equation (\ref{eq:mult-vab}), while in case (b) the {}`+' of A is closer to $q(B)>0$ leading to a repulsion and a positive sign in equation (\ref{eq:mult-vab}). (c) depicts the charge-quadrupole term $\propto R^{-3}$, for which a similar analysis based on the charge signs shows that the sign is constant when A and B are interchanged (swapped). (d), (e) an (f) correspond to various dipole-dipole terms depending on their relative orientation, thus illustrating their strongly anisotropic interaction. In (d) the interaction energy $V_{AB}(\mathbf{R})=0$ for symmetry reasons. In (e) $V_{AB}(\mathbf{R})=-d^{2}/2\pi\epsilon_{0}R^{3}$, while in (f) we have $V_{AB}(\mathbf{R})=+d^{2}/4\pi\epsilon_{0}R^{3}$. Cases (e) and (f) are often referred to as ``head-to-tail'' and ``side-by-side'' configurations.}
\label{fig:mult-inter}
\end{figure}

In the last step of our calculation, we consider that the electrostatic potential to which the distribution B is submitted is created by the distribution A. The resulting interaction energy $V_{AB}(\mathbf{R})$ depends on the vector $\mathbf{R}$ between the center-of-mass of A and B, and is obtained by applying equation (\ref{eq:mult-vb-3}) with $\Phi(\mathrm{r}_{D})=\Phi_{A}(\mathrm{r}_{D})$ given in equation (\ref{eq:mult-phiA-3}). To that end, we need to calculate the spatial derivatives of the electrostatic potential. We introduce the function
\begin{equation}
  f_{1}(x,y,z)=\left(x^{2}+y^{2}+z^{2}\right)^{-\frac{n}{2}}\,,
\end{equation}
and its partial derivative with respect to $x$
\begin{equation}
  \frac{\partial}{\partial x}f_{1}(x,y,z)=-nx\left(x^{2}+y^{2}+z^{2}\right)^{-\frac{n}{2}-1}\,.
\end{equation}
with similar expressions for the partial derivatives with respect to $y$ and $z$. The gradient vector $\mathbf{\nabla}$ writes
\begin{equation}
  \mathbf{\nabla}\left(\frac{1}{R^{n}}\right)=-\frac{n\mathbf{R}}{R^{n+2}}=-\frac{n\mathbf{u}}{R^{n+1}}\,.
\end{equation}
To calculate $\mathbf{\nabla}(\mathbf{u}\cdot\mathbf{d}(A)/R^{n})=\mathbf{\nabla}(\mathbf{R}\cdot\mathbf{d}(A)/R^{n+1})$ we introduce the function
\begin{equation}
  f_{2}(x,y,z)=\frac{xd_{x}(A)+yd_{y}(A)+zd_{z}(A)}{\left(x^{2}+y^{2}+z^{2}\right)^{\frac{n+1}{2}}}\,.
\end{equation}
As the components $d_{\alpha}(A)$ ($\alpha=x,y,z$) are independent of $x$, we obtain the $x$-partial derivative (and similar expressions for the $y$ and $z$ ones)
\begin{equation}
  \frac{\partial}{\partial x}f_{2}(x,y,z)=\frac{d_{x}(A)}{\left(x^{2}+y^{2}+z^{2}\right)^{\frac{n+1}{2}}}-\left(n+1\right)x\frac{xd_{x}(A)+yd_{y}(A)+zd_{z}(A)}{\left(x^{2}+y^{2}+z^{2}\right)^{\frac{n+3}{2}}}\,,
\end{equation}
leading to
\begin{equation}
  \mathbf{\nabla}\left(\frac{\mathbf{u}\cdot\mathbf{d}(A)}{R^{n}}\right)=\frac{\mathbf{d}(A)-\left(n+1\right)\left(\mathbf{u}\cdot\mathbf{d}(A)\right)\mathbf{u}}{R^{n+1}}\,.
\end{equation}
Finally the electric field $\mathbf{E}_{A}(\mathrm{r}_{D}=\mathbf{R})$ reads
\begin{equation}
  \mathbf{E}_{A}(\mathbf{R})=\frac{1}{4\pi\epsilon_{0}}\left(\frac{q(A)\mathbf{u}}{R^{\text{2}}}+\frac{3\left(\mathbf{u}\cdot\mathbf{d}(A)\right)\mathbf{u}-\mathbf{d}(A)}{R^{3}}\right)+\mathcal{O}(R^{-4})\,.
\end{equation}
The second partial derivatives of electrostatic potential can be calculated in a similar way. Finally the interaction energy $V_{AB}(\mathbf{R})$ is expressed as the so-called \textit{multipolar expansion}
\begin{eqnarray}
V_{AB}(\mathbf{R}) & = & \frac{1}{4\pi\epsilon_{0}}\left(\frac{q(A)q(B)}{R}+\frac{d_{u}(A)q(B)-q(A)d_{u}(B)}{R^{2}}\right.\nonumber \\
 &  & \left.+\frac{Q_{uu}(A)q(B)+q(A)Q_{uu}(B)+\mathbf{d}(A)\cdot\mathbf{d}(B)-3d_{u}(A)d_{u}(B)}{R^{3}}\right)\nonumber \\
 &  & +\mathcal{O}(R^{-4})\,,
\label{eq:mult-vab}
\end{eqnarray}
where we introduced $d_{u}(A)=\mathbf{u}\cdot\mathbf{d}(A)=\sum_{\alpha}u_{\alpha}d_{\alpha}(A)$ and $Q_{uu}(A)=\mathbf{u}\cdot\left(\mathbf{Q}(A)\cdot\mathbf{u}\right)=\sum_{\alpha\beta}u_{\alpha}u_{\beta}Q_{\alpha\beta}(A)$, and similarly for B. The interaction energy is a sum of terms proportional to $R^{-1}$. It also depends on the relative orientation of the two charge distributions through the vector $\mathbf{u}$, and the corresponding components of the dipole and quadrupole moments. Therefore the interaction can be \textit{anisotropic}. Each term depends on a specific product of multipole moments of the distributions A and B, some of them being illustrated in figure \ref{fig:mult-inter}:
\begin{itemize}
\item term $\propto R^{-1}$: the interaction between the two \textit{total} charges $q(A)$ and $q(B)$;
\item terms $\propto R^{-2}$: the interaction between one total charge and one dipole moment, which change their sign when interchanging A and B, as illustrated with panels (a) and (b) of figure \ref{fig:mult-inter}. 
\item terms $\propto R^{-3}$: one is the interaction between one total charge and one quadrupole moment which keeps its sign when interchanging A and B, and the other is the interaction between the two dipole moments. The sign and the intensity of the latter term strongly depend on the relative orientation of the dipoles, thus leading to an \textit{anisotropic interaction} (panels (d), (e) and (f) of figure \ref{fig:mult-inter}).
\item terms $\propto R^{-4}$: although we do not write them explicitly, it is easy to infer that they involve the interaction between a dipole and a quadrupole on one hand, and between a charge and an octupole on the other hand.
\end{itemize}

Note finally that equation \eqref{eq:mult-vab} does not account for the well-known van der Waals $R^{-6}$ term, which requires the use of quantum perturbation theory (see subsection \ref{ssec:pert-2nd}).

\subsection{Calculation in spherical coordinates}
\label{ssec:mult-spher}

Equation (\ref{eq:mult-vab}) above is insightful to understand the physical origin of each term of the multipolar expansion. But it is not so practical for quantum atomic and molecular systems for which the total angular momentum is a conserved quantity in the absence of an external electric field. The expression of the multipolar expansion in spherical coordinates derived in this subsection will allow us to take advantage of the the Wigner-Eckart theorem later in the chapter. We consider two different reference frames: (i) the \textit{body-fixed} (BF) $uvw$ frame where the $u$-axis joins the centers of mass of each distribution, and (ii) the \textit{space-fixed} (SF) $xyz$ frame where it is assumed that the previous $u$-axis has no preferred orientation with respect to $xyz$. An alternate equivalent way to formulate the former choice is to assume that in the BF case the $u$ axis has a fixed orientation in space, say, $u \equiv z$.


Defining $\mathbf{r}_{ij}=\mathbf{r}_{i}-\mathbf{r}_{j}$ (the difference between $\mathbf{r}_{i}$ and $\mathbf{r}_{j}$, as if those two vectors had the same origin), of norm $r_{ij}=\left|\mathbf{r}_{ij}\right|$, the distance between the charges $q_{i}$ (of A) and $q_{j}$ (of B) is
\begin{eqnarray}
\left|\mathbf{R}+\mathbf{r}_{j}-\mathbf{r}_{i}\right| & = & \sqrt{\left(\mathbf{R}-\left(\mathbf{r}_{i}-\mathbf{r}_{j}\right)\right)^{2}}=\sqrt{R^{2}-2\mathbf{R}\cdot\mathbf{r}_{ij}+r_{ij}^{2}} \nonumber \\ 
 & = & R\sqrt{1-\frac{2\mathbf{u}\cdot\mathbf{r}_{ij}}{R}+\frac{r_{ij}^{2}}{R^{2}}}\,,
\end{eqnarray}
Calling $\theta_{ij}$ the angle between $\mathbf{r}_{ij}$ and the unitary vector $\mathbf{u}$, \textit{i.e.} $\mathbf{u}\cdot\mathbf{r}_{ij}=r_{ij}\cos\theta_{ij}$, the third term of equation (\ref{eq:mult-vtot}) becomes
\begin{equation}
\frac{1}{\left|\mathbf{R}+\mathbf{r}_{j}-\mathbf{r}_{i}\right|}=\frac{1}{R\sqrt{1-\frac{2r_{ij}\cos\theta_{ij}}{R}+\frac{r_{ij}^{2}}{R^{2}}}}=\sum_{\ell=0}^{+\infty}\frac{r_{ij}^{\ell}}{R^{\ell+1}}P_{\ell}(\cos\theta_{ij})
\label{eq:mult-dist}
\end{equation}
where we used the generating series of Legendre polynomials $P_{\ell}(x)$. However it is more convenient to separate the coordinates of the charges $q_{i}$ and $q_{j}$.

\subsubsection{Calculation in the body-fixed frame}

Assuming that the vector $\mathbf{u}$ lies along the $z$ axis, $\mathbf{u}=\mathbf{u}_{z}$, the resulting interaction energy $V_{AB}^{\mathrm{BF}}(R)$ expressed in the body-fixed frame only depends on $R$, as its orientation is meaningless in this frame. The Legendre polynomial of equation (\ref{eq:mult-dist}) is proportional to the spherical harmonics $Y_{\ell 0}(\theta_{ij},\phi_{ij})$ where $\phi_{ij}$ is the azimuthal angle giving the orientation of $\mathbf{r}_{ij}$ in the BF frame
\begin{equation}
P_{\ell}(\cos\theta_{ij})=\sqrt{\frac{4\pi}{2\ell+1}}Y_{\ell0}(\theta_{ij},\phi_{ij})\,,
\label{eq:mult-pl-ylm}
\end{equation}
In order to introduce the spherical coordinates $(r_{i},\theta_{i},\phi_{i})$ of $\mathbf{r}_{i}$ and $(r_{j},\theta_{j},\phi_{j})$ of $\mathbf{r}_{j}$, we use the identity (see reference \cite{varshalovich1988}, p 167)
\begin{align}
  r_{ij}^{\ell}Y_{\ell m}(\theta_{ij},\phi_{ij}) & =  
   \sqrt{4\pi\left(2\ell+1\right)!}
   \sum_{\ell_{A},\ell_{B}=0}^{+\infty} 
   \delta_{\ell_{A}+\ell_{B},\ell} 
   \frac{\left(-1\right)^{\ell_{B}} 
         r_{i}^{\ell_{A}}r_{j}^{\ell_{B}}}
   {\sqrt{\left(2\ell_{A}+1\right)!\left(2\ell_{B}+1\right)!}}
  \nonumber \\
   & \times \sum_{m_{A}=-\ell_{A}}^{+\ell_{A}} 
   \sum_{m_{B}=-\ell_{B}}^{+\ell_{B}} 
   C_{\ell_{A}m_{A}\ell_{B}m_{B}}^{\ell m}
   Y_{\ell_{A}m_{A}}(\theta_{i},\phi_{i})
   Y_{\ell_{B}m_{B}}(\theta_{j},\phi_{j})
  \label{eq:mult-rij-expand}
\end{align}
where $C_{\ell_{A}m_{A}\ell_{B}m_{B}}^{\ell m}=\langle\ell_{A}m_{A}\ell_{B}m_{B}|\ell_{A}\ell_{B}\ell m\rangle$ is a Clebsch-Gordan (CG) coefficient, which imposes $m=m_{A}+m_{B}$ for the spherical harmonics $Y_{\ell m}(\theta_{ij},\phi_{ij})$. In the case $\ell=\ell_{A}+\ell_{B}$, the CG coefficient takes the particular expression (see reference \cite{varshalovich1988}, p.~248)
\begin{equation}
C_{\ell_{A}m_{A}\ell_{B}m_{B}}^{\ell m}=\sqrt{\frac{\left(2\ell_{A}\right)!\left(2\ell_{B}\right)!\left(\ell+m\right)!\left(\ell-m\right)!}{\left(2\ell\right)!\left(\ell_{A}+m_{A}\right)!\left(\ell_{A}-m_{A}\right)!\left(\ell_{B}+m_{B}\right)!\left(\ell_{B}-m_{B}\right)!}}\,.
\label{eq:mult-cg}
\end{equation}
Combining equations (\ref{eq:mult-pl-ylm}), (\ref{eq:mult-rij-expand}) and (\ref{eq:mult-cg}), and setting a new definition for $m=m_A=-m_B$ (due to the 0 in equation \eqref{eq:mult-pl-ylm}), we obtain a central result of this chapter, the expression of $V_{AB}^{\mathrm{BF}}(R)$ in spherical coordinates
\begin{equation}
V_{AB}^{\mathrm{BF}}(R)=\frac{1}{4\pi\epsilon_{0}}\sum_{\ell_{A},\ell_{B}=0}^{+\infty}\sum_{m=-\ell_{<}}^{+\ell_{<}}\frac{f_{\ell_{A}\ell_{B}m}}{R^{1+\ell_{A}+\ell_{B}}}\mathrm{Q}_{\ell_{A}m}^{\mathrm{BF}}(A)\mathrm{Q}_{\ell_{B},-m}^{\mathrm{BF}}(B)
\label{eq:mult-vab-bf}
\end{equation}
Equation (\ref{eq:mult-vab-bf}) displays a series of terms scaling as $R^{-1-\ell_{A}-\ell_{B}}$. The sum over $\ell=\ell_{A}+\ell_{B}$ is replaced by a sum over two positive integers $\ell_{A}$ and $\ell_{B}$ which are the ranks of the tensors associated with the multipole moments of A and B respectively, and a sum over the integer $m$, corresponding to the components of these tensors. In order to ensure that the arguments of factorials in equation (\ref{eq:mult-cg}) are not negative,  $-\ell_{<} \le m \le \ell_{<}$ where $\ell_{<}=\min(\ell_{A},\ell_{B})$ is the smaller integer between $\ell_{A}$ and $\ell_{B}$. Equation (\ref{eq:mult-vab-bf}) contains the product of multipole moments $\mathrm{Q}_{\ell_{A}m}^{\mathrm{BF}}(A)$ and $\mathrm{Q}_{\ell_{B},-m}^{\mathrm{BF}}(B)$ expressed in spherical coordinates in the BF frame (and so depending on $\theta_i$ and $\phi_i$, taken with respect to the axis $\mathbf{u}=\mathbf{u}_{z}$ joining the centers of the two charge distributions)
\begin{equation}
  \mathrm{Q}_{\ell_{A}m}^{\mathrm{BF}}(A)=\sqrt{\frac{4\pi}{2\ell_{A}+1}}\sum_{i\in A}q_{i}r_{i}^{\ell_{A}}Y_{\ell_{A}m}(\theta_{i},\phi_{i})
  \label{eq:mult-qlm-bf}
\end{equation}
and similarly for $\mathrm{Q}_{\ell_{B},-m}^{\mathrm{BF}}(B)$. Finally the numerical factor $f_{\ell_{A}\ell_{B}m}$ in equation (\ref{eq:mult-vab-bf}) is obtained by gathering the numerical factors in equations (\ref{eq:mult-pl-ylm})--(\ref{eq:mult-cg}),
\begin{eqnarray}
f_{\ell_{A}\ell_{B}m} & = & \frac{1}{\sqrt{2\ell_{A}+2\ell_{B}+1}}\times\left(-1\right)^{\ell_{B}}\sqrt{\frac{\left(2\ell_{A}+2\ell_{B}+1\right)!}{\left(2\ell_{A}\right)!\left(2\ell_{B}\right)!}}\times C_{\ell_{A}m\ell_{B}-m}^{\ell_{A}+\ell_{B},0}\nonumber \\
 & = & \frac{\left(-1\right)^{\ell_{B}}\left(\ell_{A}+\ell_{B}\right)!}{\sqrt{\left(\ell_{A}+m\right)!\left(\ell_{A}-m\right)!\left(\ell_{B}+m\right)!\left(\ell_{B}-m\right)!}}
\label{eq:mult-flalbm}
\end{eqnarray}
The sign of $f_{\ell_{A}\ell_{B}m}$ is imposed by the factor $\left(-1\right)^{\ell_{B}}$.

The multipole moments $\mathrm{Q}_{\ell_{A}m}(A)$ in spherical coordinates are linked with those in Cartesian coordinates through relations which are independent of the chosen reference frame (so that the superscript {}``BF" is dropped in the next equations). The total charge is associated with a tensor of rank 0
\begin{equation}
  \mathrm{Q}_{00}(A) = q(A)\,,
  \label{eq:mult-sph-charge}
\end{equation}
the dipole moment with a tensor of rank 1
\begin{eqnarray}
  \mathrm{Q}_{10}(A) & = & d_{z}(A)\nonumber \\
  \mathrm{Q}_{1,\pm1}(A) & = & \mp\frac{d_{x}(A)\pm id_{y}(A)}{\sqrt{2}}\,,
  \label{eq:mult-sph-dip}
\end{eqnarray}
and the quadrupole moment with a tensor of rank 2
\begin{eqnarray}
  \mathrm{Q}_{20}(A) & = & Q_{zz}(A) \nonumber\\
  \mathrm{Q}_{2,\pm1}(A) & = & \mp\sqrt{\frac{2}{3}}
    \left(Q_{xz}(A)\pm iQ_{yz}(A)\right) \nonumber\\
  \mathrm{Q}_{2,\pm2}(A) & = & \frac{1}{\sqrt{6}}
    \left(Q_{xx}(A)-Q_{yy}(A)\pm2iQ_{xy}(A)\right)\,.
  \label{eq:mult-sph-quad}
\end{eqnarray}
Generally speaking, the $2^{\ell_{A}}$-pole moment is associated with a tensor of rank $\ell_{A}$, for which equation (\ref{eq:mult-qlm-bf}) gives a general expression for any $\ell_{A}$ and $m$. In atomic and molecular systems, it is important to distinguish the tensor rank $\ell_{A}$ of the multipole moment from the orbital angular momentum of electrons which characterize quantum levels. Let us also stress that the CG coefficients in equation 
(\ref{eq:mult-rij-expand}) couple the tensor ranks, and not the angular momenta of atoms or molecules.

The influence of the interchange of A and B on the interaction energy can be easily seen with equations (\ref{eq:mult-vab-bf}) and (\ref{eq:mult-flalbm}). Due to the $\left(-1\right)^{\ell_{B}}$ factor, the sign of the energy $V_{AB}^{\mathrm{BF}}(R)$ changes if $\left(-1\right)^{\ell_{A}+\ell_{B}}=-1$, \textit{i.e.} if $\ell_{A}$ and $\ell_{B}$ have different parities. For example it is the case of the charge-dipole interaction but not of the charge-charge, charge-quadrupole and dipole-dipole interactions.

\subsubsection{Calculation in the space-fixed frame}

In this second calculation, we consider that the vector $\mathbf{R}$ joining the centers C and D has an arbitrary orientation in the SF frame, given by its spherical coordinates $(R,\Theta,\Phi)$ (we chose capital Greek letters to specify angles in the SF frame; note that for convenience, the $\Phi$ notation previously used for electrostatic potential will be uniquely used to label the angle from now on). The SF spherical coordinates of the vector $\mathbf{r}_{ij}$ are $(r_{ij},\Theta_{ij},\Phi_{ij})$. The addition theorem of the spherical harmonics (see reference \cite{varshalovich1988}, p.~165) allows for eliminating the BF angles $(\theta_{ij},\phi_{ij})$ ,
\begin{equation}
  P_{\ell}(\cos\theta_{ij}) = \frac{4\pi}{2\ell+1}
    \sum_{m=-\ell}^{+\ell}Y_{\ell m}^{*}(\Theta,\Phi)
    Y_{\ell m}(\Theta_{ij},\Phi_{ij})\,.
\end{equation}
By applying equation (\ref{eq:mult-rij-expand}) to $r_{ij}^{\ell}Y_{\ell m}(\Theta_{ij},\Phi_{ij})$, we obtain the expression of the interaction energy in the SF frame
\begin{align}
  V_{AB}^{\mathrm{SF}}(\mathbf{R}) & = \frac{1}{4\pi\epsilon_{0}}
    \sum_{\ell_{A}\ell_{B}\ell=0}^{+\infty} \delta_{\ell_{A}+\ell_{B},\ell}
    \,\frac{\left(-1\right)^{\ell_{B}}}{R^{1+\ell}}
    \sqrt{\frac{\left(2\ell\right)!}
      {\left(2\ell_{A}\right)!\left(2\ell_{B}\right)!}}
    \sum_{m=-\ell}^{+\ell}\sqrt{\frac{4\pi}{2\ell+1}}
      Y_{\ell m}^{*}(\Theta,\Phi) \nonumber \\
  & \times \sum_{m_{A}=-\ell_{A}}^{+\ell_{A}}
    \sum_{m_{B}=-\ell_{B}}^{+\ell_{B}}
      C_{\ell_{A}m_{A}\ell_{B}m_{B}}^{\ell m} 
      \mathrm{Q}_{\ell_{A}m_{A}}^{\textrm{SF}}(A)
      \mathrm{Q}_{\ell_{B}m_{B}}^{\textrm{SF}}(B)
  \label{eq:mult-vab-sf}
\end{align}
where we kept explicitly the indices $\ell$ and $m$. The latter expression will be useful for future calculations, while the following one is written in a more explicit way
\begin{align}
  V_{AB}^{\mathrm{SF}}(\mathbf{R}) & = \frac{1}{4\pi\epsilon_{0}} \sum_{\ell_{A}\ell_{B}\ell=0}^{+\infty}
    \delta_{\ell_{A}+\ell_{B},\ell}\,
    \frac{\left(-1\right)^{\ell_{B}}}{R^{1+\ell}} \sum_{m=-\ell}^{+\ell} 
    \sqrt{\frac{4\pi\left(\ell+m\right)!\left(\ell-m\right)!}{2\ell+1}}
    Y_{\ell m}^{*}(\Theta,\Phi)\nonumber \\
  & \times \sum_{m_{A}=-\ell_{A}}^{+\ell_{A}} 
    \sum_{m_{B}=-\ell_{B}}^{+\ell_{B}} 
    \frac{\mathrm{Q}_{\ell_{A}m_{A}}^{\textrm{SF}}(A)
    \mathrm{Q}_{\ell_{B}m_{B}}^{\textrm{SF}}(B)}
      {\sqrt{\left(\ell_{A}+m_{A}\right)!\left(\ell_{A}-m_{A}\right)!
             \left(\ell_{B}+m_{B}\right)!\left(\ell_{B}-m_{B}\right)!}}\,,
  \label{eq:mult-vab-sf2}
\end{align}

Unlike in the BF frame, the interaction energy contains the orientation of the $\mathbf{R}$ vector in the SF frame through $\sqrt{4\pi/\left(2\ell+1\right)}Y_{\ell m}^{*}(\Theta,\Phi)$, which is for example equal to $\cos\Theta$ when $\ell=1$, $m=0$, and to $\mp\sin\Theta\exp\left(\pm i\Phi\right)$ when $\ell=1$, $m=\pm1$. From equation (\ref{eq:mult-vab-sf}) one can retrieve the BF interaction energy of equation (\ref{fig:mult-inter}), by setting $\Theta=\Phi=0$, which imposes $\sqrt{4\pi/\left(2\ell+1\right)}Y_{\ell m}(0,0)=\delta_{m0}$ and thus $m=0$. The quantities $\mathrm{Q}_{\ell_{X}m_{X}}^{\textrm{SF}}(X)$ are the multipole moment of charge distribution $X=A,B$ expressed in the laboratory frame; they are obtained by replacing $(\theta_{i},\phi_{i})$ by $(\Theta_{i},\Phi_{i})$ and $(\theta_{j},\phi_{j})$ by $(\Theta_{j},\Phi_{j})$ in equation (\ref{eq:mult-qlm-bf}). Of course the interaction energy does not depend on the coordinate system; however, we will see in subsection \ref{sec:hetero} that the choice of the BF or the SF frame will be made for practical reasons, depending especially on the presence of an external field.

As a conclusion of this section, it is important to realize that the above derivation of electric interaction energy between two charge distributions can be extended to their magnetic interaction energy. The interaction between atoms with a magnetic dipole moment is actually an important topic in ultracold quantum gases \cite{stuhler2005,beaufils2008,aikawa2012,lu2012,frisch2015}. Assuming that the distributions A and B are composed of charges which move along closed orbits contained inside each distribution, one can show \cite{jackson1999} that the magnetostatic force exerted by A on B derives from a potential energy. For distant charge distributions, the latter can be written as a multipolar expansion formally equivalent to equations (\ref{eq:mult-vab}), (\ref{eq:mult-vab-bf}), and (\ref{eq:mult-vab-sf}), replacing the prefactor $1/4\pi\epsilon_{0}$ by $\mu_{0}/4\pi$, where $\mu_{0}$ is the vacuum permeability. In Cartesian coordinates its
potential energy reads
\begin{equation}
  V_{AB}^{\textrm{md}}(\mathbf{R})=\frac{\mu_{0}}{4\pi R^{3}}\left[\mathbf{m}(A)\cdot\mathbf{m}(B)-3\left(\mathbf{u}\cdot\mathbf{m}(A)\right)\left(\mathbf{u}\cdot\mathbf{m}(B)\right)\right]\,,
  \label{eq:mult-mag}
\end{equation}
where $\mathbf{m}(A)$ and $\mathbf{m}(B)$ are the magnetic dipole moments of the distribution A and B, respectively. For spinless charges, the magnetic dipole moment is $\mathbf{m}(A)=\frac{1}{2}\sum_{i}q_{i}\mathbf{r}_{i}\times\mathbf{v}_{i}$,
with $\mathbf{v}_{i}$ the velocity of charge $i$ with respect to C; for electrons of spin $\mathbf{s}_{i}$, $\mathbf{m}(X)$
 ($X =A, B$) reads
\begin{equation}
  \mathbf{m}(A)=-\frac{\hbar q_{e}}{2m_{e}}\sum_{i\in A}\left(\mathbf{l}_{i}+g_{s}\mathbf{s}_{i}\right)=-\mu_{B}\sum_{i\in A}\left(\mathbf{l}_{i}+g_{s}\mathbf{s}_{i}\right),
\end{equation}
where $\hbar$ is the reduced Planck constant, $m_{e}$ the electronic mass, $q_{e}$ the unsigned electronic charge, $\mu_{B}$ Bohr magneton, $g_{s}$ the spin Landé factor, and $\mathbf{l}_{i}=\mathbf{r}_{i}\times\left(m_{e}\mathbf{v}_{i}\right)/\hbar$ the orbital angular momentum of one electron in units of $\hbar$. In spherical coordinates the magnetic dipole moment is associated with a spherical tensor of rank 1, with similar relations as equations (\ref{eq:mult-sph-dip}).

\section{Perturbative calculation of long-range interactions}
\label{sec:pert-calc}

In the previous section, we derived the interaction potential energy between two charge distributions separated by a distance much larger than their spatial extension, see equation (\ref{eq:mult-cond}). This assumes that the interaction energy does not vary significantly over the distribution, validating the multipolar expansion of equations (\ref{eq:mult-vab}), (\ref{eq:mult-vab-bf}), (\ref{eq:mult-vab-sf}) and (\ref{eq:mult-mag}). This hypothesis induces a second major consequence, which will be fully exploited in the rest of the chapter. Since the inequality (\ref{eq:mult-vab}) can also be written $\left|\mathbf{r}_{i'}-\mathbf{r}_{i}\right|,\left|\mathbf{r}_{j'}-\mathbf{r}_{j}\right|\ll R$, $\forall \, i,i'\in A$ and $\forall \, j,j'\in B$, the first two terms of equation (\ref{eq:mult-vtot}), which represent the internal electrostatic energy of each charge distribution, are much larger than the third term giving the interaction energy.  Therefore, because the presence of the distribution A will only slightly affect the distribution B, and vice-versa, it is appropriate to characterize the long-range interactions between atoms and/or molecules -- which are just a special example of charge distributions, hereafter referred to as the {}``partners'' --, by using the time-independent quantum perturbation theory.

In the quantum perturbation formalism, the interaction energy becomes the \textit{perturbation operator} $\hat{\mathrm{V}}$ (see section \ref{sec:introduction}), with vanishing matrix elements as the inter-partner distance tends to infinity. The most convenient form of the perturbation operator is given in spherical coordinates (equations (\ref{eq:mult-vab-bf}) and (\ref{eq:mult-vab-sf})).
In this section, we will apply the correspondence principle of quantum mechanics, in order to replace the classical multipole moments $\mathrm{Q}_{\ell_{A}m_{A}}^{\textrm{BF/SF}}(A)$ and $\mathrm{Q}_{\ell_{B}m_{B}}^{\textrm{BF/SF}}(B)$ by quantum operators $\hat{\mathrm{Q}}_{\ell_{A}m_{A}}^{\textrm{BF/SF}}(A)$ and $\mathrm{\hat{Q}}_{\ell_{B}m_{B}}^{\textrm{BF/SF}}(B)$. The inter-partner distance $R$ will be treated as a parameter, like in the Born-Oppenheimer approximation for diatomic molecules. This yields $R$-dependent potential energy curves (PECs) which are used to characterize the collisional dynamics of the complex (see other chapters).

Calculating the matrix elements of the multipole-moment operators is thus crucial to treat long-range interactions. This will be discussed in subsection \ref{ssec:pert-qlm} both for atoms and molecules. Using the Wigner-Eckart theorem, we will emphasize the central role of the total angular momentum of each partner, whatever its orbital or spin nature. This will enable us for characterizing the interaction with an external electric field and an atom or molecule (subsection \ref{ssec:efield}). In particular we will introduce the atomic and molecular polarizabilities, and discuss their tensorial nature, that we will exploit in order to give general expressions of the first-order (see subsection \ref{ssec:pert-1st}) and second-order corrections on energy associated with long-range interactions (subsection \ref{ssec:pert-2nd}). Based on the strong angular-momentum selection rules, we will also give recipes to determine the non-vanishing terms of the multipolar expansion. In subsection \ref{ssec:pert-alk-ato}, we will illustrate such recipes in the case of two alkali-metal atoms. 

In the present section the energies of long-range interactions will be derived in the BF frame, which for two atoms, allows for calculating the asymptotic part of electronic potential-energy curves, close to their dissociation
limits. Nevertheless in subsection \ref{ssec:pert-qlm} and \ref{ssec:efield} which provide general results independent from the coordinate system, no reference is made to the BF or SF frames in the multipole moments or polarizabilities.

\subsection{Matrix elements of multipole-moment operators}
\label{ssec:pert-qlm}

Atomic and molecular energy levels are characterized by a set of quantum numbers: electronic configuration, parity, electronic or nuclear spin, etc. Even if the number and the nature of those relevant quantum numbers, denoted $\beta$ in what follows, depend on the system under consideration, in free space, there always exist two quantum numbers $J$ and $M$ which are associated with the total angular momentum $\mathbf{J}$. The corresponding ket $|JM\rangle$ is an eigenvector of $\mathbf{\hat{J}}^{2}$ and of the projection $\hat{\mathrm{J}}_{z}$ of $\mathbf{J}$ on a given quantization axis $z$, \textit{i.e.} $\mathbf{\hat{J}}^{2}|JM\rangle=\hbar^{2}J\left(J+1\right)|JM\rangle$ and $\hat{\mathrm{J}}_{z}|JM\rangle=\hbar M|JM\rangle$. Thus any atomic or molecular energy level can be generally labeled $|\beta JM\rangle$. 

The multipole-moment operator $\hat{\mathrm{Q}}_{\ell m}$ is an irreducible tensor of rank $\ell$ and component
$m$. Its matrix elements in the $\{|\beta JM\rangle\}$ basis can be written using the Wigner-Eckart theorem \cite{varshalovich1988},
\begin{equation}
\left\langle \beta'J'M'\right|\hat{\mathrm{Q}}_{\ell m}\left|\beta JM\right\rangle =\frac{C_{JM\ell m}^{J'M'}}{\sqrt{2J'+1}}\left\langle \beta'J'\right\Vert \hat{\mathrm{Q}}_{\ell}\left\Vert \beta J\right\rangle 
\label{eq:pert-wigneck}
\end{equation}
where $\left\langle \beta'J'\right\Vert \hat{\mathrm{Q}}_{\ell}\left\Vert \beta J\right\rangle $
is the \textit{reduced} matrix element of the multipole-moment operator, and $C_{JM\ell m}^{J'M'}$ a Clebsch-Gordan coefficient. To underline the general scope of this subsection, we here drop the letters "$A$'' and "$B$'' that characterize the partner as well as the superscripts "BF'' and "SF'' that specify the coordinate system. Note that throughout the rest of the chapter, the rank and the components of the tensor operator are labeled with the \textit{lowercase} indices $\ell$ and $m$, while for each particle the angular momentum and its projection  are written with the \textit{uppercase} indexes $J$ and $M$. 

The power of the Wigner-Eckart theorem is that the dependence in $M$ and $M'$ of the multipole moment matrix elements is only contained in the Clebsch-Gordan coefficient. So the knowledge of the mere reduced matrix element allows for calculating
the matrix elements for all $(M,M')$ pairs. Since these quantum numbers are associated with the operator $\hat{\mathrm{J}}_{z}$, the Clebsch-Gordan coefficient is the only term carrying the influence of the coordinate system. A non-vanishing  Clebsch-Gordan coefficient obeys strong selection rules: the so-called triangle inequality
\begin{equation}
  |J-\ell|\le J'\le J+\ell\,;
  \label{eq:pert-triangle}
\end{equation}
and the conservation of projections $M' = M + m$. For the total charge operator of a charged partner ($\ell=m=0$), only diagonal matrix element exists: $J'=J$, $M'=M$ and also $\beta'=\beta$. For the dipole moment operator ($\ell=1$), equation (\ref{eq:pert-triangle}) leads to the well-established selection rules for electric-dipole transitions $J'=J,J\pm1$ ($(J,J')=(0,0)$ being excluded).

Let us illustrate the generality of equation (\ref{eq:pert-wigneck}) with the example of an alkali-metal atom. Its (low-lying) energy levels can be accurately identified with the quantum numbers of its single valence electron: its principal quantum number $n$, its orbital angular momentum quantum number $L$ corresponding to the vector $\hat{\mathbf{L}}$, and its $z$-projection $M_{L}$. The --hydrogen-like-- electronic wave function $\Psi_{nLM_{L}}(r,\Theta,\Phi)$ depends on the valence electron coordinates $(r,\Theta,\Phi)$ with respect to the atom center
\begin{equation}
  \Psi_{nLM_{L}}(r,\Theta,\Phi)=R_{nL}(r)Y_{LM_{L}}(\Theta,\Phi),
  \label{eq:pert-wf-H}
\end{equation}
where $R_{nL}(r)$ describes the radial motion of the electron independent of $M_{L}$ due to the spherical symmetry of the core-electron potential energy, and $Y_{LM_{L}}(\Theta,\Phi)$ is a spherical harmonics describing the electron angular motion. Inserting this wave function in the definition of multipole moments (equation (\ref{eq:mult-qlm-bf})) and using the integral
\begin{align}
   & \int_{0}^{\pi}d\Theta\sin\Theta\int_{0}^{2\pi}d\Phi
    Y_{L'M'_{L}}^{*}(\Theta,\Phi)Y_{\ell m}(\Theta,\Phi)
    Y_{LM_{L}}(\Theta,\Phi)
  \nonumber \\
   & \; = \sqrt{\frac{\left(2\ell+1\right)\left(2L+1\right)}
            {4\pi\left(2L'+1\right)}} 
    C_{L0\ell0}^{L'0} C_{LM_{L}\ell m}^{L'M'_{L}}
  \label{eq:pert-3ylm}
\end{align}
we obtain the reduced matrix element
\begin{equation}
  \left\langle n'L'\right\Vert \hat{\mathrm{Q}}_{\ell}
    \left\Vert nL\right\rangle = \sqrt{2L+1}\, C_{L0\ell0}^{L'0}
    \int_{0}^{+\infty}drr^{\ell+2}R_{n'L'}(r)R_{nL}(r),
  \label{eq:pert-red-orb}
\end{equation}
where the radial integral is evaluated numerically.

The wave function of equation (\ref{eq:pert-wf-H}) does not take into account the spin $\mathbf{S}$ of the valence electron (with associated quantum numbers $S$ and $M_{S}$), which induces fine-structure splitting of the atomic energy levels \cite{cowan1981}. The spin-orbit interaction couples $\mathbf{L}$ and $\mathbf{S}$ into the total electronic angular momentum $\mathbf{J}=\mathbf{L}+\mathbf{S}$. The resulting fine-structure energy levels are now labeled $|nLSJM\rangle$, and are related to the $|nLM_{L}\rangle$ states associated with the wave function of equation (\ref{eq:pert-wf-H}) according to
\begin{equation}
\left|nLSJM\right\rangle =\sum_{M_{L}=-L}^{+L}\sum_{M_{S}=-S}^{+S}C_{LM_{L}SM_{S}}^{JM}\left|nLM_{L}\right\rangle \left|SM_{S}\right\rangle \,.
\label{eq:pert-wf-sf}
\end{equation}
As the multipole moment operator $\hat{\mathrm{Q}}_{\ell m}$ does not act on spin coordinates, its matrix elements in the $\{|nLSJM\rangle\}$ basis read
\begin{align}
  \left\langle n'L'S'J'M'\right| \hat{\mathrm{Q}}_{\ell m}
   \left|nLSJM\right\rangle & =  
   \sum_{M'_{L}M'_{S}} \sum_{M_{L}M_{S}} 
   C_{L'M'_{L}S'M'_{S}}^{J'M'}C_{LM_{L}SM_{S}}^{JM}
   \left\langle S'M'_{S}\left|SM_{S}\right\rangle \right.
  \nonumber \\
  & \times \frac{ C_{LM_{L}\ell m}^{L'M'_{L}}} {\sqrt{2L'+1}}
   \left\langle n'L'\right\Vert \hat{\mathrm{Q}}_{\ell}
   \left\Vert nL\right\rangle 
\end{align}
where we assumed that the radial wave functions of the fine-structure levels do not depend on $J$. The sum on $M'_{S}$ is eliminated since $\langle S'M'_{S}|SM_{S}\rangle=\delta_{SS'}\delta_{M_{S}M'_{S}}$. The useful relation for the product of CG coefficients (see reference \cite{varshalovich1988}, p.~260)
\begin{eqnarray}
\sum_{m_{L}M'_{L}M_{S}}C_{LM_{L}SM_{S}}^{JM}C_{L'M'_{L}SM_{S}}^{J'M'}C_{LM_{L}\ell m}^{L'M'_{L}} & = & \left(-1\right)^{\ell+J+S+L'}\sqrt{\left(2J+1\right)\left(2L'+1\right)}\nonumber \\
 & \times & \left\{ \begin{array}{ccc}
L & S & J\\
J' & \ell & L'
\end{array}\right\} C_{JM\ell m}^{J'M'}\,,
\label{eq:prod-3Clebsch}
\end{eqnarray}
where $\{ \}$ denotes a Wigner $6j$ symbol, yields a remarkable result: the matrix elements of the $\hat{\mathrm{Q}}_{\ell m}$ operator in the $\{|nLSJM\rangle\}$ basis also satisfy the Wigner-Eckart theorem, see equation (\ref{eq:pert-wigneck}), involving the reduced matrix element
\begin{align}
  \left\langle n'L'S'J'\right\Vert \hat{\mathrm{Q}}_{\ell}
   \left\Vert nLSJ\right\rangle & =\delta_{SS'} 
   \left(-1\right)^{S+J+\ell+L'} 
   \sqrt{\left(2J+1\right)\left(2J'+1\right)}
  \nonumber \\
  & \times \sixj{L}{S}{J}{J'}{\ell}{L'}
   \left\langle n'L'\right\Vert \hat{\mathrm{Q}}_{\ell} 
   \left\Vert nL\right\rangle .
\label{eq:pert-red-sf}
\end{align}

Equation (\ref{eq:pert-red-sf}) illustrates that the Wigner-Eckart theorem, and the ensuing formalism of long-range interactions, can be applied for any type of angular momentum. For instance equation (\ref{eq:pert-red-sf}) can be extended to the hyperfine-structure levels of the atoms. This statement is also valid for diatomic molecules, regardless of their Hund's case character, and even for polyatomic molecules.

\subsection{Potential energy in an external electric field}
\label{ssec:efield}

Since the long-range electrostatic interactions between two partners are due to the electric field created by one partner and exerted on the other one, it is insightful to calculate the potential energy of an atom or a molecule in an external electric field, giving rise to the \textit{Stark shift} of the energy levels. Thus the energies of long-range and Stark interactions depend on the same quantities: the permanent multipole moments and polarizabilities.

We consider a homogeneous electric field $\mathbf{E}=\mathcal{E}\mathbf{u}_{z}$ oriented along the $z$ direction. The amplitude $\mathcal{E}$ of the field is low enough to treat its influence using perturbation theory. In this subsection we calculate the energy corrections to the first order and second order in $\mathcal{E}$. The Wigner-Eckart theorem will allow us to expand the second-order Stark shift as an isotropic and an anisotropic contribution.

At the first-order of perturbation, the Stark Hamiltonian $\hat{\mathrm{V}}_{\mathrm{Stark}}=-\mathcal{E}\hat{\mathrm{d}}_{z}=-\mathcal{E}\hat{\mathrm{Q}}_{10}$ brings non-zero energy corrections only for polar molecules, \textit{i.e.} for molecules with a permanent dipole moment in their own frame. For instance, if we consider a diatomic molecule in a given electronic state $e$ and a vibrational level $v$ ($\beta \equiv (ev)$), the electric field $\mathbf{E}$ couples the different rotational levels $J$. Only diagonal matrix elements with respect to $M$ exist because the electric field is oriented along $z$. The corresponding matrix element of the Stark Hamiltonian reads
\begin{equation}
  \left\langle \beta J'M\right|\hat{\mathrm{V}}_{\mathrm{Stark}}
    \left|\beta JM\right\rangle =-\mathcal{E}
    \left\langle \beta J'M\right|\hat{\mathrm{Q}}_{10}
    \left|\beta JM\right\rangle = -\mathcal{E}
    \frac{C_{JM10}^{J'M}}{\sqrt{2J'+1}}
    \left\langle \beta J'\right\Vert \hat{\mathrm{Q}}_{1}
    \left\Vert \beta J\right\rangle .
\end{equation}

At the second order, there exists a non-zero Stark shift $W_{S}$ for all atomic and molecular systems, which can be written as $W_{S}=-\alpha_{1010}\mathcal{E}^{2}/2$, with 
\begin{equation}
  \alpha_{1010}\equiv\alpha_{zz} = 2\sum_{(\beta''J'')\ne(\beta J)}
    \frac{\left\langle \beta JM\right|\hat{\mathrm{Q}}_{10}
      \left|\beta''J''M\right\rangle \left\langle \beta''J''M\right|
      \hat{\mathrm{Q}}_{10}\left|\beta JM\right\rangle }
    {E_{\beta''J''}-E_{\beta J}}\, .
  \label{eq:pert-alpha-zz}
\end{equation}
The indices $(1010)$ appearing next to $\alpha$ refer to those appearing next to the operators $\hat{\mathrm{Q}}$ in the perturbation expansion: $\alpha_{1010}$, or $\alpha_{zz}$, is the $zz$ component of the electric dipole polarizability which expresses the response of the system to the electric field. As above, only matrix elements diagonal in $M$ are involved in the summation. 

Similarly one can show \cite{manakov1986} that the response of the $\left|\beta JM\right\rangle$ level to an oscillating electric field of angular frequency $\omega$ far for any resonance of the system (\textit{i.e} $\hbar \omega \ne |E_{\beta''J''}-E_{\beta J}|\, \forall \beta''J''$), is characterized by the component $\alpha_{1010}(\omega)$ of the dynamic dipole polarizability
\begin{equation}
  \alpha_{1010}(\omega) = 2\sum_{(\beta''J'')\ne(\beta J)} 
   \frac{\left(E_{\beta''J''}-E_{\beta J}\right)
    \left\langle \beta JM\right|\hat{\mathrm{Q}}_{10}
    \left|\beta''J''M\right\rangle
    \left\langle \beta''J''M\right|\hat{\mathrm{Q}}_{10}
    \left|\beta JM\right\rangle }
   {\left(E_{\beta''J''}-E_{\beta J}\right)^{2}
    -\hbar^{2}\omega^{2}}.
  \label{eq:pert-alpha-zz-dyn}
\end{equation}
One can immediately see that setting $\omega=0$ in equation (\ref{eq:pert-alpha-zz-dyn}) yields the static polarizability of equation (\ref{eq:pert-alpha-zz}). For generality we will keep the explicit dependence in $\omega$ in the following.

In subsection \ref{ssec:pert-qlm} we applied the Wigner-Eckart theorem, see equation (\ref{eq:pert-wigneck}), to express all the matrix elements of the multipole-moment operator $\hat{\mathrm{Q}}_{\ell m}$ as functions of the related reduced matrix element. In the case of polarizability, applying the Wigner-Eckart theorem to the two dipole-moment operators of equation (\ref{eq:pert-alpha-zz-dyn}) yields
\begin{align}
  \alpha_{1010}(\omega) & = 2\sum_{\beta''J''}
    \frac{C_{J''M10}^{JM}C_{JM10}^{J''M}}
      {\sqrt{\left(2J+1\right)\left(2J''+1\right)}} 
  \nonumber \\
   & \quad\times \frac{\left(E_{\beta''J''}-E_{\beta J}\right)
      \left\langle \beta J\right\Vert \hat{\mathrm{Q}}_{1}
      \left\Vert \beta''J''\right\rangle 
      \left\langle \beta''J''\right\Vert 
      \hat{\mathrm{Q}}_{1}\left\Vert \beta J\right\rangle}
    {\left(E_{\beta''J''}-E_{\beta J}\right)^{2} 
      -\hbar^{2}\omega^{2}} .
  \label{eq:pert-alpha-zz-dyn-1}
\end{align}
Because the products of Clebsch-Gordan coefficients and the energies of intermediate levels $\beta''J''\rangle$ depend on $J''$, it is not possible to express equation (\ref{eq:pert-alpha-zz-dyn-1}) as a function of a single reduced matrix element. In theoretical calculations, we can overcome that problem by partitioning equation (\ref{eq:pert-alpha-zz-dyn-1}) into sums for each possible value of $J''$ \cite{zhu2004, lepers2011a}. However we can eliminate the $J''$-dependence of the Clebsch-Gordan coefficients by expanding their product as (see Ref.~\cite{varshalovich1988}, p.~261, equation (35) and p.~245, equation (10)),
\begin{equation}
  C_{f\varphi a\alpha}^{d\delta}
   C_{d\delta b\beta}^{e\varepsilon} = 
   \sum_{c\gamma} \left(-1\right)^{c+e+f} 
   \sqrt{\left(2c+1\right)\left(2d+1\right)}
   C_{b\beta a\alpha}^{c\gamma} 
   C_{f\varphi c\gamma}^{e\varepsilon} \sixj{a}{b}{c}{e}{f}{d}
  \label{eq:pert-2cg}
\end{equation}
where the quantity in curly brackets is a Wigner $6j$ symbol. Applying equation (\ref{eq:pert-2cg}) to $a=b=1$, $e=f=J$, $d=J''$, $c=k$ and the corresponding projections, we obtain for equation (\ref{eq:pert-alpha-zz-dyn-1})
\begin{eqnarray}
\alpha_{1010}(\omega) & = & 2\sum_{\beta''J''}\frac{\left(E_{\beta''J''}-E_{\beta J}\right)\left\langle \beta J\right\Vert \hat{\mathrm{Q}}_{1}\left\Vert \beta''J''\right\rangle \left\langle \beta''J''\right\Vert \hat{\mathrm{Q}}_{1}\left\Vert \beta J\right\rangle }{\left(E_{\beta''J''}-E_{\beta J}\right)^{2}-\hbar^{2}\omega^{2}}\nonumber \\
 & \times & \left(-1\right)^{2J}\sum_{k=0,2}\sqrt{\frac{2k+1}{2J+1}}\, C_{1010}^{k0}C_{JMk0}^{JM}\left\{ \begin{array}{ccc}
1 & 1 & k\\
J & J & J''
\end{array}\right\} 
\label{eq:pert-alpha-zz-dyn-2}
\end{eqnarray}
which comprises a new sum over the integer $k$. Generally speaking, $k$ is limited by the triangle inequality with rank-1 dipolar tensor: $k$ can range from 0 to 2. Since its projection is limited to 0 due to the $z$-polarization of the electric field, the value $k=1$ is excluded as $C_{1010}^{10}=0$. By expanding $\alpha_{1010}(\omega)$ as
\begin{equation}
  \alpha_{1010}(\omega) = \sum_{k=0,2}
   \frac{C_{1010}^{k0} C_{JMk0}^{JM}}
    {\sqrt{2J+1}} \alpha_{(11)k}(\omega)
  \label{eq:pert-alpha-zz-dyn-3}
\end{equation}
we recognize the Wigner-Eckart theorem applied to tensors of rank $k=0$ and 2, with reduced matrix elements $\alpha_{(11)k}(\omega)$. Therefore the dipole polarizability $\alpha_{1010}(\omega)$ comes out as a sum of two terms:
\begin{itemize}
\item the first one is a tensor of rank $k=0$. Because $C_{JM00}^{JM}=1$, the $(k=0)$-term does not depend on the azimuthal quantum number $M$; it is \textit{isotropic} and proportional to the so-called scalar polarizability $\alpha_{\mathrm{scal}}(\omega)$ in the case of atoms, or isotropic polarizability $\bar{\alpha}(\omega)$ in the case of molecules. 
\item The second one is a tensor of rank $k=2$ which depends on $M$. It is \textit{anisotropic} and proportional to the so-called tensor polarizability $\alpha_{\mathrm{tens}}(\omega)$ in the case of atoms, or anisotropic polarizability $\Delta\alpha(\omega)$ in the case of molecules. The $(k=2)$-term is non-zero whenever $J\ge1$.
\end{itemize}
The relationships between $\alpha_{(11)0}(\omega)$, $\alpha_{\mathrm{scal}}(\omega)$ or $\bar{\alpha}(\omega)$ on the one hand, and $\alpha_{(11)2}(\omega)$, $\alpha_{\mathrm{tens}}(\omega)$ or $\Delta\alpha(\omega)$ on the other hand, are given for example in Refs.~\cite{manakov1986, angel1968, lepers2012}.

Therefore, in order to characterize the second-order response to a (static or linearly-polarized oscillating) electric field of all the sublevels $M$ of a given level $|\beta J\rangle$, one only needs two quantities: $\alpha_{(11)0}(\omega)$ and $\alpha_{(11)2}(\omega)$. For a circularly-polarized oscillating field, the rank-1 tensor $\alpha_{(11)1}(\omega)$, proportional to the so-called axial or vector polarizability also comes into play; see \cite{manakov1986} for details. The Stark shift for a given sublevel can then be calculated by applying equation (\ref{eq:pert-alpha-zz-dyn-3}), which illustrates the extraordinary convenience of the Wigner-Eckart theorem. The approach presented in this subsection can be generalized from several points of view. Firstly if the electric field has an arbitrary orientation (different from $z$), the Stark Hamiltonian is likely to couple different sublevels $M$ and $M'$. In this case, it is convenient to express the polarizability as an effective operator $\hat{\alpha}_{1m'1m}(\omega)$, with matrix elements $\left\langle \beta JM'\right|\hat{\alpha}_{1m'1m}\left|\beta JM\right\rangle =\sum_{k}\alpha_{(11)k} C_{1m'1m}^{k,m+m'} C_{JMk,m+m'}^{JM'} /\sqrt{2J+1}$. Secondly, the dipole polarizability introduced here can be extended to any type of multipole moments. Since the long-range interactions
between two partners is intrinsically due to the interaction of one partner in the electric field created by the other, such polarizabilities will come into play in the calculation of second-order energy corrections (see subsection \ref{ssec:pert-2nd}).

\subsection{First-order energy correction from long-range interactions}
\label{ssec:pert-1st}

We want to calculate the energy correction due to the perturbation operator $\hat{\mathrm{V}}(R)\equiv\hat{\mathrm{V}}_{AB}^{\mathrm{BF}}(R)$, see equation (\ref{eq:mult-vab-bf}), on the unperturbed energies $E_{q}^{(0)}=E_{\beta_{A}J_{A}}+E_{\beta_{B}J_{B}}$ of the $\left(2J_{A}+1\right)\left(2J_{B}+1\right)$-time degenerate states of the complex $|\Psi_{q}^{(0)}\rangle\equiv|\beta_{A}J_{A}M_{A}\beta_{B}J_{B}M_{B}\rangle$. Therefore, unless $J_{A}=J_{B}=0$, perturbation theory for degenerate levels must be employed. The first-order energy corrections $E_{q,i}^{(1)}$ are the eigenvalues of the perturbation operator $\hat{\mathrm{V}}_{AB}^{\mathrm{BF}}(R)$ restricted to the ``subspace of degeneracy'' of dimension $\left(2J_{A}+1\right)\left(2J_{B}+1\right)$ spanned by the possible values of $M_{A}$ and $M_{B}$ for given $\beta_{A}$, $J_{A}$, $\beta_{B}$ and $J_{B}$. The corresponding matrix elements are
\begin{eqnarray}
 &  & \left\langle \beta_{A}J_{A}M'_{A}\beta_{B}J_{B}M'_{B}\right|\hat{\mathrm{V}}_{AB}^{\mathrm{BF}}(R)\left|\beta_{A}J_{A}M_{A}\beta_{B}J_{B}M_{B}\right\rangle \nonumber \\
 & = & \frac{1}{4\pi\epsilon_{0}}\sum_{\ell_{A},\ell_{B}=0}^{+\infty}\sum_{m=-\ell_{<}}^{+\ell_{<}}\frac{f_{\ell_{A}\ell_{B}m}}{R^{1+\ell_{A}+\ell_{B}}}\frac{C_{J_{A}M_{A}\ell_{A}m}^{J_{A}M'_{A}}C_{J_{B}M_{B}\ell_{B}-m}^{J_{B}M'_{B}}}{\sqrt{\left(2J_{A}+1\right)\left(2J_{B}+1\right)}}\nonumber \\
 &  & \times\left\langle \beta_{A}J_{A}\right\Vert \hat{\mathrm{Q}}_{\ell_{A}}\left\Vert \beta_{A}J_{A}\right\rangle \left\langle \beta_{B}J_{B}\right\Vert \hat{\mathrm{Q}}_{\ell_{B}}\left\Vert \beta_{B}J_{B}\right\rangle 
\label{eq:pert-1st}
\end{eqnarray}
where $f_{\ell_{A}\ell_{B}m}$ is given by equation (\ref{eq:mult-flalbm}). This expression contains only diagonal reduced matrix elements of the multipole-moment operators, namely the \textit{permanent multipole moments} of partners A and B in their state $\left| \beta_{A}J_{A}\right\rangle$ and  $\left| \beta_{B}J_{B}\right\rangle$, respectively. The Clebsch-Gordan coefficients do not vanish for a single value of $m=M'_{A}-M_{A}=M_{B}-M'_{B}$, hence 
\begin{equation}
  M'_{A}+M'_{B}=M_{A}+M_{B}\,.
  \label{eq:pert-M-const}
\end{equation}
The projection of the total angular momentum of the complex on the inter-partner axis $z$ is thus unaffected by the first-order energy correction. At this stage, the first-order perturbed states of the complex has the same axial symmetry properties than a diatomic molecule. It is worth mentioning that equation (\ref{eq:pert-1st}) also possesses the other symmetries of the diatomic molecule: the matrix elements may be odd or even after (i) a reflection with respect to any plane containing the inter-partner axis (``+/-'' symmetries); and (ii) inversion with respect to the center-of-mass of
the complex if A and B are identical (``g/u'' symmetries).

A particular case deserves attention. For neutral partners, \textit{i.e.}~$\langle\hat{\mathrm{Q}}_{00}(A,B)\rangle=0$, and if $J_{A}=0$ or $J_{B}=0$, at least one of the Clebsch-Gordan coefficients of equation (\ref{eq:pert-1st}) is zero, and so the energies will be modified only at the second order of perturbation. This will be the case if one partner at least is an atom in a $S$ state ($L=0$) or in the $J=0$ component of a $^{3}P_{J}$ or $^{5}D_{J}$ multiplets, or a molecule in its $J=0$ rotational level (regardless of its Hund's case).

\subsection{Second-order energy correction from long-range interactions}
\label{ssec:pert-2nd}

Due to the $\left(2J_{A}+1\right)\left(2J_{B}+1\right)$-fold degeneracy of the unperturbed levels, these corrections must be calculated in the framework of the degenerate second-order perturbation theory \cite{landau1967}. To that end, one has to diagonalize the effective operator $\hat{\mathrm{W}}_{AB}^{\mathrm{BF}}(R)$, see equation (\ref{eq:W_AB}), in the subspaces of degeneracy already mentioned in subsection \ref{ssec:pert-1st}. The $\hat{\mathrm{W}}_{AB}^{\mathrm{BF}}(R)$ operator reads
\begin{align}
\hat{\mathrm{W}}_{AB}^{\mathrm{BF}}(R) & =  -\sum_{\beta''_{A}J''_{A}M''_{A}}\sum_{\beta''_{B}J''_{B}M''_{B}}\frac{\hat{\mathrm{V}}_{AB}^{\mathrm{BF}}\left|\beta''_{A}J''_{A}M''_{A}\beta''_{B}J''_{B}M''_{B}\right\rangle \left\langle \beta''_{A}J''_{A}M''_{A}\beta''_{B}J''_{B}M''_{B}\right|\hat{\mathrm{V}}_{AB}^{\mathrm{BF}}}{E_{\beta''_{A}J''_{A}}-E_{\beta_{A}J_{A}}+E_{\beta''_{B}J''_{B}}-E_{\beta_{B}J_{B}}}\nonumber \\
 & = -\frac{1}{16\pi^{\text{2}}\epsilon_{0}^{2}}\sum_{\ell'_{A}\ell'_{B}m'}\sum_{\ell_{A}\ell_{B}m}\frac{f_{\ell'_{A}\ell'_{B}m'}f_{\ell_{A}\ell_{B}m}}{R^{2+\ell'_{A}+\ell'_{B}+\ell_{A}+\ell_{B}}}\sum_{\beta''_{A}J''_{A}M''_{A}}\sum_{\beta''_{B}J''_{B}M''_{B}}  \nonumber \\
 & \quad \frac{\hat{\mathrm{Q}}_{\ell'_{A}m'}^{\textrm{BF}}\left|\beta''_{A}J''_{A}M''_{A}\right\rangle \left\langle \beta''_{A}J''_{A}M''_{A}\right|\hat{\mathrm{Q}}_{\ell_{A}m}^{\textrm{BF}}\times\hat{\mathrm{Q}}_{\ell'_{B}-m'}^{\textrm{BF}}\left|\beta''_{B}J''_{B}M''_{B}\right\rangle \left\langle \beta''_{B}J''_{B}M''_{B}\right|\hat{\mathrm{Q}}_{\ell_{B}-m}^{\textrm{BF}}}{E_{\beta''_{A}J''_{A}}-E_{\beta_{A}J_{A}}+E_{\beta''_{B}J''_{B}}-E_{\beta_{B}J_{B}}} \nonumber \\
 & & \label{eq:pert-2nd-oper}
\end{align}
where the levels of the complex appearing in the sum are such that $|\beta''_{A}J''_{A}\beta''_{B}J''_{B}\rangle \ne |\beta_{A}J_{A}\beta_{B}J_{B}\rangle$, to ensure that the denominators do not vanish. Note that in the last line, we have separated the intermediate levels of A and those of B.

We see that the second-order effective operator imposes the same selection rule as the first-order one, see equation
(\ref{eq:pert-1st}), in particular $M_{A}+M_{B}=M'_{A}+M'_{B}$. We already identify in equation (\ref{eq:pert-2nd-oper}) the dependence on inverse $R$ powers of the second-order energy correction, which can be illustrated on two particular cases:
\begin{itemize}
\item For two neutral partners, the leading term comes from $\ell_{A}=\ell'_{A}=\ell_{B}=\ell'_{B}=1$. It scales as $R^{-6}$, and it is usually referred to as the \textit{van der Waals interaction} (as invoked in the introduction of the chapter). It involves all the levels of the partners A and B that are coupled by electric-dipole transitions with the level for which we calculate the energy correction. Moreover, it is effective for any pair of (charged or neutral) partners, except for point-like ones such as single electrons and nuclei H$^{+}$, He$^{2+}$, etc, which cannot present an induced dipole. 
\item If one of the partners, say A, is an ion, $\ell_{A}=\ell'_{A}=0$, and the other a neutral particle, then the leading term comes from $\ell_{B}=\ell'_{B}=1$, and it scales as $R^{-4}$. 
\end{itemize}

The sum in equation (\ref{eq:pert-2nd-oper}) runs over intermediate levels $|\beta''_{A}J''_{A}M''_{A} \beta''_{B}J''_{B}M''_{B}\rangle$ of the complex with different unperturbed energies from the level $|\beta_{A}J_{A}M_{A} \beta_{B}J_{B}M_{B}\rangle$, which can result in two situations. Either one partner lies in an intermediate level, say $(\beta''_{A}J''_{A})\ne(\beta_{A}J_{A})$, while the other does not, $(\beta''_{B}J''_{B})=(\beta_{B}J_{B})$, which corresponds to \textit{induction} interaction; or both partners lie in intermediate levels, $(\beta''_{A}J''_{A})\ne(\beta_{A}J_{A})$ and $(\beta''_{B}J''_{B})\ne(\beta_{B}J_{B})$, which corresponds to \textit{dispersion} interaction. In the dipole-dipole case $\ell_A = \ell'_A = \ell_B = \ell'_B = 1$, the induction and dispersion interactions are respectively called Keesom and London interactions. If both types of interactions are present in the sum for a given set of indices $(\ell'_{A},\ell_{A},\ell'_{B},\ell_{B})$, the dispersion interaction is often much stronger than the induction interaction  \cite{kaplan2006, stone1996}. A simple explanation is the following: if we assume that there are $N$ levels $|\beta''_{A}J''_{A}\rangle$ and $|\beta''_{B}J''_{B}\rangle$ for each partner, the induction energy comprises $N$ terms, while the dispersion one comprises $N^{2}$ terms.

\subsubsection{The induction energy}

If $|\beta''_{B}J''_{B}\rangle = |\beta_{B}J_{B}\rangle$, the denominator in equation (\ref{eq:pert-2nd-oper}) is equal to $E_{\beta''_{A}J''_{A}}-E_{\beta_{A}J_{A}}$, so that the contributions from the two partners can be factorized, yielding the matrix elements of the corresponding $\hat{\mathrm{W}}_{B\to A}^{\mathrm{BF}}(R)$ operator
\begin{eqnarray}
 &  & \left\langle \beta_{A}J_{A}M'_{A}\beta_{B}J_{B}M'_{B}\right|\hat{\mathrm{W}}_{B\to A}^{\mathrm{ind,BF}}(R)\left|\beta_{A}J_{A}M_{A}\beta_{B}J_{B}M_{B}\right\rangle \nonumber \\
 & = & -\frac{1}{16\pi^{\text{2}}\epsilon_{0}^{2}}\sum_{\ell'_{A}\ell'_{B}m'}\sum_{\ell_{A}\ell_{B}m}\frac{f_{\ell'_{A}\ell'_{B}m'}f_{\ell_{A}\ell_{B}m}}{R^{2+\ell'_{A}+\ell'_{B}+\ell_{A}+\ell_{B}}}\nonumber \\
 & \times & \sum_{\beta''_{A}J''_{A}M''_{A}}\frac{\left\langle \beta_{A}J_{A}M'_{A}\right|\hat{\mathrm{Q}}_{\ell'_{A}m'}^{\textrm{BF}}\left|\beta''_{A}J''_{A}M''_{A}\right\rangle \left\langle \beta''_{A}J''_{A}M''_{A}\right|\hat{\mathrm{Q}}_{\ell_{A}m}^{\textrm{BF}}\left|\beta_{A}J_{A}M_{A}\right\rangle }{E_{\beta''_{A}J''_{A}}-E_{\beta_{A}J_{A}}}\nonumber \\
 & \times & \sum_{M''_{B}}\left\langle \beta_{B}J_{B}M'_{B}\right|\hat{\mathrm{Q}}_{\ell'_{B}-m'}^{\textrm{BF}}\left|\beta_{B}J_{B}M''_{B}\right\rangle \left\langle \beta_{B}J_{B}M''_{B}\right|\hat{\mathrm{Q}}_{\ell_{B}-m}^{\textrm{BF}}\left|\beta_{B}J_{B}M_{B}\right\rangle .
\label{eq:pert-ind}
\end{eqnarray}
Note that the sum over $M''_{B}$ is still present, as the unperturbed energies do not depend on $M''_{B}$. The last line of the equation contains a product of permanent multipole moments of the partner B, and the last but one line the transition multipole moments from $|\beta_{B}J_{B}\rangle$ towards the $|\beta''_{A}J''_{A}\rangle$ levels. Generalizing the definition of the static dipole polarizability (subsection \ref{ssec:efield}) to the case $\ell_{A}$ and $\ell'_{A}\neq 1$, we can introduce the effective operator for static multipole polarizability $\hat{\alpha}_{\ell'_{A}m'\ell_{A}m}^{\textrm{BF}}(A;0)$ 
\begin{equation}
  \hat{\alpha}_{\ell'_{A}m'\ell_{A}m}^{\textrm{BF}}(A;0)
    = 2\sum_{\beta''_{A}J''_{A}M''_{A}} 
    \frac{\hat{\mathrm{Q}}_{\ell'_{A}m'}^{\textrm{BF}}\left|
      \beta''_{A}J''_{A}M''_{A}\right\rangle 
      \left\langle \beta''_{A}J''_{A}M''_{A}\right|
      \hat{\mathrm{Q}}_{\ell_{A}m}^{\textrm{BF}}}
    {E_{\beta''_{A}J''_{A}}-E_{\beta_{A}J_{A}}} \,.
\end{equation}
which leads to a more compact form of equation (\ref{eq:pert-ind})
\begin{eqnarray}
 &  & \left\langle \beta_{A}J_{A}M'_{A}\beta_{B}J_{B}M'_{B}\right|\hat{\mathrm{W}}_{B\to A}^{\mathrm{ind,BF}}(R)\left|\beta_{A}J_{A}M_{A}\beta_{B}J_{B}M_{B}\right\rangle \nonumber \\
 & = & -\frac{1}{16\pi^{\text{2}}\epsilon_{0}^{2}}\sum_{\ell'_{A}\ell'_{B}m'}\sum_{\ell_{A}\ell_{B}m}\frac{f_{\ell'_{A}\ell'_{B}m'}f_{\ell_{A}\ell_{B}m}}{R^{2+\ell'_{A}+\ell'_{B}+\ell_{A}+\ell_{B}}}\nonumber \\
 & \times & \frac{1}{2}\left\langle \beta_{A}J_{A}M'_{A}\right|\hat{\alpha}_{\ell'_{A}m'\ell_{A}m}^{\textrm{BF}}(A;0)\left|\beta_{A}J_{A}M_{A}\right\rangle \nonumber \\
 & \times & \sum_{M''_{B}}\left\langle \beta_{B}J_{B}M'_{B}\right|\hat{\mathrm{Q}}_{\ell'_{B}-m'}^{\textrm{BF}}\left|\beta_{B}J_{B}M''_{B}\right\rangle \left\langle \beta_{B}J_{B}M''_{B}\right|\hat{\mathrm{Q}}_{\ell_{B}-m}^{\textrm{BF}}\left|\beta_{B}J_{B}M_{B}\right\rangle .
\label{eq:pert-ind-pol}
\end{eqnarray}
Using a classical image, the induction energy in equation (\ref{eq:pert-ind}) is due to the influence of the permanent multipoles of B, which distorts the electronic cloud of A, hence inducing multipoles. This justifies the symbol $B\to A$ in the name of the operator $\hat{\mathrm{W}}_{B\to A}^{\mathrm{ind,BF}}(R)$. Obviously there also exists the reverse interaction $\hat{\mathrm{W}}_{A\to B}^{\mathrm{ind,BF}}(R)$, obtained by interchanging A and B in equation (\ref{eq:pert-ind}).

\subsubsection{Dispersion energy}

Unlike the previous case, the operator $\hat{\mathrm{W}}_{AB}^{\mathrm{BF}}(R)$ cannot \textit{a priori} be factorized using A and B properties. However one can use the following mathematical identity obtained with the theorem of residuals,
\begin{equation}
\frac{1}{a+b}=\frac{2}{\pi}\int_{0}^{+\infty}du\frac{ab}{\left(a^{2}+u^{2}\right)\left(b^{2}+u^{2}\right)}
\label{eq:pert-resid}
\end{equation}
valid for $a,b>0$. Plugging equation (\ref{eq:pert-resid}) into equation (\ref{eq:pert-2nd-oper}), with $a=E_{\beta''_{A}J''_{A}}-E_{\beta_{A}J_{A}}$, $b=E_{\beta''_{B}J''_{B}}-E_{\beta_{B}J_{B}}$ and $u=\hbar\omega$, one recognizes a generalization of equation (\ref{eq:pert-alpha-zz-dyn}) with operators for the dynamic multipole polarizabilities $\hat{\alpha}_{\ell'_{A} m'\ell_{A} m}^{\textrm{BF}}(A;i\omega)$ and $\hat{\alpha}_{\ell'_{B}-m'\ell_{B}-m}^{\textrm{BF}}(B;i\omega)$, so that the dispersion effective operator takes the compact expression
\begin{eqnarray}
  \hat{\mathrm{W}}_{AB}^{\mathrm{disp,BF}}(R) & = &
    -\frac{\hbar}{32\pi^{3}\epsilon_{0}^{2}} 
    \sum_{\ell'_{A}\ell'_{B}m'} \sum_{\ell_{A}\ell_{B}m}
    \frac{f_{\ell'_{A}\ell'_{B}m'}f_{\ell_{A}\ell_{B}m}}
      {R^{2+\ell'_{A}+\ell'_{B}+\ell_{A}+\ell_{B}}}\nonumber \\
   & \times & \int_{0}^{+\infty}d\omega
    \hat{\alpha}_{\ell'_{A} m'\ell_{A} m}^{\textrm{BF}}(A;i\omega)
    \hat{\alpha}_{\ell'_{B}-m'\ell_{B}-m}^{\textrm{BF}}(B;i\omega).
  \label{eq:pert-disp-1}
\end{eqnarray}
Note that the plus signs in the denominator of equation (\ref{eq:pert-resid}) are responsible for the \textit{imaginary} frequencies $i\omega$.

The dispersion energy again appears as a product of the individual properties of the partners. However, unlike permanent multipole moments or dynamic multipole polarizabilities at \textit{real} frequencies, the dynamic multipole polarizabilities at \textit{imaginary} frequencies do not have a real physical meaning and they cannot be measured experimentally. They represent a useful quantity which can only be calculated theoretically, and which are widely used in order to compute dispersion energies.

We imposed an important restriction in the identity (\ref{eq:pert-resid}): $a>0$ and $b>0$. This means that equation (\ref{eq:pert-disp-1}) is valid only if the two partners are in their lowest energy level. If for instance B lies in an excited level $|\beta_{B}J_{B}\rangle$, at least one level $|\beta''_{B}J''_{B}\rangle$ is such that $E_{\beta''_{B}J''_{B}}-E_{\beta_{B}J_{B}}<0$. Thus for $a>0$ and $b<0$, one can write the simple identity
\begin{equation}
\frac{1}{a+b}=\frac{1}{\left|a\right|-\left|b\right|}=-\frac{\left|a\right|+\left|b\right|}{\left(\left|a\right|+\left|b\right|\right)\left(\left|b\right|-\left|a\right|\right)}=-\frac{1}{\left|a\right|+\left|b\right|}-\frac{2a}{b^{2}-a^{2}}.\label{eq:pert-resid-2}
\end{equation}
The first term of the right-hand side can be rewritten using equation (\ref{eq:pert-resid}), with the numerator of the integrand $-\left|a\right|\left|b\right|=ab$. Assuming again $a=E_{\beta''_{A}J''_{A}}-E_{\beta_{A}J_{A}}$ and $b=E_{\beta''_{B}J''_{B}}-E_{\beta_{B}J_{B}}$, one retrieves the product of the multipole polarizabilities at imaginary frequencies. The term $-2a/(b^2-a^2) = 2a/(a^2-b^2)$ results in the dynamic polarizability of partner A, taken at the \textit{real} frequencies corresponding to the transitions from $|\beta_{B}J_{B}\rangle$ to lower-energy levels of partner B \cite{zhu2004, lepers2011a}. Finally the dispersion effective operator reads
\begin{eqnarray}
  \hat{\mathrm{W}}_{AB}^{\mathrm{disp,BF}}(R) & = & 
    - \frac{1}{16\pi^{2}\epsilon_{0}^{2}} 
    \sum_{\ell'_{A}\ell'_{B}m'}\sum_{\ell_{A}\ell_{B}m}
    \frac{f_{\ell'_{A}\ell'_{B}m'}f_{\ell_{A}\ell_{B}m}}
         {R^{2+\ell'_{A}+\ell'_{B}+\ell_{A}+\ell_{B}}} \nonumber \\
   & \times & \left[\frac{\hbar}{2\pi}\int_{0}^{+\infty}d\omega
    \hat{\alpha}_{\ell'_{A} m'\ell_{A} m}^{\textrm{BF}}(A;i\omega)
    \hat{\alpha}_{\ell'_{B}-m'\ell_{B}-m}^{\textrm{BF}}(B;i\omega) \phantom{\sum_{\omega''_{B}}}
  \right.\nonumber \\
 &  & \left.+\sum_{B'',\,\omega''_{B}<0} 
    \hat{\alpha}_{\ell'_{A}m'\ell_{A}m}^{\textrm{BF}}(A;\omega''_{B}) 
    \hat{\mathrm{Q}}_{\ell'_{B}-m'}^{\textrm{BF}}
    \left|B''\right\rangle \left\langle B''\right|
    \hat{\mathrm{Q}}_{\ell_{B}-m}^{\textrm{BF}}\right]
  \label{eq:pert-disp-2}
\end{eqnarray}
where $B''\equiv(\beta''_{B}J''_{B}M''_{B})$ and $\hbar\omega''	_{B}=E_{\beta''_{B}J''_{B}}-E_{\beta_{B}J_{B}}$. Obviously if A lies in an excited level and B in the ground level, A and B should be interchanged in equation (\ref{eq:pert-disp-2}).

\subsubsection{Second-order energy correction and irreducible tensors}

In subsection \ref{ssec:efield}, we demonstrated that the second-order Stark shift due to an electric field applied on a single partner could be expanded as a sum of tensorial terms with a well-defined rank (see equation \eqref{eq:pert-alpha-zz-dyn-2}), which allowed for the use of the Wigner-Eckart theorem. This expansion is also applicable to the effective operators defining the second-order long-range interactions between partners A and B, leading to a sum over three indices: the tensor rank $k_{A}$ ($k_{B}$) associated to the partner A (B), and the one $k$ associated to the complex. Here we will work out equation (\ref{eq:pert-2nd-oper}), in order to give a general expression of the matrix elements of the second-order effective operator $\hat{\mathrm{W}}_{AB}^{\mathrm{BF}}(R)$. It may be extended in a straightforward manner to the  dispersion or induction interactions.

Staring from equation (\ref{eq:pert-2nd-oper}), we apply the Wigner-Eckart theorem to the four multipole-moment matrix elements, and we introduce the expressions for $f_{\ell'_A\ell'_Bm'}$ and $f_{\ell_A\ell_Bm}$ (see the first two line of equation (\ref{eq:mult-flalbm})), which gives
\begin{eqnarray}
  &  & \langle\beta_{A}J_{A}M'_{A}\beta_{B}J_{B}M'_{B}|
    \hat{\mathrm{W}}_{AB}^{\mathrm{BF}}(R)|
    \beta_{A}J_{A}M_{A}\beta_{B}J_{B}M_{B}\rangle \nonumber \\
  & = & -\frac{1}{16\pi^{\text{2}}\epsilon_{0}^{2}}
    \sum_{\ell'_{A}\ell'_{B}m'}\sum_{\ell_{A}\ell_{B}m}
    \frac{\left(-1\right)^{\ell'_{B}+\ell_{B}}}
      {R^{2+\ell'_{A}+\ell'_{B}+\ell_{A}+\ell_{B}}}
    \sqrt{\frac{\left(2\ell'_{A}+2\ell'_{B}\right)!
      \left(2\ell_{A}+2\ell_{B}\right)!}
      {\left(2\ell'_{A}\right)!\left(2\ell'_{B}\right)!
       \left(2\ell_{A}\right)!\left(2\ell_{B}\right)!}}
  \nonumber \\
  & \times & C_{\ell'_{A}m'\ell'_{B}-m'}^{\ell'_{A}+\ell'_{B},0}
    C_{\ell_{A}m\ell_{B}-m}^{\ell_{A}+\ell_{B},0}
    \sum_{\beta''_{A}J''_{A}} \sum_{\beta''_{B}J''_{B}}
    \left\langle \beta_{A}J_{A}\right\Vert
    \hat{\mathrm{Q}}_{\ell'_{A}}^{\textrm{BF}}
    \left\Vert\beta''_{A}J''_{A}\right\rangle 
  \nonumber \\
  & \times & \frac{ \left\langle \beta''_{A}J''_{A}\right\Vert
    \hat{\mathrm{Q}}_{\ell_{A}}^{\textrm{BF}}
    \left\Vert\beta_{A}J_{A}\right\rangle 
    \left\langle\beta_{B}J_{B}\right\Vert 
    \hat{\mathrm{Q}}_{\ell'_{B}}^{\textrm{BF}}
    \left\Vert \beta''_{B}J''_{B}\right\rangle
    \left\langle \beta''_{B}J''_{B}\right\Vert
    \hat{\mathrm{Q}}_{\ell_{B}}^{\textrm{BF}}
    \left\Vert\beta_{B}J_{B}\right\rangle}
    {E_{\beta''_{A}J''_{A}}-E_{\beta_{A}J_{A}}	
    +E_{\beta''_{B}J''_{B}}-E_{\beta_{B}J_{B}}}
  \nonumber \\
  & \times & \sum_{M''_{A}} \frac{
    C_{J''_A M''_A\ell'_A m'}^{J_A M'_A}
    C_{J_A M_A\ell_A m}^{J''_A M''_A}}
      {\sqrt{\left(2J_A+1\right)\left(2J''_A+1\right)}}
    \sum_{M''_{B}}
    \frac{ C_{J''_B M''_B\ell'_B -m'}^{J_B M'_B}
    C_{J_B M_B\ell_B -m}^{J''_B M''_B}}
      {\sqrt{\left(2J_B+1\right)\left(2J''_B+1\right)}}. 
  \label{eq:pert-2nd-oper-1}
\end{eqnarray}
The last line contains two products of Clebsch-Gordan coefficients similar to the one appearing in the polarizability, see equation (\ref{eq:pert-alpha-zz-dyn-1}). Therefore we can apply equation (\ref{eq:pert-2cg}) to: $a=\ell_{A}$, $b=\ell'_{A}$, $e=J'_{A}$, $f=J_{A}$, $d=J''_{A}$, $c=k_{A}$ on the one hand, and to the same indices, except replacing
A by B, on the other hand, which gives
\begin{eqnarray}
  & & \frac{C_{J''_A M''_A\ell'_A m'}^{J_A M'_A}
    C_{J_A M_A\ell_A m}^{J''_A M''_A}
    C_{J''_B M''_B\ell'_B -m'}^{J_B M'_B}
    C_{J_B M_B\ell_B -m}^{J''_B M''_B}}
  {\sqrt{\left(2J_A+1\right)\left(2J''_A+1\right)
         \left(2J_B+1\right)\left(2J''_B+1\right)}}
  \nonumber \\
  & \times & \sum_{k_Ak_B} 
    \left(-1\right)^{2J_A+k_A+2J_B+k_B}
    \sqrt{\frac{\left(2k_A+1\right)\left(2k_B+1\right)}
      {\left(2J_A+1\right)\left(2J_B+1\right)}}
    \sixj{\ell_A}{\ell'_A}{k_A}{J_A}{J_A}{J''_A}
  \nonumber\\
  & \times & \sixj{\ell_B}{\ell'_B}{k_B}{J_B}{J_B}{J''_B} 
    \sum_{q_Aq_B}
    C_{\ell'_Am'\ell_Am}^{k_Aq_A}C_{\ell'_B-m'\ell_B-m}^{k_Bq_B}
    C_{J_AM_Ak_Aq_A}^{J_AM'_A}C_{J_BM_Bk_Bq_B}^{J_BM'_B}
  \label{eq:pert-2nd-oper-1-last}
\end{eqnarray}
We obtain sums over the two indices $k_A$ and $k_B$ which contain the coefficients $C_{J_AM_Ak_Aq_A}^{J_AM'_A}$ and $C_{J_BM_Bk_Bq_B}^{J_BM'_B}$, characteristic of the Wigner-Eckart theorem applied for tensors of ranks $k_A$ and $k_B$ respectively. The Clebsch-Gordan coefficients $C_{\ell'_Am'\ell_Am}^{k_Aq_A}$ and $C_{\ell'_B-m'\ell_B-m}^{k_Bq_B}$ show that $k_A$ and $k_B$ are constructed by angular-momentum additions of $\ell'_A$, $\ell_A$ and $\ell'_B$, $\ell_B$.

By introducing equation (\ref{eq:pert-2nd-oper-1-last}) into (\ref{eq:pert-2nd-oper-1}), we can see that the indices $m$ and $m'$ only appear in the sum of products of four Clebsch-Gordan coefficients $\sum_{mm'} = C_{\ell'_{A}m'\ell'_{B}-m'}^{\ell'_{A}+\ell'_{B},0} C_{\ell_{A}m\ell_{B}-m}^{\ell_{A}+\ell_{B},0} C_{\ell'_Am'\ell_Am}^{k_Aq_A} C_{\ell'_B-m'\ell_B-m}^{k_Bq_B}$. We can calculate the latter by using the relationship (see \cite{varshalovich1988}, p.~260, equation (20)): 
\begin{align}
  \sum_{\beta\gamma\varepsilon\varphi}
   C_{b\beta c\gamma}^{a\alpha}
   C_{e\varepsilon f\varphi}^{d\delta}
   C_{e\varepsilon b\beta}^{g\eta}
   C_{f\varphi c\gamma}^{j\mu}
   & = \sqrt{\left(2a+1\right)\left(2d+1\right)
    \left(2g+1\right)\left(2j+1\right)}
  \nonumber \\
   & \times \sum_{k\kappa} C_{g\eta j\mu}^{k\kappa}
   C_{d\delta a\alpha}^{k\kappa}
   \ninej{c}{b}{a}{f}{e}{d}{j}{g}{k}
  \label{eq:pert-4cg}
\end{align}
where the number between curly brackets is a Wigner $9j$ symbol, with $a=k_{B}$, $b=\ell'_{B}$, $c=\ell_{B}$, $d=k_{A}$, $e=\ell'_{A}$, $f=\ell_{A}$, $g=\ell'_{A}+\ell'_{B}$ and $j=\ell_{A}+\ell_{B}$. Equation (\ref{eq:pert-4cg}) allows us for getting rid of $\varepsilon=-\beta=-m'$ and $\varphi=-\gamma=-m$; here $\eta=\mu=0$ which imposes $\kappa=0$, and so $\delta=-\alpha\equiv q$. Finally we get to the expression for the matrix elements
\begin{eqnarray}
  &  & \langle\beta_{A}J_{A}M'_{A}\beta_{B}J_{B}M'_{B}|
    \hat{\mathrm{W}}_{AB}^{\mathrm{BF}}(R)|
    \beta_{A}J_{A}M_{A}\beta_{B}J_{B}M_{B}\rangle \nonumber \\
  & = & -\frac{1}{16\pi^{\text{2}}\epsilon_{0}^{2}}
    \sum_{\ell'_{A}\ell'_{B}}\sum_{\ell_{A}\ell_{B}}
    \frac{\left(-1\right)^{\ell'_B+\ell_B+2J_A+2J_B}}
      {R^{2+\ell'_{A}+\ell'_{B}+\ell_{A}+\ell_{B}}}
    \sqrt{\frac{\left(2\ell'_{A}+2\ell'_{B}+1\right)!
      \left(2\ell_{A}+2\ell_{B}+1\right)!}
      {\left(2\ell'_{A}\right)!\left(2\ell'_{B}\right)!
       \left(2\ell_{A}\right)!\left(2\ell_{B}\right)!}}\nonumber \\
  & \times & \sum_{k_{A}k_{B}kq} \left(-1\right)^{k_{A}+k_{B}}
    \left(2k_{A}+1\right)\left(2k_{B}+1\right)
    C_{(\ell'_{A}+\ell'_{B})0(\ell_{A}+\ell_{B})0}^{k0}
    C_{k_{A}qk_{B}-q}^{k0}
    \ninej{\ell'_A        }{\ell_A       }{k_A}
          {\ell'_B        }{\ell_B       }{k_B}
          {\ell'_A+\ell'_B}{\ell_A+\ell_B}{k} \nonumber \\
  & \times & \sum_{\beta''_{A}J''_{A}}\sum_{\beta''_{B}J''_{B}}
    \frac{\left\langle \beta_{A}J_{A}\right\Vert
      \hat{\mathrm{Q}}_{\ell'_{A}}^{\textrm{BF}}
      \left\Vert \beta''_{A}J''_{A} \right\rangle
      \left\langle \beta''_{A}J''_{A} \right\Vert
      \hat{\mathrm{Q}}_{\ell_{A}}^{\textrm{BF}}
      \left\Vert \beta_{A}J_{A}\right\rangle
      \left\langle \beta_{B}J_{B}\right\Vert
      \hat{\mathrm{Q}}_{\ell'_{B}}^{\textrm{BF}}
      \left\Vert \beta''_{B}J''_{B}\right\rangle 
      \left\langle \beta''_{B}J''_{B}\right\Vert 
      \hat{\mathrm{Q}}_{\ell_{B}}^{\textrm{BF}}
      \left\Vert \beta_{B}J_{B}\right\rangle }
    {E_{\beta''_{A}J''_{A}}-E_{\beta_{A}J_{A}}
    +E_{\beta''_{B}J''_{B}}-E_{\beta_{B}J_{B}}} \nonumber \\
  & \times & \sixj{\ell_A}{\ell'_A}{k_A}{J_A}{J_A}{J''_A}
    \sixj{\ell_B}{\ell'_B}{k_B}{J_B}{J_B}{J''_B}
    \frac{C_{J_{A}M_{A}k_{A}q}^{J_{A}M'_{A}}
      C_{J_{B}M_{B}k_{B}-q}^{J_{B}M'_{B}}}
    {\sqrt{\left(2J_{A}+1\right)\left(2J_{B}+1\right)}},
  \label{eq:pert-2nd-oper-2}
\end{eqnarray}
where we used the invariance of $9j$ symbols with respect to a line permutation followed by a column permutation.

Obviously,	 equation (\ref{eq:pert-2nd-oper-2}) exhibits the same dependence in inverse powers of $R$ as equation (\ref{eq:pert-2nd-oper}), with exponents depending on the ranks $\ell'_{A}$, $\ell_{A}$, $\ell'_{B}$ and $\ell_{B}$ of the multipole-moment tensor operators. By contrast, the tensors on which the Wigner-Eckart theorem is applied have ranks equal to $k_{A}$ and $k_{B}$, constructed by angular-momentum-type additions of $\ell'_{A}$ and $\ell_{A}$ on the one hand, $\ell'_{B}$ and $\ell_{B}$ on the other hand. If $k_{A}=k_{B}=0$, the effective operator $\hat{\mathrm{W}}_{AB}^{\mathrm{BF}}(R)$ is diagonal, which means that the corresponding interaction is isotropic, exactly like the charge-charge interaction.

Table \ref{tab:pert-2nd} displays some examples of commonly met second-order interactions with the possible indices $k_{A}$, $k_{B}$ and $k$. The Clebsch-Gordan coefficients of equation (\ref{eq:pert-2nd-oper-2}) imply that $\ell'_{A}+\ell_{A}+\ell'_{B}+\ell_{B}+k$ and $k_{A}+k_{B}+k$ are both even. The induction interaction in $R^{-4}$ occurs between a charge and a dipole-polarizable neutral system: it is usually the dominant interaction in ion-neutral complexes. The term in $R^{-6}$ describes the second-order correction originating from the dipole-dipole interaction between neutral partners, including the dispersion interaction, and the induction interaction if one of the partners is a polar molecule.

\begin{table}
\begin{centering}
\begin{tabular}{|c|c|c|c|c|}
\hline 
$R$-dependence & interaction name & $(\ell'_{A},\ell_{A},\ell'_{B},\ell_{B})$ & $(k_{A},k_{B})$ & $k$\tabularnewline
\hline 
$R^{-4}$ & charge-induced dipole & $(0,0,1,1)$ & $(0,0)$ & 0 \\
         &                       &             & $(0,2)$ & 2 \\
         & induced dipole-charge & $(1,1,0,0)$ & $(0,0)$ & 0 \\
         &                       &             & $(2,0)$ & 2 \\
$R^{-6}$ & van der Waals & $(1,1,1,1)$ & $(0,0)$          & 0 \\
         &               &             & $(0,2)$, $(2,0)$ & 2 \\
         &               &             & $(1,1)$          & 0, 2 \\
         &               &             & $(2,2)$          & 0, 2, 4 \\
\hline 
\end{tabular}
\par\end{centering}
\caption{\label{tab:pert-2nd}Most commonly used terms of the second-order long-range interactions, with their $R$ dependence, and the ranks ($\ell'_{A},\ell_{A},\ell'_{B},\ell_{B}$) of the relevant tensor operators. The pairs $(k_{A},k_{B})$, and the values of $k$ denote the possible ranks of tensorial terms in equation \eqref{eq:pert-2nd-oper-2} as constrained by the values of ($\ell'_{A},\ell_{A},\ell'_{B},\ell_{B}$).}
\end{table}

With a similar method than above, the matrix elements of the effective second-order operator $\hat{\mathrm{W}}_{AB}^\mathrm{SF}(\mathbf{R})$ in the space-fixed frame can be derived. We only display below the final expression, obtained by starting from equation (\ref{eq:mult-vab-sf}) and applying equations (\ref{eq:pert-2nd-oper-1})--(\ref{eq:pert-4cg}). The main difference comes from the presence of the indices $\ell$  and $\ell'$ related to the orientation of the molecular axis in the SF frame, involving the following change in the product of Clebsch-Gordan coefficients with respect to equation (\ref{eq:pert-2nd-oper-2})
\begin{eqnarray}
  & & C_{(\ell'_A+\ell'_B)0(\ell_A+\ell_B)0}^{k0} 
    C_{k_Aqk_B-q}^{k0} C_{J_AM_Ak_Aq}^{J_AM'_A}
    C_{J_BM_Bk_B-q}^{J_BM'_B}
    \,\,[\mathrm{in\,\,Eq.~}(\ref{eq:pert-2nd-oper-2})]
  \nonumber \\
  & \to & \sum_{\ell\ell'} \delta_{\ell'_A+\ell'_B,\ell'}
    \delta_{\ell_A+\ell_B,\ell} \, C_{k_Aq_Ak_Bq_B}^{kq}
    C_{J_AM_Ak_Aq_A}^{J_AM'_A} C_{J_BM_Bk_Bq_B}^{J_BM'_B}
  \nonumber \\
  & \times & \sum_{mm'} C_{\ell'm'\ell m}^{kq}
    \frac{4\pi Y_{\ell'm'}^*(\Theta,\Phi) 
          Y_{\ell m}^*(\Theta,\Phi)}{\sqrt{(2\ell'+1)(2\ell+1)}}.
  \label{eq:w-bf-2-sf}
\end{eqnarray}
The factor on the last line of equation (\ref{eq:w-bf-2-sf}) equals $\sqrt{4\pi/(2k+1)}Y_{kq}^*(\Theta,\Phi)$ (see Ref.~\cite{varshalovich1988}, p.~144, Eq.~(10)), so that we get to the final expression
\begin{eqnarray}
  &  & \langle\beta_{A}J_{A}M'_{A}\beta_{B}J_{B}M'_{B}|
    \hat{\mathrm{W}}_{AB}^{\mathrm{SF}}(R)|
    \beta_{A}J_{A}M_{A}\beta_{B}J_{B}M_{B}\rangle \nonumber \\
  & = & -\frac{1}{16\pi^{\text{2}}\epsilon_{0}^{2}}
    \sum_{\ell'_A\ell'_B\ell'}\sum_{\ell_A\ell_B\ell}
    \delta_{\ell'_A+\ell'_B,\ell'}\delta_{\ell_A+\ell_B,\ell}\,
    \frac{\left(-1\right)^{\ell'_B+\ell_B+2J_A+2J_B}}
      {R^{2+\ell'+\ell}}
    \sqrt{\frac{\left(2\ell'+1\right)!\left(2\ell+1\right)!}
      {\left(2\ell'_{A}\right)!\left(2\ell'_{B}\right)!
       \left(2\ell_{A}\right)!\left(2\ell_{B}\right)!}}\nonumber \\
  & \times & \sum_{k_{A}k_{B}k} \sum_{q_{A}q_{B}q}
    C_{\ell'0\ell 0}^{k0}
    \sqrt{\frac{4\pi}{2k+1}} Y_{kq}^*(\Theta,\Phi)
    \left(-1\right)^{k_{A}+k_{B}}
    \left(2k_{A}+1\right)\left(2k_{B}+1\right)
    C_{k_Aq_Ak_Bq_B}^{kq} \nonumber \\
  & \times & \sum_{\beta''_{A}J''_{A}}\sum_{\beta''_{B}J''_{B}}
    \frac{\left\langle \beta_{A}J_{A}\right\Vert
      \hat{\mathrm{Q}}_{\ell'_{A}}^{\textrm{BF}}
      \left\Vert \beta''_{A}J''_{A} \right\rangle
      \left\langle \beta''_{A}J''_{A} \right\Vert
      \hat{\mathrm{Q}}_{\ell_{A}}^{\textrm{BF}}
      \left\Vert \beta_{A}J_{A}\right\rangle
      \left\langle \beta_{B}J_{B}\right\Vert
      \hat{\mathrm{Q}}_{\ell'_{B}}^{\textrm{BF}}
      \left\Vert \beta''_{B}J''_{B}\right\rangle 
      \left\langle \beta''_{B}J''_{B}\right\Vert 
      \hat{\mathrm{Q}}_{\ell_{B}}^{\textrm{BF}}
      \left\Vert \beta_{B}J_{B}\right\rangle }
    {E_{\beta''_{A}J''_{A}}-E_{\beta_{A}J_{A}}
    +E_{\beta''_{B}J''_{B}}-E_{\beta_{B}J_{B}}} \nonumber \\
  & \times & \ninej{\ell'_A}{\ell_A}{k_A}
                   {\ell'_B}{\ell_B}{k_B}
                   {\ell'  }{\ell  }{k}
    \sixj{\ell_A}{\ell'_A}{k_A}{J_A}{J_A}{J''_A}
    \sixj{\ell_B}{\ell'_B}{k_B}{J_B}{J_B}{J''_B}
    \frac{C_{J_{A}M_{A}k_{A} q}^{J_{A}M'_{A}}
          C_{J_{B}M_{B}k_{B}-q}^{J_{B}M'_{B}}}
    {\sqrt{\left(2J_{A}+1\right)\left(2J_{B}+1\right)}},
  \label{eq:pert-2nd-oper-sf}
\end{eqnarray}

\subsection{Example: long-range interaction between two alkali-metal atoms}
\label{ssec:pert-alk-ato}

Alkali-metal atoms have been at the heart of the successful developments of research on ultracold quantum gases for the last three decades: their strong $^2S \to\ ^2P$ resonant transition with wavelengths in the range of easily available narrow-band continuous lasers make them convenient for laser-cooling. Thus the knowledge of their long-range interactions attracted a lot of attention, as their simple electronic structure -- a single valence electron in the field of a closed-shell $^1S$ ionic core-- allows for very accurate calculations by various methods \cite{bussery1987, merawa1994, marinescu1994a, marinescu1995, marinescu1997a, merawa1997, merawa1998, rerat1998, marinescu1999, derevianko1999, derevianko2001, porsev2003, zhang2007, zhang2007a, derevianko2010}, which can be compared among each other and with experimental measurements. For instance, the so-called \textit{photoassociation spectroscopy} \cite{stwalley1999} consists in exciting a pair of ultracold ground-state atoms into high-lying rovibrational levels of an electronically excited state of the associated diatomic molecule. This method allows for characterizing the long-range interactions between a $^2S$ ground state atom and a $^2P$ excited one, which determines the radiative lifetime of the $^2P$ atom \cite{bouloufa2009}.

Various cases will be considered in what follows: two alkali-metal atoms of the same species (\textit{homonuclear} case) or of different species (\textit{heteronuclear} case), both in their $^2S$ ground state, or one being in its first $^2P$ excited state. We will only consider electrostatic interactions, since the magnetostatic interactions due to the spin $1/2$ are several orders of magnitude smaller. Therefore the electronic and nuclear spins of the atoms are considered spectators of the interactions. The atomic quantum levels are characterized by the principal $n$, orbital $L$ and magnetic $M_{L}$ quantum numbers of the valence electron. The unperturbed energy levels are written as $|\Psi_{q,i}^{(0)}\rangle=|n_{A}L_{A}M_{L_{A}}n_{B}L_{B}M_{L_{B}}\rangle$. These levels are also commonly referred to as the dissociation limit or asymptote A($|n_{A}L_{A}M_{L_{A}}\rangle$) + B($|n_{B}L_{B}M_{L_{B}}\rangle$) of the AB molecule when $R \rightarrow \infty$, with energy equal to the sum of the respective atomic energy levels. In the following sections, we determine long-range potential-energy curves (PECs) for each asymptote in the BF frame, which must be matched around the LeRoy radius (defined in the introduction) to PECs for shorter distances $R$ either extracted from molecular spectroscopic investigations or from elaborate methods of quantum chemistry. 

The goal of the present section is to demonstrate how the values of $L_{A}$ and $L_{B}$ influence the nature of the interactions, \textit{i.e.} the exponent, the amplitude, and the sign of the terms of the multipolar expansion. The numerical values of the associated energies strongly depends on the quality of the unperturbed atomic wave functions of A and B, and details on their evaluation are available in the references quoted above.

\subsubsection{Interaction between two ground-state atoms}

The unperturbed level of the pair of atoms in their lowest $^2S$ state is non-degenerate and is $|\Psi_{q}^{(0)}\rangle=|n_{A}00n_{B}00\rangle$, with $n_{A}$ and/or $n_{B}$ equal to 2, 3, 4, 5, 6 for lithium (Li), sodium (Na), potassium (K), rubidium (Rb), and cesium (Cs), respectively. The first-order energy correction $\langle\hat{\mathrm{V}}_{AB}^{\mathrm{BF}}(R)\rangle$ is the diagonal matrix elements of the operator given by equation (\ref{eq:pert-1st})
\begin{equation}
\langle\hat{\mathrm{V}}_{AB}^{\mathrm{BF}}(R)\rangle=\frac{1}{4\pi\epsilon_{0}}\sum_{\ell_{A}\ell_{B}m}\frac{f_{\ell_{A}\ell_{B}m}}{R^{1+\ell_{A}+\ell_{B}}}C_{00\ell_{A}m}^{00}C_{00\ell_{B}-m}^{00}\left\langle n_{A}0\right\Vert \hat{\mathrm{Q}}_{\ell_{A}}^{\textrm{BF}}\left\Vert n_{A}0\right\rangle \left\langle n_{B}0\right\Vert \hat{\mathrm{Q}}_{\ell_{B}}^{\textrm{BF}}\left\Vert n_{B}0\right\rangle .
\label{eq:pert-s-1st}
\end{equation}
Since $C_{00b\beta}^{00}=\delta_{b0}\delta_{\beta0}$, the Clebsch-Gordan coefficients of equation (\ref{eq:pert-s-1st}) are zero except for $\ell_{A}=\ell_{B}=0$, see selection rules (\ref{eq:pert-triangle}), which corresponds to the charge-charge interaction. As the two partners are neutral, \textit{the first-order energy correction between two ground-state alkali-metal atoms is equal to zero}. This result is straightforwardly generalized to alkali-metal atoms in excited $^2S$ levels, and even to any pair of atoms in a $S$ state such as alkaline-earth atoms.

The leading term of the multipolar expansion thus comes from the second-order energy correction. For the two neutral atoms, it is associated with the second-order dipole-dipole interaction (sometimes called also induced dipole-induced dipole interaction), given by $\ell'_{A}=\ell_{A}=\ell'_{B}=\ell_{B}=1$, and scaling as $R^{-6}$. According to the inequality (\ref{eq:pert-triangle}), the intermediate levels appearing in the sum of equation (\ref{eq:pert-2nd-oper}) are characterized by $L''_{A}=L''_{B}=1$, namely, levels of $P$ symmetry: they can be excited $np$ bound levels, continuum states with a free $p$ electron, or levels for which one electron of the ionic core is excited.

By writing the energy correction as $E_{q}^{(2)}=C_{6}/R^{6}$, we can extract the $C_{6}$ coefficient from equation (\ref{eq:pert-2nd-oper})
\begin{align}
  C_{6} & = -\frac{1}{16\pi^{\text{2}}\epsilon_{0}^{2}} 
   \sum_{\beta''_{A}\beta''_{B}} \sum_{M''_{L_{A}}M''_{L_{B}}}
   \frac{1}{E_{\beta''_{A}1}-E_{n_{A}0}+E_{\beta''_{B}1}-E_{n_{B}0}}
  \nonumber \\
   & \times \left(\sum_{m=-1}^{+1}
   \frac{2}{\left(1+m\right)!\left(1-m\right)!}
   \left\langle \beta''_{A}1M''_{L_{A}}\left|
   \hat{\mathrm{Q}}_{1m}^{\textrm{BF}} 
   \right|n_{A}00\right\rangle 
   \left\langle \beta''_{B}1M''_{L_{B}}\left|
   \hat{\mathrm{Q}}_{1-m}^{\textrm{BF}}
   \right|n_{B}00\right\rangle \right)^{2}.
  \label{eq:pert-s-c6}
\end{align}
Note that we take the square of the matrix element $\langle\beta''_{A}1M''_{L_{A}}\beta''_{B}1M''_{L_{B}}|\hat{\mathrm{V}}_{AB}^{\textrm{BF}}(R)|n_{A}00n_{B}00\rangle$ as in the first term of equation (\ref{eq:Eq2}), hence the absence of sum over $m'$. The transition dipole moments $\langle \beta''_{A}1M''_{L_{A}}| \hat{\mathrm{Q}}_{1m}^{\textrm{BF}} |n_{A}00\rangle$ (and the same for B) are non-zero for $m=M''_{L_{A}}=-M''_{L_{B}}$, and are equal to $\langle\beta''_{A}1\Vert\hat{\mathrm{Q}}_{1}^{\textrm{BF}}\Vert n_{A}0\rangle/\sqrt{3}$ (and the same for B) such that
\begin{equation}
  C_{6} = -\frac{1}{12\pi^{\text{2}}\epsilon_{0}^{2}}
   \sum_{\beta''_{A}\beta''_{B}}
   \frac{\left|\left\langle \beta''_{A}1\left\Vert 
    \hat{\mathrm{Q}}_{1}^{\textrm{BF}}
    \right\Vert n_{A}0\right\rangle 
    \left\langle \beta''_{B}1\left\Vert 
    \hat{\mathrm{Q}}_{1}^{\textrm{BF}}
    \right\Vert n_{B}0\right\rangle \right|^{2}}
   {E_{\beta''_{A}1}-E_{n_{A}0}+E_{\beta''_{B}1}-E_{n_{B}0}}.
  \label{eq:pert-s-c6-2}
\end{equation}
Note that equation (\ref{eq:pert-s-c6-2}) may also be obtained from equation (\ref{eq:pert-2nd-oper-2}), with $k_{A}=k_{B}=k=0$ since we deal with $S$ atoms which interact isotropically. 

Being of dispersive nature, the $C_{6}$ coefficient is conveniently written using the dynamic polarizabilities at imaginary frequencies, see equation (\ref{eq:pert-disp-1}). Its value is often given in atomic units, \textit{i.e.} in units of $E_{h}a_{0}^{6}$, with $E_{h}$ the Hartree energy and $a_{0}$ the Bohr radius, as
\begin{equation}
  C_{6} = -\frac{3}{\pi} \int_{0}^{+\infty}d\omega
    \alpha_{\mathrm{scal}}(A;i\omega)\alpha_{\mathrm{scal}}(B;i\omega),
  \label{eq:pert-s-c6-3}
\end{equation}
where $\alpha_{\mathrm{scal}}(A(B);i\omega) = -\alpha_{(11)0}^{A(B)}(i\omega)/\sqrt{3}$ stands for the scalar dipole polarizability of atom A (B) expressed in atomic units (equation (\ref{eq:pert-alpha-zz-dyn-2})). Being isotropic, those polarizabilities as well as the $C_{6}$ coefficient do not depend on the coordinate system, hence the omitted ``BF''
label. Table \ref{tab:pert-c6-s} displays the calculated $C_{6}$ coefficients for all pairs of ground-state alkali-metal atoms from Li to Cs \cite{derevianko2010}. They increase with the size of the electronic cloud, \textit{i.e.} the size of the atom.

\begin{table}
\begin{centering}
\begin{tabular}{|c|ccccc|}
\hline 
 & Li & Na & K & Rb & Cs\tabularnewline
\hline 
Li & -1389 & -1467 & -2322 & -2545 & -3065 \tabularnewline
Na &       & -1556 & -2447 & -2683 & -3227 \tabularnewline
K  &       &       & -3897 & -4274 & -5159 \tabularnewline
Rb &       &       &       & -4690 & -5663 \tabularnewline
Cs &       &       &       &       & -6846 \tabularnewline
\hline 
\end{tabular}
\par\end{centering}
\caption{\label{tab:pert-c6-s}Theoretical $C_{6}$ coefficients in atomic units, characterizing the van der Waals interaction between two ground-state alkali-metal atoms \cite{derevianko2010}. The minus signs denote an attractive interaction.}
\end{table}

Beyond the $R^{-6}$ contribution, the next terms of the multipolar expansion come from the induced dipole-induced
quadrupole interaction scaling as $R^{-8}$, and from the induced quadrupole-induced quadrupole interaction scaling as $R^{-10}$. Those three terms allow for the very accurate calculation of ground-state long-range PECs of alkali-metal dimers \cite{porsev2003, docenko2004b}.

\subsubsection{Interaction between one ground state atom and one atom of a different species in the first-excited state}

We consider two alkali-metal atoms $A$ and $B$ of different species. A is in its ground state $n_{A}S$, while B is in its first-excited state $n_{B}P$ (ignoring the fine structure in the following). The unperturbed levels are $|\Psi_{q,i}^{(0)}\rangle=|n_{A}00n_{B}1M_{L_{B}}\rangle$. In the BF frame, the projection $M_{L_{A}}+M_{L_{B}} \equiv M_{L_{B}}$ of the total orbital angular momentum on the interatomic axis is a good quantum number, see equation (\ref{eq:pert-M-const}). The values $M_{L_{B}}=0$ and $\pm1$ correspond to $\Sigma$ and $\Pi$ electronic states of the AB molecule, respectively. Since atom A is in an $S$ state, all its multipole moments are equal to zero, and so the first-order energy correction is zero as well, see equation (\ref{eq:pert-1st}). As above, the leading term of the multipolar expansion is the van der Waals  contribution varying as $R^{-6}$, with a given value of the coefficient $C_{6}^{\text{BF}}(M_{L_{B}})$ for each symmetry in the BF frame. The atomic levels participating to the sums of equation (\ref{eq:pert-2nd-oper}) are $P$ states for atom A, and $S$ and $D$ states for atom B.

It is also convenient to represent the van der Waals interactions in terms of isotropic and anisotropic coefficients $C_{6}^{k_{A}k_{B}k}$ which do not depend on the coordinate system. The isotropic coefficient $C_{6}^{000}$ can be calculated using equation (\ref{eq:pert-s-c6-3}); it turns out to be an arithmetic average of the BF coefficients, $C_{6}^{000}=\left(C_{6}^{\textrm{BF}}(\Sigma)+2C_{6}^{\textrm{BF}}(\Pi)\right)/3$, the $\Pi$ state being doubly-degenerate (see e.g.~\cite{chu2005}). As for the anisotropic coefficients, since $L_{A}=0$ and $L_{B}=1$, the only possible value of $k_{A}$ and $k_{B}$ are 0 and 2, respectively, which imposes $k=2$ (see Table \ref{tab:pert-2nd}). The corresponding anisotropic coefficient is given by \cite{chu2005} $C_{6}^{022}=5\left(C_{6}^{\textrm{BF}}(\Sigma)-C_{6}^{\textrm{BF}}(\Pi)\right)/3$.

\subsubsection{Interaction between one ground-state atom and one atom of the same species in the first-excited state}
\label{sssec:pert-alk-ato-same}

The case of identical atoms, $n_{A}=n_{B}=n$, dramatically differs from the previous one, because the unperturbed levels $|n00n1M_{L}\rangle$ and $|n1M_{L}n00\rangle$ are degenerate, and also because they are coupled by the dipole-dipole interaction $\hat{\mathrm{V}}_{dd}^{\textrm{BF}}(R)$. Indeed setting $\ell_{A}=\ell_{B}=1$, we rewrite equation (\ref{eq:pert-1st}) as
\begin{equation}
\left\langle n1M_{L}n00\right|\hat{\mathrm{V}}_{dd}^{\textrm{BF}}(R)\left|n00n1M_{L}\right\rangle = \frac{\left(-1\right)^{1-M_{L}}\left(1+\delta_{M_{L}0}\right)}{12\pi\epsilon_{0}R^{3}}\left|\left\langle n1\right\Vert \hat{\mathrm{Q}}_{1}\left\Vert n0\right\rangle \right|^{2}
\label{eq:pert-res-dip}
\end{equation}
where we replaced $m$ by its only possible value $M_{L}$. In this equation we also accounted for the following identities: 
$\langle n1\Vert\hat{\mathrm{Q}}_{1}\Vert n0\rangle=-\langle n0\Vert\hat{\mathrm{Q}}_{1}\Vert n1\rangle$, $C_{00b\beta}^{b\beta}=1$, $C_{a\alpha a-\alpha}^{00}=\left(-1\right)^{a-\alpha}/\sqrt{2a+1}$, and  $f_{110}=-2$ and $f_{11\pm1}=-1$. Therefore, with each $M_{L}$-value is associated a two-dimensional subspace built on the two vectors $\{|n00n1M_{L}\rangle$ and $|n1M_{L}n00\rangle\}$, with an unperturbed energy $E_{n1}$ of the $nP$ state (assuming the $nS$-state
energy $E_{n0}=0$) and coupled by the matrix element of equation (\ref{eq:pert-res-dip}). Applying the first-order degenerate perturbation theory, and setting $E_{q,i}^{(1)}=C_{3}^{\textrm{BF}}(M_{L})/R^{3}$, one finds (in atomic units) $C_{3}^{\textrm{BF}}(\Sigma)=\pm2|\langle n1\Vert\hat{\mathrm{Q}}_{1}\Vert n0\rangle|^{2}/3$ and $C_{3}^{\textrm{BF}}(\Pi)=\pm|C_{3}^{\textrm{BF}}(\Sigma)|/2$. A single $C_{3}^{\textrm{BF}}(M_{L})$ coefficient is necessary to fully characterize this so-called \textit{resonant dipolar} interaction, due by \textit{exchange of dipolar excitation}. This phenomenon can also take place for pairs of levels with different principal quantum numbers $n$ and $n'$  (see table \ref{tab:pert-c3-sp}), including when one atom is in a Rydberg state \cite{gallagher1994}. Note that in addition there is also a van der Waals contribution to the interaction energy, involving states of $S$, $P$ and $D$ symmetries in the sums of the relevant equations for second-order perturbation theory in the above sections.

\begin{table}
\begin{centering}
\begin{tabular}{|c|ccccc|}
\hline 
$n'$  & Li & Na & K & Rb & Cs \\
\hline 
$n  $ & 11.01 & 12.26 & 17.33 & 18.40 & 20.95 \\
$n+1$ & 0.03364 & 0.08432 & 0.09225 & 0.1428 & 0.1482 \\
\hline 
\end{tabular}
\par\end{centering}
\caption{\label{tab:pert-c3-sp} Calculated $|C_{3}^{\textrm{BF}}(\Sigma)|$ coefficients (in atomic units) \cite{marinescu1995} characterizing the resonant dipolar interaction between two alkali-metal atoms of the same species, one being in the ground state $nS$, and one in the first ($n'=n$)  and second ($n'=n+1$) $^2P$ excited state.}
\end{table}

\section{Long-range interactions between two heteronuclear alkali-metal diatomic molecules in an external electric field}
\label{sec:hetero}

An important ongoing quest in the field of ultracold chemistry is the formation of ultracold molecules possessing a permanent electric dipole moment in their ground state, which allows for studying so-called quantum dipolar gases, \textit{i.e.}~where particles interact through \textit{anisotropic} dipole-dipole forces. Indeed, such systems are expected to provide fantastic tools for novel applications like quantum simulation of condensed-matter phase, for quantum information, and for metrology \cite{micheli2007}.

Nowadays, there is a method of choice to reach such low temperatures while ensuring the maximal population of the absolute ground level of the molecule: the association of a pair of different ultracold atoms into an ultracold molecule with the help of an external magnetic field, and of a pair of laser pulses to control the population at the single quantum level. Due to their intrinsic magnetic moment, in addition to the existence of efficient cooling schemes to reach quantum degeneracy, alkali-metal atoms have been the only suitable species up to now to achieve this goal, with the formation of heteronuclear alkali-metal diatomics. Following first attempts with LiCs \cite{deiglmayr2008a}, KRb \cite{banerjee2012} and RbCs \cite{bruzewicz2014}, experimentalists were successful to create a pure gas of dipolar molecules and their absolute ground state with KRb \cite{ni2008, ospelkaus2010b}, RbCs \cite{takekoshi2014, molony2014, shimasaki2015}, NaK \cite{park2015} and NaRb \cite{guo2016}.

It is thus crucial to fully characterize the long-range interactions between two bialkali molecules in their electronic ground state $X^{1}\Sigma^{+}$, in order to determine under which conditions the electric dipole-dipole interaction is the dominant one. Several studies have been devoted to this topic \cite{quemener2011a, byrd2012a, byrd2012b, buchachenko2012, zuchowski2013}, including by the authors of the present chapter \cite{lepers2013, perez-rios2014, vexiau2015}. Out of the 10 possible diatomic species that can be assembled from Li, Na, K, Rb, and Cs atoms, 8 species posses a strong (say, larger than 1~Debye) permanent electric dipole moment (PEDM) in their own frame, while two species (LiNa and KRb) have a much smaller PEDM. An important result was that for the former ones, the potential energy generated by the interaction between the PEDM $d_{0}$ of their lowest vibrational level accounted for at least 85 \% of the total interaction potential energy at large distances, reaching up to 99.4 \% for $^{23}$Na$^{87}$Rb, with $d_{0}=3.2$~D. In this section, we will limit our discussion to this electric dipole-dipole interaction.

We consider two molecules A and B in the lowest vibrational level $v=0$ of the ground electronic state $X^{1}\Sigma^{+}$. For each molecule, say species A in the following, we denote this state $\beta_{A}=(X^{1}\Sigma^{+},\, v_{A}=0)$. The molecule lies in a given rotational level $|J_{A}M_{A}\rangle$ of energy $B_{0}J_{A}(J_{A}+1)$, with $B_{0}$ the rotational constant of the vibrational level. The unperturbed levels in our formalism are thus written $|\Psi_{q,i}^{(0)}\rangle=|J_{A}M_{A}J_{B}M_{B}\rangle$. For $^{1}\Sigma^+$ molecules, the rotational wave function is the spherical harmonics $Y_{J_{A}M_{A}}(\Theta_{A},\Phi_{A})$ where the angles $(\Theta_{A},\Phi_{A})$ characterize the orientation of the molecular axis with respect to the quantization axis $z$. Using $\hat{\textrm{Q}}_{1m}(A)=\sqrt{4\pi/3}\, Y_{1m}(\Theta_{A},\Phi_{A})\hat{\mathrm{d}}_{z_{A}}$, where $\hat{\mathrm{d}}_{z_{A}}$ is the dipole-moment operator along
the molecular axis $z_{A}$, and applying equation (\ref{eq:pert-3ylm}), we write the matrix element of the dipole-moment operator as
\begin{equation}
\left\langle J'_{A}M'_{A}\right|\hat{\mathrm{Q}}_{1m}\left|J_{A}M_{A}\right\rangle =\sqrt{\frac{2J_{A}+1}{2J'_{A}+1}}C_{J_{A}M_{A}1m}^{J'_{A}M'_{A}}C_{J_{A}010}^{J'_{A}0}d_{0}\label{eq:mol-dip}
\end{equation}
where $d_{0}=\langle\hat{\mathrm{d}}_{z_{A}}\rangle$ is the mean value of the dipole moment averaged over the radial wave function of the lowest vibrational level. Equation (\ref{eq:mol-dip}) satisfies the Wigner-Eckart theorem (\ref{eq:pert-wigneck}) with $\langle J'_{A}\Vert\hat{\mathrm{Q}}_{1}\Vert J_{A}\rangle=\sqrt{2J_{A}+1}\, C_{J_{A}010}^{J'_{A}0}d_{0}$. Again, we did not specify the coordinate system to highlight the generality of equation (\ref{eq:mol-dip}). 

External fields have fixed orientations in the laboratory. It is in principle more appropriate to characterize long-range interactions between molecules in the presence of a field in the SF frame, so that the resulting PECs can be used for the modeling of the scattering dynamics of a molecule pair. However, the PECs in the BF frame turn out to be very insightful to understand the dynamics of the complex. Therefore in subsection \ref{ssec:mol-bf}, we characterize the inter-molecular long-range interactions in the BF frame, in the absence of electric field. In particular we identify the PECs that will play an important role for the calculations in the SF frame and in the presence of an electric field, treated in subsection \ref{ssec:mol-sf}.

\subsection{Calculation in the body-fixed frame without an external field}
\label{ssec:mol-bf}

In this section, we  follow the same approach as reference \cite{lepers2013} to characterize the long-range interactions in the BF frame. We apply the methodology of the previous section: to each pair of rotational levels $(J_{A},J_{B})$ corresponds an unperturbed energy $E_{q}^{(0)}=B_{0}[J_{A}(J_{A}+1)+J_{B}(J_{B}+1)]$, associated to the unperturbed state $|\Psi_{q,i}^{(0)}\rangle=|J_{A}M_{A}J_{B}M_{B}\rangle$. By making an analogy between the molecular rotational and the atomic orbital quantum numbers, we will obtain similar physical pictures as in section \ref{ssec:pert-alk-ato}.

\subsubsection{Calculation for each individual rotational level}

For $J_{A}=J_{B}=0$, the matrix element of the dipole moment (and all other multipole moments) operator is zero, and so is the first-order correction, see equation (\ref{eq:pert-1st}). In other words, $J=0$ molecules are spherically symmetric and behave as $S$ atoms. The leading term of the multipolar expansion comes from the van der Waals interaction. A useful approximation consists in restricting the relevant sum to the rotationally excited level $J''_{A}=J''_{B}=1$ (due to the selection rules imposed by equation (\ref{eq:mol-dip})) which is the closest in energy from the unperturbed level. This yields the simple formula in SI units
\begin{equation}
\left\langle 0000\right|\hat{\mathrm{W}}_{AB}^{\mathrm{BF}}(R)\left|0000\right\rangle =-\frac{d_{0}^{4}}{96\pi^{\text{2}}\epsilon_{0}^{2}B_{0}R^{6}},
\label{eq:mol-vdw-00}
\end{equation}
from which we can extract the coefficient $C_{6}^g=-d_{0}^{4}/6B_{0}$ in atomic units. Due to its $d_{0}^{4}$ dependence, this $C_{6}^g$ coefficient can vary over several orders of magnitude of the series of bialkali molecular species (see table \ref{tab:C6}). Beyond this approximation the total van der Waals interaction energy also comprises the contribution from the electronically excited states, which changes less strongly from one molecule to another as the typical energy gap between the ground state and the lowest excited electronic state is similar for all species. Table \ref{tab:C6} shows that that there are two classes of bialkali species, as quoted earlier, linked to the magnitude of the "rotational'' $C_{6}^g$ coefficient (as expressed in equation (\ref{eq:pert-alpha-zz-dyn-2}) and below): due to their very low value of $d_0$, it accounts for only 6.7 \% for LiNa and 20.9\% for KRb, in strong contrast with the 8 other species. The eight other species  exhibit $C_{6}$ coefficients with a huge magnitude in comparison to the typical values for interatomic van der Waals coefficients (Table \ref{tab:pert-c6-s}).

\begin{table}
\begin{centering}
\begin{tabular}{rrrr}
\hline
           Molecule &    $C_6$ &  $C_6^g$ &  $C_6^g/C_6$ \\
                    &   (a.u.) &   (a.u.) &         (\%) \\
\hline
$^{23}$Na$^{133}$Cs & -7323100 & -7311100 &         99.8 \\
   $^7$Li$^{133}$Cs & -4585400 & -4574400 &         99.8 \\
 $^{23}$Na$^{87}$Rb & -1524900 & -1515800 &         99.4 \\
    $^7$Li$^{87}$Rb & -1252300 & -1244205 &         99.4 \\ 
     $^7$Li$^{39}$K &  -570190 &  -563500 &         98.8 \\
  $^{23}$Na$^{39}$K &  -561070 &  -553520 &         98.7 \\
 $^{39}$K$^{133}$Cs &  -345740 &  -329510 &         95.3 \\
$^{87}$Rb$^{133}$Cs &  -147260 &  -129250 &         87.8 \\
\hline
  $^{39}$K$^{87}$Rb &   -15972 &    -3336 &         20.9 \\
    $^7$Li$^{23}$Na &    -3583 &     -241 &          6.7 \\
		\hline
\end{tabular}
\par \end{centering}
\caption{The $C_6$ coefficient between two like ground-level bialkali heteronuclear molecules, and the $C_6^g$ part with the associated percentage compared to $C_6$, as computed in reference \cite{lepers2013}. }
\label{tab:C6}
\end{table}

For rotationally-excited levels none of the rotational levels have a permanent dipole moment, because the Clebcsh-Gordan coefficient $C_{J_{A}010}^{J'_{A}0}$ in equation (\ref{eq:mol-dip}) vanishes when $J'_{A}=J_{A}$. Therefore, the first-order dipole-dipole interaction energy also vanishes for $J'_{A}=J_{A}=J'_{B}=J_{B}$. Rotationally excited levels do possess a non-zero quadrupole moment, but its magnitude is small and the related first-order quadrupole-quadrupole interaction is not expected to play a significant role for most of the molecules by comparison with the van der Waals interaction \cite{lepers2013}. 

Let us first consider the case $(J_{A},J_{B})=(1,1)$, for which the unperturbed energy $E_{q}^{(0)}=4B_{0}$ is 9-fold degenerate. It is convenient here to express the van der Waals energy using its expression in terms of irreducible tensors. To that end, we write the operator $\hat{\mathrm{W}}_{AB}^{\mathrm{BF}}(R)$, see equation (\ref{eq:pert-2nd-oper-2}), by writing the explicit values of Clebsch-Gordan coefficients, $6j$ and $9j$ symbols \cite{varshalovich1988, stone-369j}
\begin{eqnarray}
 &  & \langle1M'_{A}1M'_{B}|\hat{\mathrm{W}}_{AB}^{\mathrm{BF}}(R)|1M_{A}1M_{B}\rangle\nonumber \\
 & = & \frac{d_{0}^{4}}{16\pi^{\text{2}}\epsilon_{0}^{2}B_{0}R^{6}}\left[\frac{1}{6}\delta_{M_{A}M'_{A}}\delta_{M_{B}M'_{B}}\left(-1+\sqrt{\frac{2}{5}}\left(C_{1M_{A}20}^{1M_{A}}+C_{1M_{B}20}^{1M_{B}}\right)\right)\right.\nonumber \\
 &  & \left.+\frac{45}{4}\sum_{q=-1}^{1}A_{1}(q)C_{1M_{A}1q}^{1M'_{A}}C_{1M_{B}1-q}^{1M'_{B}}+\frac{5}{4}\sum_{q=-2}^{2}A_{2}(q)C_{1M_{A}2q}^{1M'_{A}}C_{1M_{B}2-q}^{1M'_{B}}\right]\label{eq:mol-vdw-11}
\end{eqnarray}
where
\begin{equation}
A_{n}(q)=\sum_{p=0}^{n}C_{2020}^{2p,0}C_{nqn-q}^{2p,0}\left\{ \begin{array}{ccc}
1 & 1 & 2\\
1 & 1 & 2\\
n & n & 2p
\end{array}\right\} \label{eq:mol-vdw-11-a}
\end{equation}
is a polynomial of degree $2n$ in $q$. In equation (\ref{eq:mol-vdw-11}), the three terms of the first line correspond to $(k_{A},k_{B})=(0,0)$, $(2,0)$ and $(0,2)$ in equation (\ref{eq:pert-2nd-oper-2}), which imposes $q=0$ and so $M'_{A}=M_{A}$, $M'_{B}=M_{B}$. Note that the  $(k_{A},k_{B})=(0,0)$ term is equal to the interaction energy between two molecules in their lowest rotational level $J_{A}=J_{B}=0$, see equation (\ref{eq:mol-vdw-00}), and is referred to as \textit{isotropic}. The second line of equation (\ref{eq:mol-vdw-11}) contains diagonal ($q=0$) as well as off-diagonal ($q\ne0$) matrix elements of the operator $\hat{\mathrm{W}}_{AB}^{\mathrm{BF}}(R)$. They obey the selection rule $M_{A}+M_{B}=M'_{A}+M'_{B}$. By accounting for the conservation of the total angular momentum projection $M_{A}+M_{B}$, and for the permutation symmetry, which gives rise to symmetric or antisymmetric eigenvectors, we can divide the $9\times9$ matrix of equation (\ref{eq:mol-vdw-11}) into smaller matrices of dimension 1 or 2. The corresponding eigenvalues and eigenvectors, calculated by hand, are presented for the NaRb species in Table \ref{tab:mol-c6-11}, both in rescaled units and physical units. Being proportional to $d_{0}^{4}/B_{0}$, these $C_{6}$ coefficients are also very large in absolute value, and all but one induce attractive interactions. Indeed, the asymptotes $(J_A,J_B) = (0,2)$ and $(2,0)$, of energy $6B_0$, is closer to $(1,1)$, of energy $4B_0$, than the $(0,0)$ and $(2,2)$ ones, respectively at 0 and $12B_0$. Therefore, due to the denominator of equation \eqref{eq:pert-2nd-oper}, the $C_6$ tend to be negative. On the other hand, the off-diagonal matrix elements of equation \eqref{eq:mol-vdw-11-a}, combine in such a way that we obtain one repulsive coefficient.

\begin{table}
\begin{centering}
\begin{tabular}{|c|c|c|c|}
\hline 
$|M_{A}+M_{B}|$ & Eigenvector & $C_{6}$ (units of  & $C_{6}$ \\
                &             & $d_{0}^{4}/(16\pi^{2}\epsilon_{0}^{2}B_{0})$) & ($10^{6}$ a.u.) \\
\hline
2 & $|1\pm1,1\pm1\rangle$ & $-\frac{4}{25}$ & -1.25 \\
1 & $\frac{1}{\sqrt{2}}(|1\pm1,10\rangle+|10,1\pm1\rangle)$ & $-\frac{17}{200}$ & -0.66 \\
 & $\frac{1}{\sqrt{2}}(|1\pm1,10\rangle-|10,1\pm1\rangle)$ & $-\frac{13}{40}$ & -2.55 \\
0 & $\frac{1}{\sqrt{3(1-\sqrt{3})}}[|10,10\rangle+\frac{1-\sqrt{3}}{2}(|11,1-1\rangle+|1-1,11\rangle)]$ & $-\frac{13\sqrt{3}+16}{100}$ & -3.02 \\
 & $\frac{1}{\sqrt{3(1+\sqrt{3})}}[|10,10\rangle+\frac{1+\sqrt{3}}{2}(|11,1-1\rangle+|1-1,11\rangle)]$ & $\frac{13\sqrt{3}-16}{100}$ & 0.51 \\
 & $(|11,1-1\rangle-|1-1,11\rangle)/\sqrt{2}$ & $-\frac{1}{25}$ & -0.31 \\
\hline 
\end{tabular}
\par\end{centering}
\caption{\label{tab:mol-c6-11}Van der Waals coefficients $C_{6}$ and corresponding eigenvectors, expressed as sums of the kets $|J_AM_A,J_BM_B\rangle$, characterizing the interaction between two NaRb molecules in the first rotationally-excited level $J_A=J_B=1$ of the ground vibrational
level $v_A=v_B=0$.}
\end{table}

A remarkable case occurs with the levels characterized by $(J_{A},J_{B})=(1,0)$ and $(0,1)$ with the same unperturbed energy $E_{q}^{(0)}=2B_{0}$. Just like for the case of an $S$ atom and a $P$ atom, as explained in section \ref{sssec:pert-alk-ato-same}, the molecules are directly coupled by the dipole-dipole interaction that varies as $C_{3}/R^{3}$ through the exchange of dipolar excitation. The $C_3$ values and the corresponding eigenvectors are summarized below for each value of $|M_{A}+M_{B}|$:
\begin{eqnarray}
|M_{A}+M_{B}|=0:&C_{3}=\pm2d_{0}^{2}/3&(|1000\rangle\mp|0010\rangle)/\sqrt{2} \nonumber \\
|M_{A}+M_{B}|=1:&C_{3}=\pm d_{0}^{2}/3&(|1\pm100\rangle\pm|001\pm1\rangle)/\sqrt{2}.
\label{eq:c3-1001}
\end{eqnarray}

Each pair of rotational levels $(J_{A},J_{B})$ can be treated independently provided that the calculated energy corrections are perturbations compared to difference between unperturbed energies. By imposing $|C_{3}|/R^{3}$ and $|C_{6}|/R^{6}$ for each relevant pair of rotational levels that are much smaller than the typical energy spacing $B_{0}$ between unperturbed rotational levels (and dropping out all numerical factors), one can define the characteristic inter-partner distance 
\begin{equation}
R^{*}=\left(\frac{d_{0}^{2}}{4\pi\epsilon_{0}B_{0}}\right)^{1/3}
\label{eq:mol-rstar}
\end{equation}
beyond which such an independent treatment is applicable. For pairs of bialkali molecules, it is found that $R^{*}$ is always larger than the LeRoy radius $R_{LR}\sim20$ a.u.. For instance, $R^{*}=31$, 175 and 234~a.u.~for LiNa, NaRb and NaCs respectively \cite{lepers2013}. Thus there exists a range of inter-partner distances $R_{LR}<R<R^{*}$, for which the perturbation approach  must be reformulated to take in account several pairs of rotational levels.

\subsubsection{Calculation for coupled rotational levels}
\label{sssec:coupled-rot-levels}

The different pairs of rotational levels are \textit{coupled by the dipole-dipole interaction}, as it can be deduced from equations (\ref{eq:pert-1st}) and (\ref{eq:mol-dip}). The pair $(J_{A},J_{B})=(0,0)$ is coupled to $(1,1)$,which in turn is coupled to $(2,0)$ and $(0,2)$ and so on. Generally speaking, each $(J_{A},J_{B})$ pair is coupled to the $(J_{A}\pm1,J_{B}\pm1)$ ones. The rotational Hamiltonian of individual molecules must therefore be included in the perturbation operator $\hat{\mathrm{V}}$,
\begin{equation}
  \hat{\mathrm{V}}(R) \equiv  
    B_{0}\left(\hat{\mathbf{J}}_{A}^{2}+\hat{\mathbf{J}}_{B}^{2}\right) + 
    \hat{\mathrm{V}}_{\textrm{dd}}^{\textrm{BF}}(R),
  \label{eq:mol-oper-new}
\end{equation}
which is processed using perturbation theory of \textit{quasi-degenerate} levels \cite{cohen-tannoudji2006}. There is a single unperturbed level associated to the two molecules in their vibrational ground level, with an energy arbitrarily put to $E_{q}^{(0)}=0$. It is crucial to note that the perturbation subspace associated with this unperturbed level is of \textit{infinite dimension} corresponding to all possible $(J_{A},J_{B})$ coupled pairs. In practical calculations, we take into account rotational quantum numbers $(J_{A},J_{B})$ up to a maximum $J_{\mathrm{max}}$ such that the desired convergence is reached on eigenvalues and/or eigenvectors of the operator $\hat{\mathbf{V}}$, see equation (\ref{eq:mol-oper-new}). This convergence is accelerated by the fact that the rotational splittings increase with $J_{A}$ and $J_{B}$.

To get a first insight into the effect of rotational couplings, we start with considering a two-channel model, including $(J_{A},J_{B})=(0,0)$ and $(1,1)$, for $M_{A}+M_{B}=0$. The corresponding basis of the perturbation space is $\{|0000\rangle$, $|1010\rangle$, $(|111-1\rangle\pm|1-111\rangle)\sqrt{2}\}$. The operator $\hat{\mathrm{V}}$ can thus be diagonalized analytically, leading to four PECs. If the label $q$ refers to the energy ordering of the eigenvalues, the lowest and highest ones, $V_{q=0}^{\textrm{BF}}(R)$ and $V_{q=3}^{\textrm{BF}}(R)$, read
\begin{eqnarray}
  V_{0}^{\textrm{BF}}(R) &=& 2B_{0}\left(1-\sqrt{1+\frac{R^{*6}}{6R^{6}}}\right)\, \nonumber \\
  V_{3}^{\textrm{BF}}(R) &=& 2B_{0}\left(1+\sqrt{1+\frac{R^{*6}}{6R^{6}}}\right)\,
  \label{eq:mol-bf-v0}
\end{eqnarray}
where $R^{*}$ is defined in equation (\ref{eq:mol-rstar}), while $V_{1}^{\textrm{BF}}(R) = V_{2}^{\textrm{BF}}(R) = 0$. As $R \rightarrow \infty$, the $V_{0}^{\textrm{BF}}(R)$ curve  converges towards $(J_{A},J_{B})=(0,0)$ with zero energy; its $R^{-6}$ asymptotic behavior, obtained by a Taylor expansion of equation (\ref{eq:mol-bf-v0}) assuming $R\gg R^{*}$, is the same as in equation (\ref{eq:mol-vdw-00}). The three other PECs have an asymptotic energy of $4B_{0}$. Their $R^{-6}$ asymptotic behavior is expected to be inaccurate as compared with Table \ref{tab:mol-c6-11}, because the pairs $(J_{A},J_{B})=(2,0)$ and $(0,2)$ are not considered in this simplified model. For $R\lesssim R^{*}$, equation (\ref{eq:mol-bf-v0}) reveals that the lowest PEC turns from a $R^{-6}$ into a $R^{-3}$ variation, namely $V_{0}^{\textrm{BF}}(R)\approx-\sqrt{2/3}\times B_{0}(R^{*}/R)^{3}=-\sqrt{2/3}\times d_{0}^{2}/R^{3}$, expressed in atomic units. In this region, the corresponding states are so strongly mixed by the dipole-dipole interaction that $(J_{A},J_{B})$  cannot be considered as {}``good'' quantum numbers. 

\begin{figure}
\begin{centering}
\includegraphics[width=8cm]{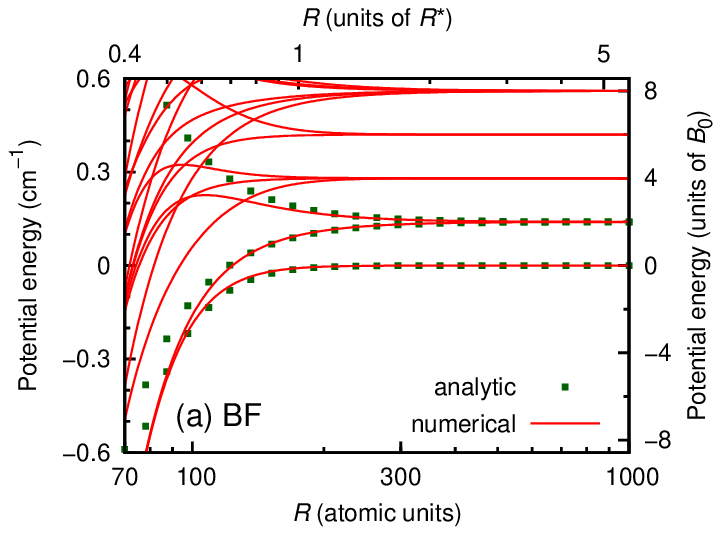}\includegraphics[width=8cm]{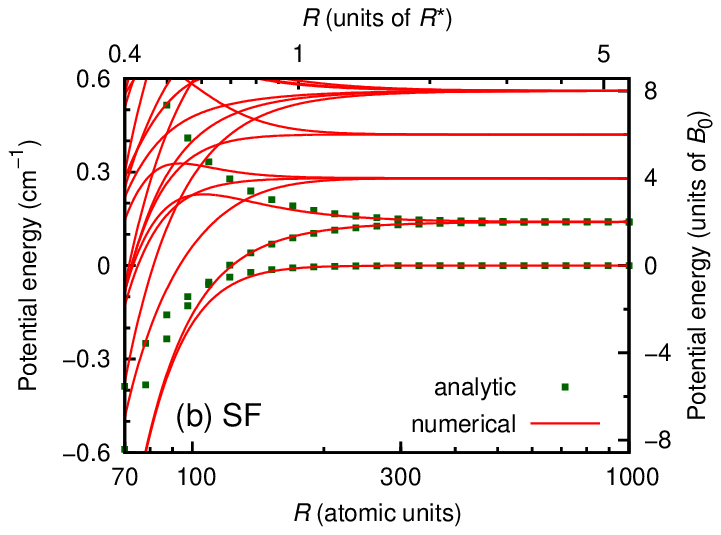}
\par\end{centering}
\caption{\label{fig:mol-pecs-bf-sf} Potential energy curves characterizing the interaction between two NaRb molecules in their vibrational ground level: (a) in the body-fixed frame with $M_A+M_B=0$; (b) in the space-fixed frame with $M=0$. The solid curves are obtained by numerical diagonalization while the symbols denote the estimates from the simplified model (see text).}
\end{figure}

Beyond this simplified model, accurate PECs are deduced from the numerical diagonalization of the matrix of the $\hat{\mathrm{V}}$ operator of equation (\ref{eq:mol-oper-new}), including the states built from $J_{A}$ and $J_{B}$ equal to 0 up to $J_{\textrm{max}}=6$. This value ensures a satisfactory convergence of the lowest PECs, which are displayed in figure \ref{fig:mol-pecs-bf-sf} (a). The qualitative picture established with the above simplified model is probed to be valid for the two lowest PECs. However the corresponding prefactor $-\sqrt{2/3}\times d_{0}^{2}$ predicted in equation (\ref{eq:mol-bf-v0}) is changed to $\approx -2/3\times d_{0}^{2}$, as given by a numerical fit of the PEC. For $R\ll R^{*}$, the two lowest PEC become degenerate with each other, which will be important in the presence of an external electric field.

One may wonder about the significance of the PECs plotted in figure \ref{fig:mol-pecs-bf-sf}. The analogy to electronic PECs for atom pairs can be invoked. However, the present PECs include the energy of the rotational motion of the individual molecules. They are so close in energy that, unlike electronic PECs, the dynamics of the complex at large distances may imply more than a single PEC. We would actually reach such a situation with interacting atom pairs if we included the atomic hyperfine structure in the hamiltonian \cite{tiesinga2005, orban2015}.

Moreover, the physical picture underlying the PECs of figure \ref{fig:mol-pecs-bf-sf} is valid only at sufficiently large distances. It has been successfully used to describe the structure and dynamics of weakly-bound, so-called \textit{van der Waals}, molecules \cite{vanderavoird1987, dubernet1994, dubernet1994b, wormer2000}, and more recently the collisions
of ultracold molecules \cite{idziaszek2010a, idziaszek2010b, julienne2011, quemener2011a}. In the latter case, the BF-frame PECs of figure \ref{fig:mol-pecs-bf-sf}, e.g.~their potential wells or barriers, can yield a very good physical insight into the dynamics. For $R\ll R^{*}$, although the multipolar expansion is still applicable according to the LeRoy criterion invoked in the introduction of the chapter, there are so many coupled $(J_{A},J_{B})$ pairs that the value of $J_{\mathrm{max}}$ necessary for a good convergence might be too high for a tractable numerical diagonalization of the matrix of the perturbation operator of equation (\ref{eq:mol-oper-new}). This reflects the fact the basis $\{|J_{A}M_{A}J_{B}M_{B}\rangle\}$ is not appropriate due to the strong anisotropy of the molecule-molecule interaction which favors a specific geometry of the complex. Therefore, in order to obtain reliable quantitative results for a wide range of distances, and also in the presence of external electromagnetic fields, it is more appropriate to describe the long-range interactions in the SF frame.

\subsection{Calculation in the space-fixed frame}
\label{ssec:mol-sf}

The perturbation operator $\hat{\mathrm{V}}$ in the SF frame,
\begin{equation}
  \hat{\mathrm{V}}(R) \equiv 
    B_{0}\left(\hat{\mathbf{J}}_{A}^{2}+\hat{\mathbf{J}}_{B}^{2}\right)
    + \hat{\mathrm{V}}_{\textrm{dd}}^{\textrm{SF}}(R)
    + \frac{\hbar^{2}\hat{\mathbf{L}}^{2}}{2\mu R^{2}}
    - \mathcal{E}\left(\hat{\mathrm{Q}}_{10}^{\mathrm{SF}}(A)
      + \hat{\mathrm{Q}}_{10}^{\mathrm{SF}}(B)\right)
  \label{eq:mol-oper-sf}
\end{equation}
involves the dipole-dipole interaction energy $\hat{\mathrm{V}}_{\textrm{dd}}^{\textrm{SF}}(R)$ written in the SF frame and obtained by setting $\ell_{A}=\ell_{B}=1$ in equation (\ref{eq:mult-vab-sf}). The third term accounts for the angular part of the intermolecular kinetic energy involving the end-over-end orbital angular momentum $\hat{\mathbf{L}}$ of the molecule complex with reduced mass $\mu$. The last term is the Stark energy due to an electric field of amplitude $\mathcal{E}$ polarized along the $z$-axis in the SF frame.

\subsubsection{The choice of basis for the unperturbed space}

The choice of an adequate basis is often crucial in molecular physics and in scattering physics. Up to now we naturally chose (symmetrized) products of states of each partner, adapted to the BF frame. But in fact, several choices are possible. The most obvious choice is to complete the set of individual quantum numbers $J_{A}$, $M_{A}$, $J_{B}$ and $M_{B}$ with the quantum numbers $L$ and $M_{L}$ associated with the orbital angular momentum $\hat{\mathbf{L}}$ of the complex (hereafter refereed to as \textit{partial wave}, as in scattering theory). This leads to the so-called \textit{uncoupled} basis $|J_{A}M_{A}J_{B}M_{B}LM_{L}\rangle$ in which each individual angular momentum of the problem is considered with its $z$-projection. Recalling that the wave function associated with $L$ and $M_{L}$ is the spherical harmonics $Y_{LM_{L}}(\Theta,\Phi)$, one can express the matrix elements of the SF perturbation operator $\hat{\mathrm{V}}_{AB}^{\textrm{SF}}(R)$, see equation (\ref{eq:mult-vab-sf}), as
\begin{align}
  & \left\langle \beta'_{A}J'_{A}M'_{A} \beta'_{B}J'_{B}M'_{B} 
    L'M'_{L} \right|
    \hat{\mathrm{V}}_{AB}^{\textrm{SF}}(R) 
    \left|\beta_{A}J_{A}M_{A} \beta_{B}J_{B}M_{B} 
    LM_{L} \right\rangle \nonumber \\
  & = \frac{1}{4\pi\epsilon_{0}} \sum_{\ell_{A}\ell_{B}\ell} 
    \frac{\left(-1\right)^{\ell_{B}}\delta_{\ell_A+\ell_B,\ell}}
      {R^{1+\ell}} 
    \sqrt{\frac{\left(2\ell\right)!}
               {\left(2\ell_{A}\right)!\left(2\ell_{B}\right)!}}
    \sqrt{\frac{2L+1}{2L'+1}} C_{L0\ell0}^{L'0}
    \frac{\left\langle \beta'_{A}J'_{A}\right\Vert 
          \hat{\mathrm{Q}}_{\ell_{A}} 
          \left\Vert \beta_{A}J_{A}\right\rangle}
         {\sqrt{2J'_A+1}} \nonumber \\
  & \times \frac{\left\langle \beta'_{B}J'_{B}\right\Vert 
      \hat{\mathrm{Q}}_{\ell_{B}}
      \left\Vert \beta_{B}J_{B}\right\rangle}
      {\sqrt{2J'_B+1}} 
    \sum_{mm_{A}m_{B}} \left(-1\right)^{m} 
    C_{\ell_{A}m_{A}\ell_{B}m_{B}}^{\ell m} 
    C_{LM_{L}\ell-m}^{L'M'_{L}}
    C_{J_{A}M_{A}\ell_{A}m_{A}}^{J'_{A}M'_{A}}
    C_{J_{B}M_{B}\ell_{B}m_{B}}^{J'_{B}M'_{B}}
  \label{eq:mol-vsf-elem}
\end{align}
where we used the property $Y_{\ell m}^{*}(\Theta,\Phi)=\left(-1\right)^{m}Y_{\ell-m}(\Theta,\Phi)$, and the identity of equation (\ref{eq:pert-3ylm}) to integrate the product of three spherical harmonics. In order to remain general, we restored the full notations $|\beta_{A} J_{A} M_{A} \rangle$, $|\beta_{B} J_{B} M_{B}\rangle$ and the reduced matrix elements $\langle \beta'_{A} J'_{A} \Vert \hat{\mathrm{Q}}_{1} \Vert \beta_{A} J_{A} \rangle$ and $\langle \beta'_{B} J'_{B} \Vert \hat{\mathrm{Q}}_{1} \Vert \beta_{B} J_{B} \rangle$ of the dipole-moment operators. It is important to note that $\hat{\mathrm{V}}_{AB}^{\textrm{SF}}$ couples the partial waves $L$ and $L'$ such that: $|L-\ell|\le L'\le\ell+L$ and $L+\ell+L'$ is even. In the dipole-dipole case $\ell_A=\ell_B=1$, this yields $L'=L$ or $L\pm 2$, excluding $(L,L')=(0,0)$.

An alternate choice of basis consists in accounting for the symmetries of the problem. In the absence of an electric field, $\mathcal{E}=0$, the total angular momentum $\hat{\mathbf{J}}$ of the complex,
\begin{equation}
  \hat{\mathbf{J}} = \hat{\mathbf{J}}_{A} + \hat{\mathbf{J}}_{B} + 
    \hat{\mathbf{L}} = \hat{\mathbf{J}}_{AB} + \hat{\mathbf{L}} ,
  \label{eq:mol-tot-am}
\end{equation}
is a constant of motion, and is characterized by the two quantum numbers $J$ and $M$. The resulting \textit{fully coupled}
angular-momentum basis set is composed of vectors 
\begin{align}
  \left|\left( \left(J_{A}J_{B}\right) J_{AB}L\right) 
   JM \right\rangle & =  
    \sum_{M_{AB}M_{L}} C_{J_{AB}M_{AB}LM_{L}}^{JM}
    \left|\left(J_{A}J_{B}\right)J_{AB}M_{AB}LM_{L}\right\rangle 
  \nonumber \\
  & = \sum_{M_{A}M_{B}M_{L}}C_{J_{AB}(M_{A}+M_{B})LM_{L}}^{JM} 
    C_{J_{A}M_{A}J_{B}M_{B}}^{J_{AB}(M_{A}+M_{B})}
    \left|J_{A}M_{A}J_{B}M_{B}LM_{L}\right\rangle . \nonumber \\
  \label{eq:mol-cpl-bas}
\end{align}
The basis built from $\{|(J_{A}J_{B})J_{AB}M_{AB}LM_{L}\rangle\}$ is referred to as \textit{partially coupled}. Note that the choice of the intermediate angular momentum $\hat{\mathbf{J}}_{AB}$ is not unique (see reference \cite{dubernet1994} for a detailed discussion of the atom-diatom case).
By combining equations (\ref{eq:prod-3Clebsch}), (\ref{eq:pert-4cg}), (\ref{eq:mol-vsf-elem}) and (\ref{eq:mol-cpl-bas}), as well as particular values of a $9j$ symbol with a vanishing argument (see \cite{varshalovich1988}, p~.357, Eq.~(2)), we can evaluate the matrix elements of the SF-frame dipole-dipole operator $\hat{\mathrm{V}}_{\textrm{dd}}^{\textrm{SF}}(R)$ in the fully coupled basis
\begin{eqnarray}
 &  & \left\langle \beta'_{A}\beta'_{B}\left(\left(J'_{A}J'_{B}\right)J'_{AB}L'\right)J'M'\right|\hat{\mathrm{V}}_{\textrm{dd}}^{\textrm{SF}}(R)\left|\beta_{A}\beta_{B}\left(\left(J_{A}J_{B}\right)J_{AB}L\right)JM\right\rangle \nonumber \\
 & = & -\frac{\sqrt{30}}{4\pi\epsilon_{0}R^{3}}\delta_{JJ'}\delta_{MM'}\left(-1\right)^{J+J_{AB}+L'}C_{L0\ell0}^{L'0}\left\{ \begin{array}{ccc}
J'_{A} & J'_{B} & J'_{AB}\\
J_{A} & J_{B} & J_{AB}\\
1 & 1 & 2
\end{array}\right\} \left\{ \begin{array}{ccc}
J_{AB} & L & J\\
L' & J'_{AB} & 2
\end{array}\right\} \nonumber \\
 & \times & \sqrt{\left(2L+1\right)\left(2J_{AB}+1\right)\left(2J'_{AB}+1\right)}\,\left\langle \beta'_{A}J'_{A}\right\Vert \hat{\mathrm{Q}}_{1}\left\Vert \beta_{A}J_{A}\right\rangle \left\langle \beta'_{B}J'_{B}\right\Vert \hat{\mathrm{Q}}_{1}\left\Vert \beta_{B}J_{B}\right\rangle \,. \nonumber \\
\label{eq:mol-vdd-sf}
\end{eqnarray}
Again, in order to remain general, we kept the full notation $|\beta_{A}J_{A}\rangle$ abd $|\beta_{B}J_{B}\rangle$. Note that, to obtain the interaction energy between any multipole moment, one can replace in equation (\ref{eq:mol-vdd-sf}) the tensor rank values 1 and 2, by $\ell_{A}$, $\ell_{B}$ and $\ell$. Also, the kinetic operator matrix is diagonal in this basis and its elements are equal to $\hbar^{2}L(L+1)/2\mu R^{2}$.

When the electric field is present along the $z$ axis, the total angular momentum projection $M$ is still a good quantum number, due to the cylindrical symmetry around the field direction. Although $J$ is not anymore a good quantum number, the field only couples $J$ and $J\pm1$ states, so that the use of the coupled basis remains appropriate for a weak field amplitude. By combining equations (\ref{eq:prod-3Clebsch}), (\ref{eq:mol-dip}) and (\ref{eq:mol-cpl-bas}), we obtain
\begin{eqnarray}
 &  & \left\langle \beta'_{A}\beta'_{B}\left(\left(J'_{A}J'_{B}\right)J'_{AB}L'\right)J'M'\right|\hat{\mathrm{Q}}_{10}^{\mathrm{SF}}(A)\left|\beta_{A}\beta_{B}\left(\left(J_{A}J_{B}\right)J_{AB}L\right)JM\right\rangle \nonumber \\
 & = & -\delta_{\beta_{B}\beta'_{B}}\delta_{J_{B}J'_{B}}\delta_{LL'}\delta_{MM'}\left(-1\right)^{J'_{A}+J_{B}+J_{AB}+J'_{AB}+L+J}\left\{ \begin{array}{ccc}
J_{A} & J_{B} & J_{AB}\\
J'_{AB} & 1 & J'_{A}
\end{array}\right\} \left\{ \begin{array}{ccc}
J_{AB} & L & J\\
J' & 1 & J'_{AB}
\end{array}\right\} \nonumber \\
 & \times & \sqrt{\left(2J_{AB}+1\right)\left(2J'_{AB}+1\right)\left(2J+1\right)}\, C_{JM10}^{J'M}\left\langle \beta'_{A}J'_{A}\right\Vert \hat{\mathrm{Q}}_{1}\left\Vert \beta_{A}J_{A}\right\rangle 
\label{eq:mol-dip-sf}
\end{eqnarray}
where again we restored the general notation $|\beta_{A}J_{A}\rangle$, $|\beta_{B}J_{B}\rangle$ and the reduced matrix element $\langle\beta'_{A}J'_{A}\Vert\hat{\mathrm{Q}}_{1}\Vert\beta_{A}J_{A}\rangle$ of the dipole-moment operator for molecule A. Obviously, the latter does not act on the quantum numbers of molecule B, nor on the partial-wave quantum numbers $L$ and $M_{L}$. The Wigner $6j$ symbols of equation (\ref{eq:mol-dip-sf}) impose the selection rules $|J_{AB}-1|\le J'_{AB}\le J_{AB}+1$ and $|J-1|\le J'\le J+1$.

\subsubsection{Dipole-dipole interaction without electric field}

We start from equation (\ref{eq:mol-oper-sf}) in the framework of the two-channel model achieved in section \ref{sssec:coupled-rot-levels}. In the fully coupled basis, the first channel is characterized by $J_{A}=J_{B}=0$ and $L=0$ ($s$-wave collision), which imposes $J_{AB}=J=M=0$. Being directly coupled to the first channel by dipole-dipole interaction, the second channel is therefore characterized by $J_{A}=J_{B}=1$ and $L=2$, following equation (\ref{eq:mol-vdd-sf}). Because the dipole-dipole interaction conserves $J$ and $M$ (equal to zero), the only possible value of $J_{AB}$
is 2. In the fully coupled basis, our two-channel model thus involves the two-dimensional subspace $\{|((00)00)00\rangle,|((11)22)00\rangle\}$. The lowest PEC obtained after diagonalization of the matrix of the operator (\ref{eq:mol-oper-sf}) reads
\begin{equation}
V_{0}^{\textrm{SF}}(R)=\left(2B_{0}+\frac{3\hbar^{2}}{2\mu R^{2}}\right)\left(1-\sqrt{1+\frac{2d_{0}^{4}}{3\left(2B_{0}+\frac{3\hbar^{2}}{2\mu R^{2}}\right)^{2}R^{6}}}\right)
\label{eq:mol-sf-v0}
\end{equation}
which is similar to its BF-frame counterpart visible in equation (\ref{eq:mol-bf-v0}), apart from the centrifugal term $L(L+1)\hbar^{2}/2\mu R^{2}=3\hbar^{2}/\mu R^{2}$ (for $L=2$). Here, this centrifugal term is much smaller than the rotational splittings of individual molecules. One can indeed evaluate the ratio $\hbar^{2}/B_{0}\mu R^{2}$ between the centrifugal term and the typical rotational spacing of individual molecules $B_{0}\approx\hbar^{2}/2\mu_{A}R_{e}^{2}$ depending on the reduced mass $\mu_{A}$ and $\mu_{B} = \mu_{A}$ of each molecule and on their equilibrium distance $R_{e}$. As $\mu_{A}^{-1} = \mu_{B}^{-1} = m_{1}^{-1}+m_{2}^{-1}$, where $m_{1}$ and $m_{2}$ are the masses of each atom composing the A and B diatomic molecules, and $\mu^{-1} = m_A^{-1} + m_B^{-1} = 2(m_1+m_2)^{-1}$, where $m_A = m_B = m_1+m_2$ is the masse of one molecule, this ratio reduces to
\begin{equation}
\frac{\hbar^{2}}{B_{0}\mu R^{2}}\approx\frac{4m_{1}m_{2}}{\left(m_{1}+m_{2}\right)^{2}}\left(\frac{R_{e}}{R}\right)^{2}=\left[1-\left(\frac{m_{1}-m_{2}}{m_{1}+m_{2}}\right)^{2}\right]\left(\frac{R_{e}}{R}\right)^{2}.
\label{eq:mol-ratio}
\end{equation}
One has necessarily $R\gg R_{e}$ in the long-range region since $R_{e}$ is by definition a distance at which chemical forces are important. The prefactor tends to 1 when $m_{1}\approx m_{2}$, while it remains much smaller than 1 when $m_{1}\ll m_{2}$ or $m{}_{2}\ll m_{1}$ (which tends to be the case for strongly polar molecules. So the ratio of equation (\ref{eq:mol-ratio}) is generally much smaller than unity. 

We plot in figure \ref{fig:mol-pecs-bf-sf} (b) the PECs resulting from the numerical diagonalization of equation (\ref{eq:mol-oper-sf}). These SF-frame PECs look like their BF counterparts of figure \ref{fig:mol-pecs-bf-sf} (a), so that the BF curves provide an enlightening view of the dynamics of the complex in absence of external electric field.

Another case which will be relevant when an electric field is present is $(J_{A},J_{B})=(1,0)$ and $(0,1)$. The basis vector $|((00)00)00\rangle$ is coupled to the vectors $|((10)10)10\rangle$ and $|((01)10)10\rangle$ by the Stark effect, see equation (\ref{eq:mol-dip-sf}). In turn, the latter are coupled to $|((10)12)10\rangle$ and $|((01)12)10\rangle$ by the dipole-dipole interaction (equation (\ref{eq:mol-vdd-sf})). The matrix of the perturbation operator can then be restricted to the two distinct two-dimensional subspaces built from the symmetric (+) and antisymmetric (-) combinations $\{[|((10)10)10\rangle\pm|((01)10)10\rangle]/\sqrt{2}$, $[|((10)12)10\rangle\pm|((01)12)10\rangle]/\sqrt{2}\}$. This yields
\begin{equation}
  \mathrm{V}^{\pm} = \left(\begin{array}{cc}
    2B_{0} & 0 \\
         0 & 2B_{0} \end{array}\right) 
    + \frac{d_{0}^{2}}{12\pi\epsilon_{0}R^{3}} \left(\begin{array}{cc}
              0 & \pm\sqrt{2} \\
    \pm\sqrt{2} & \mp1 \end{array}\right)
  \label{eq:mol-sf-10}
\end{equation}
where $\mathrm{V}^{+}$ and $\mathrm{V}^{-}$ correspond to the symmetric and antisymmetric combination, respectively. As announced above, we neglected the centrifugal term for the states with $L=2$. Remarkably, the eigenvalues of equation (\ref{eq:mol-sf-10}) are identical to the ones in the BF frame (see subsection \ref{ssec:mol-bf}), \textit{i.e.} $2B_{0}\pm2d_{0}^{2}/3R^{3}$
and $2B_{0}\pm d_{0}^{2}/3R^{3}$ in atomic units; the lowest one is symmetric upon the permutation of the two molecules.

\subsubsection{Application of the external electric field}

\begin{figure}
\begin{centering}
\includegraphics[width=8cm]{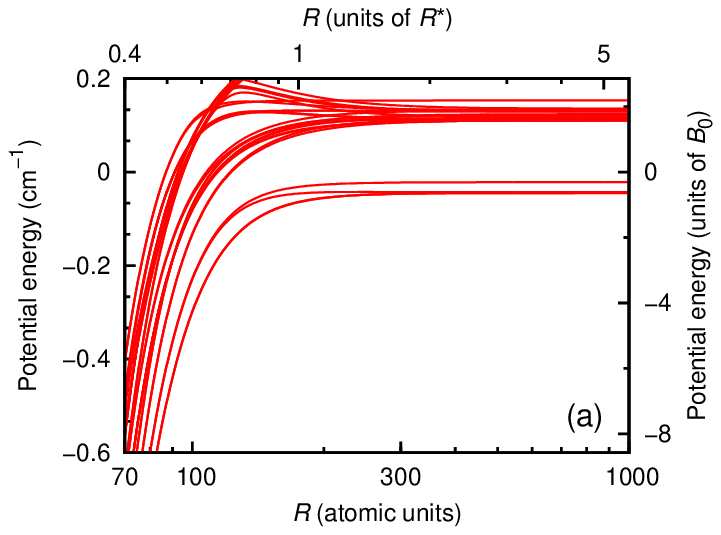}\includegraphics[width=8cm]{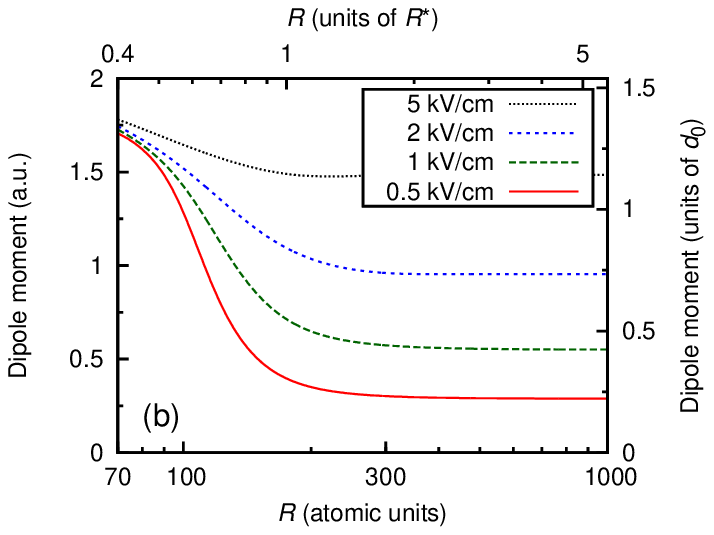}
\par\end{centering}
\caption{\label{fig:mol-field} Two NaRb molecules in an external electric field. (a) Potential-energy curves obtained for $M=0$ and $\mathcal{E}=2$~kV/cm; (b) Total induced dipole moment of the two molecules as a function of the intermolecular distance, and for various field amplitudes. It is shown both in atomic units for NaRb, and in units of $d_0$.}
\end{figure}

An electric field couples the PECs discussed above, \textit{i.e.} the PECs for $J=0$ are directly coupled to those for $J=1$, themselves coupled to those for $J=2$, and so on. In Figure \ref{fig:mol-field} (a), we plot the 20 lowest PECs obtained after the numerical diagonalization of equation (\ref{eq:mol-oper-sf}) for an experimentally realistic field amplitude $\mathcal{E}=2$~kV/cm. A satisfactory convergence was obtained with a basis characterized by $J=0$ to 3, $L=0$ to 8, and $J_1,\,J_2=0$ to $J_\mathrm{max}=6$. The {}``zero" of energy is the lowest dissociation energy in the field-free case, see figure \ref{fig:mol-pecs-bf-sf} (b).

In comparison with the field-free curves, the ones of figure \ref{fig:mol-field} (a) possess shifted asymptotes, which correspond to the sum of Stark energies of individual molecules. The curves are also denser, due to the large number of $L$ and $J$ values included in the basis. Like their field-free counterparts, the PECs of figure \ref{fig:mol-field} (a) show two distinct regions of intermolecular distances. For $R>R^*$, the curves conserved their individual character related to the quantum numbers of their dissociation limit, being either attractive or repulsive. In contrast, they all become attractive for $R<R^*$, due to the large dipole-dipole interaction between the molecules which dominate their interaction with the external electric field. This is confirmed by figure \ref{fig:mol-field} (b), where we plot the total induced dipole moment along the field direction $\langle\hat{\mathrm{Q}}_{10}(A)+\hat{\mathrm{Q}}_{10}(B)\rangle$ as a function of $R$ for the lowest PEC and for various electric-field amplitudes. These two regions are even more visible: for $R>R^*$ we obtain the sum of individual induced dipole moments, which of course increases with the electric field. Remarkably, for $R<R^*$, all the curves converge to the same value which does not depend anymore on the electric field. This strong induced dipole moment is related to the quasi-degeneracy between the two lowest PECs of figure \ref{fig:mol-pecs-bf-sf} (a) that we evoked earlier. Due to their very efficient coupling, their respective $|((00)00)00\rangle$ and $[|((10)10)10\rangle+|((01)10)10\rangle]/\sqrt{2}$ characters are strongly mixed  even at low field amplitudes. The molecules are strongly ''locked'' to each other and with the external field. This ''locking'' phenomenon, which was already reported in the BF frame and for field polarized along the inter-molecular axis \cite{lepers2013}, is thus also visible in the SF frame.

The strong $R$-variation of the induced dipole moment (see Figure \ref{fig:mol-field} (b)) is also interesting. It is likely to result in a significant transition dipole moment between a scattering state of the two molecules and a bound level of the tetramer whose rovibrational wave function is maximal around 100 to 200 atomic units. This opens the possibility to form ultracold polar tetramers by stimulated one-photon radiative association, along the lines proposed in Ref.~\cite{juarros2006}.

\section{Conclusion}
\label{sec:conclusion}

In this chapter we have exposed the general formalism for calculating the long-range interaction between atoms and molecules. This is clearly a widely studied topic of atomic and molecular physics, which recently attracted new interest as these interactions dominate the dynamics of ultracold quantum gases. In order to accurately evaluate the magnitude of such interactions, it is crucial to benefit from a detailed knowledge of the unperturbed individual particles, \textit{i.e.} their energy levels and their wave functions. This is the reason why we illustrated our derivation with two examples relevant for current ongoing research: (i) the long-range interaction between a pair of ultracold alkali-metal atoms with spin $1/2$, and (ii) between a pair of polar molecules in their ground $^1\Sigma^+$ state composed of two different alkali-metal atoms. In both cases, very precise spectroscopy and theoretical models are available for such species. The extension to more complex cases like the interaction between open-shell high-angular momentum atoms or between open-shell molecules in arbitrary rotational levels, while achievable starting from the present chapter, could easily fill in a complete book, and still deserves researches articles devoted to particular aspects. For instance the long-range interaction between a $^3P$ ground state oxygen atom and a $^3\Sigma^-$ ground state oxygen dimer, relevant for the understanding of ozone formation, has been investigated by the authors \cite{lepers2012}. This long-range data has been recently included in a global potential energy surface for the O$_3$ molecule \cite{dawes2013}. Another example concerns the long-range interaction between C($^3P$) or Si($^3P$) and OH($^2\Pi$) \cite{bussery-honvault2008, bussery-honvault2009}.

Despite their huge importance at very low collision energies, the long-range interactions do not tell the entire story to understand the dynamics of ultracold molecular collisions, as it is largely argued all along the present book. Indeed, due to their numerous degrees of freedom, the description of the dynamics at short intermolecular distances most often remains an open problem with a poor knowledge, except in a very limited number of cases like the examples above. Of particular interest are, again, the collisions between two polar alkali-metal diatomics \cite{byrd2010, byrd2012c}. It is admitted that among the series of ten possible heteronucler bialkali molecules, four of them do not react when colliding together in their absolute ground state, namely NaK, NaRb, KCs, RbCs \cite{zuchowski2010}. However one observes experimentally that the loss rate of such trapped ultracold dipolar molecules is unexpectedly much larger that the estimated one for elastic collisions \cite{takekoshi2014, molony2014, park2015, guo2016}, reaching the same order of magnitude as for reactive molecules \cite{ospelkaus2010a}. Several theoretical approaches have been proposed to embrace the complexity of their dynamics, based on long-range capture models and statistical arguments \cite{mayle2012,mayle2013,gonzalez-martinez2014}.

It should be noted finally that the validity of the present formalism is limited by the finite time taken by the electromagnetic field to propagate from one particle to the other, as pointed out by the seminal work of Casimir and Polder \cite{casimir1946, casimir1948}. Such a \textit{retardation} effect brings an energy correction at large distances proportional to $R^{-7}$ for two ground-state atoms. This correction motivated numerous theoretical investigations to evaluate its magnitude. One can cite for instance the work by Marinescu and You for alkali-metal-atom pairs \cite{marinescu1999a}.



\end{document}